\documentstyle[11pt]{article}
\newfont{\blackb}{msbm10 scaled\magstep1}
\def\Bbb#1{\hbox{\blackb #1}}
\newtheorem{one}{Proposition}

\newtheorem{three}{Lemma}

\newtheorem{five}{Theorem}

\newtheorem{six}{Theorem}

\newtheorem{seventh}{Proposition}

\newtheorem{nine}{Lemma}

\newtheorem{ten}{Lemma}

\newtheorem{twelve}{Proposition}

\newtheorem{thirteen}{Lemma}

\newtheorem{fourteen}{Proposition}

\newtheorem{fifteen}{Lemma}

\newtheorem{sixteen}{Lemma}

\newtheorem{nineteen}{Proposition}

\newtheorem{twenty}{Lemma}

\newtheorem{odeen}{Proposition}

\newtheorem{dva}{Proposition}

\newtheorem{cheterye}{Corollary}

\newtheorem{pyaty}{Corollary}

\newtheorem{tree}{Theorem}

\newtheorem{meg}{Proposition}

\newtheorem{yergoo}{Proposition}

\newtheorem{yereg}{Proposition}

\newtheorem{chorss}{Lemma}

\newtheorem{seep}{Proposition}

\newtheorem{rov}{Definition}

\newtheorem{eeh}{Definition}

\newtheorem{hme}{Lemma}

\newtheorem{tya}{Definition}

\newtheorem{lna}{Proposition}

\newtheorem{vly}{Lemma}

\newtheorem{pyt}{Proposition}

\newtheorem{eng}{Lemma}

\begin{document}
\baselineskip=14pt
\title{Leading Order Temporal Asymptotics of the 
Modified Non-Linear Schr\"{o}dinger Equation: 
Solitonless Sector}
\author{A. V. Kitaev\thanks{Supported by the 
Alexander Von Humboldt Foundation. E-mail: 
kitaev@gauss.pdmi.ras.ru.} \, 
and A. H. Vartanian\thanks{Partially supported 
by the Russian Academy of Sciences. E-mail: 
arthur@gauss.pdmi.ras.ru.} \\
Steklov Mathematical Institute \\
Fontanka 27 \\
St.~Petersburg 191011 \\
Russia}
\date{22 June 1997}
\maketitle 
\begin{abstract}
\noindent
Using the matrix Riemann-Hilbert factorisation approach 
for non-linear evolution equations (NLEEs) integrable in 
the sense of the inverse scattering method, we obtain, in 
the solitonless sector, the leading-order asymptotics as 
$t \! \rightarrow \! \pm \infty$ of the solution to the 
Cauchy initial-value problem for the modified non-linear 
Schr\"{o}dinger equation, $i \partial_{t} u \! + \! 
\frac{1}{2} \partial_{x}^{2} u \! + \! \vert u \vert^{2} u 
\! + \! i s \partial_{x} (\vert u \vert^{2} u) \! = \! 0$, 
$s \! \in \! \Bbb R_{> 0}$: also obtained are analogous 
results for two gauge-equivalent NLEEs; in particular, the 
derivative non-linear Schr\"{o}dinger equation, $i 
\partial_{t} q \! + \! \partial_{x}^{2} q \! - \! i 
\partial_{x} (\vert q \vert^{2} q) \! = \! 0$.

\vspace{0.95cm}
PACS (1996): 02.30.Jr, 02.30.Mv, 42.65.Tg, 42.81.-i. 

\vspace{0.55cm}
1991 Mathematics Subject Classification: Primary 35Q15, 
35Q55; Secondary 58F07, 

78A60.
\end{abstract}
\clearpage
\section{Introduction}
The study of the fundamental dynamical processes associated 
with the propagation of high-power ultrashort pulses in 
optical fibres is of paramount importance: the non-linear 
soliton(ic), or near-soliton(ic), operating mode(s) in such 
systems are very promising; in particular, very high data 
transmission rates, high noise immunity, and the accessibility 
of new frequency bands \cite{a1}. The classical, mathematical 
model for non-linear pulse propagation in the picosecond time 
scale in the anomalous dispersion regime in an isotropic, 
homogeneous, lossless, non-amplifying, polarisation-preserving
single-mode optical fibre is the non-linear Schr\"{o}dinger
equation, NLSE \cite{a1,a2}; however, in the 
subpicosecond-femtosecond time scale, experiments and theories 
on the propagation of high-power ultrashort pulses in long 
monomode optical fibres have shown that the NLSE is no longer 
valid and that additional non-linear terms (dispersive and 
dissipative) and higher-order linear dispersion should be taken 
into account \cite{a1}. In this case, subpicosecond-femtosecond 
pulse propagation is described (in dimensionless and normalised 
form) by the following non-linear evolution equation (NLEE) 
\cite{a1},
\begin{eqnarray}
i \partial_{\xi} u + \frac{1}{2} \partial^{2}_{\tau} u + 
\vert u \vert^{2} u + i s \partial_{\tau}(\vert u \vert^{2} 
u) = - i \widetilde{\Gamma} u + i \widetilde{\delta} 
\partial^{3}_{\tau} u + \frac{\tau_{n}}{\tau_{0}} u 
\partial_{\tau} (\vert u \vert^{2}),
\end{eqnarray}
where $u$ is the slowly varying amplitude of the complex 
field envelope, $\xi$ is the propagation distance along 
the fibre length, $\tau$ is the time measured in a frame 
of reference moving with the pulse at the group velocity 
(the retarded frame), $s$ $(> 0)$ governs the effects due 
to the intensity dependence of the group velocity 
(self-steepening), $\widetilde{\Gamma}$ is the intrinsic 
fibre loss, $\widetilde{\delta}$ governs the effects of 
the third-order linear dispersion, and $\tau_{n} / \tau_{0}$, 
where $\tau_{0}$ is the normalised input pulsewidth and 
$\tau_{n}$ is related to the slope of the Raman gain curve 
(assumed to vary linearly in the vicinity of the mean 
carrier frequency, $\omega_{0}$), governs the soliton 
self-frequency shift (SSFS) effect \cite{a3}. 

Our strategy to the study of (1) will be the following: 
(i) by setting the right-hand side of (1) equal to zero, 
we obtain the following equation (integrable in the sense 
of the inverse scattering method (ISM) \cite{a4}),
\begin{eqnarray}
i \partial_{t} u  + \frac{1}{2} \partial_{x}^{2} u + \vert 
u \vert^{2} u + i s \partial_{x} (\vert u \vert^{2} u) = 0,
\end{eqnarray}
which we hereafter call the modified non-linear Schr\"{o}dinger 
equation, MNLSE (note that, the physical variables, $\xi$ and 
$\tau$, have been mapped isomorphically onto the mathematical 
$t$ and $x$ variables, which are standard in the ISM context); 
and (ii) since $\widetilde{\Gamma}$, $\widetilde{\delta}$, and 
$\tau_{n} / \tau_{0}$ are small parameters \cite{a1}, we treat 
(1) as a non-integrable perturbation of the MNLSE.  

{}From the above discussion, it is clear that (multi-) 
soliton solutions of the MNLSE play a pivotal role in 
the physical context related to optical fibres. However, 
since practical lasers cannot be designed to excite only 
the soliton(ic) mode(s), but also excite an entire continuum 
of linear-like dispersive (radiative) waves, to have 
physically meaningful and practically representative results, 
it is necessary to investigate solutions of the MNLSE under 
general initial (launching, in the optical fibre literature 
\cite{a1}) conditions, without any artificial restrictions 
and/or constraints, which have both soliton(ic) and 
non-soliton(ic) (continuum) components; physically, it is 
towards such a solution that the initial pulse launched into 
an optical fibre is evolving asymptotically \cite{a5}. Also, 
one may think of general initial conditions (Cauchy data) 
as a perturbation of the so-called reflectionless potential 
(soliton(ic)) initial conditions, and, since the action of 
any perturbation is always more or less non-adiabatic, the 
soliton(ic) component of the solution will be accompanied 
by the appearance of a non-soliton(ic) component \cite{a6}. 
{}From the physical point of view, therefore, it is seminal 
to understand how the continuum and the soliton(s) interact. 
For several soliton-bearing equations, e.g., KdV, 
Landau-Lifshitz, and NLS, and the RMB (reduced Maxwell-Bloch) 
system, it is well known that the dominant $({\cal O}(1))$ 
asymptotic $(t \! \rightarrow \! \infty)$ effect of the 
continuous spectra on the multi-soliton solutions is a shift 
in phase and position of their constituent solitons \cite{a7}. 
The purpose of our studies is to derive an explicit functional 
form for the next-to-leading-order $({\cal O}(t^{-1/2}))$ 
term of the effect of this interaction for the MNLSE. 
Although there have been several papers devoted to studying 
soliton solutions of the MNLSE \cite{a8}, to the best of our 
knowledge, very little, if anything, is still known about 
its solution(s) for the class of non-reflectionless initial 
data. As is well known \cite{a9}, the investigation of 
multi-soliton solutions on the continuum background requires 
more sophisticated techniques, e.g., asymptotic, than, say, 
the algebraic methods for the construction of pure soliton 
solutions. An asymptotic investigation of the aforementioned 
solution can be divided into two stages: (i) the investigation 
of the continuum (solitonless) component of the solution 
\cite{a10}; and (ii) the inclusion of the soliton component 
via the application of a ``dressing'' procedure \cite{a11} 
to the continuum background. The purpose of this paper is to 
carry out, systematically, stage (i) of the above-mentioned 
asymptotic paradigm (since this phase of the asymptotic 
procedure is rather technical and long in itself, the 
completed results for stage (ii) are the subject of a 
forthcoming article \cite{a12}). The results obtained in 
this paper are formulated as Theorems~3.2 and 3.3.  

This paper is organised as follows. In Sec.~2, we transform 
(2) into a gauge-equivalent NLEE (see the first of~(4)), and, 
using only some basic facts concerning its direct scattering 
theory \cite{a13}, pose the inverse scattering problem as a 
matrix oscillatory Riemann-Hilbert (RH) factorisation problem. 
In Sec.~3, we: (i) briefly review the Beals and Coifman 
formulation \cite{a14} for the solution of a RH problem on an 
oriented contour; (ii) give a general outline of the Deift and 
Zhou procedure \cite{a15} for obtaining the long-time asymptotics 
of the solution to the RH problem; and (iii) state our final 
results as Theorems~3.2 and 3.3. In Sec.~4, the initial 
oscillatory RH problem is formulated as an auxiliary RH 
problem on an augmented contour. In Sec.~5, the auxiliary 
RH problem is reformulated as an equivalent RH problem on 
a truncated contour. In Sec.~6, it is shown that, to leading 
order as $t \! \rightarrow \! + \infty$, the solution of the 
equivalent RH problem on the truncated contour is equal, 
modulo decreasing terms, to the solution of an explicitly 
solvable model RH problem on a contour which consists of the 
disjoint union of three crosses. Finally, in Secs.~7 and 8, 
the model RH problem is solved, and the asymptotic solution 
of related, auxiliary NLEEs and the MNLSE are obtained.  
\section{The Riemann-Hilbert (RH) Problem} 
We begin by introducing some notation which is used throughout 
the paper: (i) for $D$ an unbounded domain of $\Bbb R \! \cup 
\! i \Bbb R$, let ${\cal S}(D)$ denote the Schwartz class on 
$D$, i.e., the class of smooth scalar-valued functions $f(x)$ 
on $D$ which together with all derivatives tend to zero faster 
than any positive power of $\vert x \vert^{-1}$ for $\vert x 
\vert \! \rightarrow \! \infty$; (ii) $\sigma_{3} \! = \! {\rm 
diag}(1,-1)$, $\sigma_{+} =$ 
$\left(\! \begin{array}{cc}
           0 & 1 \\
           0 & 0 
        \end{array} \! \right)$, $\sigma_{-} = 
\left(\! \begin{array}{cc}
           0 & 0 \\
           1 & 0 
        \end{array} \! \right)$, and $\sigma_{1} = \sigma_{-} 
+ \sigma_{+}$; (iii) for a scalar $\varpi$ and a $2 \times 2$ 
matrix $\Upsilon$, $\varpi^{{\rm ad}(\sigma_{3})} \Upsilon 
\equiv \varpi^{\sigma_{3}} \Upsilon \varpi^{-\sigma_{3}}$; (iv) 
for $k \in \{1,2\}$, ${\cal L}_{k}^{(2 \times 2)}(D) \equiv 
\{ \mathstrut F(\lambda); \, \lambda \in D, \, \mathstrut 
F_{ij}(\lambda) \in {\cal L}^{k} (D)$, $i,j \in \{1,2 \}\}$, 
where ${\cal L}^{k}(D) \equiv \{ \mathstrut f(\lambda); 
\mathstrut \lambda \in D, \, \vert \vert f \vert \vert_{{\cal 
L}^{k}(D)} \equiv (\int_{D} \vert f (\lambda) \vert^{k} \vert 
d \lambda \vert)^{1/k} \! < \infty \}$, and ${\cal L}_{
\infty}^{(2 \times 2)}(D) \equiv \{ \mathstrut G (\lambda); \, 
\lambda \in D, \, \vert \vert G_{ij} \vert \vert_{ {\cal L}^{
\infty}(D)} \equiv \sup\limits_{\lambda \in D} \vert G_{ij}
(\lambda) \vert < \infty, \, i,j \in \{1,2\}\}$, with the 
norms taken as follows, $\vert \vert (\cdot) \vert \vert_{
{\cal L}^{(2 \times 2)}_{n} (D)} \equiv \max\limits_{i,j \in 
\{1,2\}} \vert \vert (\cdot)_{ij} \vert \vert_{{\cal L}^{n}
(D)}$, $n \in \{1,2,\infty\}$; (v) $\overline{(\bullet)}$ 
denotes complex conjugation of $(\bullet)$; and (vi) ${\rm 
I}$ denotes the $2 \times 2$ identity matrix.

\setcounter{odeen}{0}
\begin{odeen}[\cite{a13}]
The necessary and sufficient condition for the compatibility 
of the following system of linear ODEs (the Lax pair) for 
arbitrary $\lambda \in \Bbb C$,
\begin{eqnarray}
&\partial_{x} \Psi(x,t;\lambda) = U(x,t;\lambda) \Psi(x,t;
\lambda), \, \, \, \, \, \, \, \partial_{t} \Psi(x,t;\lambda) 
= V(x,t;\lambda) \Psi(x,t;\lambda),& 
\end{eqnarray}
where
\begin{eqnarray*}
&U(x,t;\lambda) = \lambda (- i \lambda \sigma_{3} + P 
\sigma_{-} + Q \sigma_{+}) - \frac{i}{2} P Q \sigma_{3},& \\
&V(x,t;\lambda) \! = \! 2 \lambda^{2} U(x,t;\lambda) \! - \!
i \lambda ((\partial_{x} P) \sigma_{-} \! - \! (\partial_{x}
Q) \sigma_{+}) \! + \! (\frac{i}{4}(P Q)^{2} \! + \! \frac{
1}{2} (P \partial_{x} Q \! - \! Q \partial_{x} P)) \sigma_{
3},& 
\end{eqnarray*}
with $P(x,t) = \overline{Q(x,t)}$, is that $Q(x,t)$ $($resp. 
$P(x,t))$ satisfies the following NLEE,
\begin{eqnarray}\label{QP}
&i {\partial}_{t} Q + {\partial}_{x}^{2} Q + i Q^{2} 
{\partial}_{x} \overline{Q} + \frac{1}{2} Q \vert Q \vert^{4} 
= 0 \, \, \, \, \, \, \, \, (i {\partial}_{t} P - {\partial}_{
x}^{2} P + i P^{2} {\partial}_{x} \overline{P} - \frac{1}{2} P 
\vert P \vert^{4} = 0).&
\end{eqnarray}
\end{odeen}

{\em Proof.\/} (4) are the Frobenius compatibility conditions 
for (3). \ \ \rule{6pt}{6pt}

\setcounter{meg}{1}
\begin{meg}
Let $Q(x,t)$ be a solution of the first of Eqs.~(4). Then 
there exists a corresponding solution of system~(3) such 
that $\Psi(x,t;0)$ is a diagonal matrix.
\end{meg}

{\em Proof.\/} For given $Q(x,t)$, let $\widehat{\Psi}(x,t;
\lambda)$ be a solution of (3) which exists in accordance with 
Proposition~2.1. Setting $\lambda = 0$ in (3), we get that 
$\widehat{\Psi}(x,t;0) = \exp \{- \frac{i \sigma_{3}}{2} \! 
\int_{x_{0}}^{x} \vert Q(\xi,t) \vert^{2} d \xi\} 
\linebreak[4] \cdot \widehat{{\cal K}}$, for some $x_{0} \! 
\in \! \Bbb R$ and non-degenerate matrix $\widehat{{\cal K}}$ 
which is independent of $x$ and $t$. The function $\Psi(x,t;
\lambda) \equiv \widehat{\Psi}(x,t;\lambda) \widehat{{\cal 
K}}^{-1}$ is the solution of (3) which is diagonal at 
$\lambda = 0$. \ \ \rule{6pt}{6pt}

\setcounter{yergoo}{2}
\begin{yergoo}[\cite{a16}]
Let $Q(x,t)$ be a solution of the first of Eqs.~(4) and $\Psi
(x,t;\lambda)$ the corresponding solution of system~(3) given 
in Proposition~2.2. Set $\Psi_{q}(x,t;\lambda) \! \equiv \! 
\Psi^{-1}(x,t;0) \linebreak[4] \cdot \Psi(x,t;\lambda)$. 
Then
\begin{eqnarray}
\partial_{x} \Psi_{q}(x,t;\lambda) = {\cal U}_{q}(x,t;\lambda) 
\Psi_{q}(x,t;\lambda), \, \, \, \, \, \, \,
\partial_{t} \Psi_{q}(x,t;\lambda) = {\cal V}_{q}(x,t;\lambda) 
\Psi_{q}(x,t;\lambda),& 
\end{eqnarray}
where
\begin{eqnarray}
&{\cal U}_{q}(x,t;\lambda) = - i \lambda^{2} \sigma_{3} + 
\lambda (\overline{q} \sigma_{-} + q \sigma_{+}),& 
\end{eqnarray}
\begin{eqnarray} 
&{\cal V}_{q}(x,t;\lambda) = \left(\! \begin{array}{cc}
- 2 i \lambda^{4} - i \lambda^{2} \overline{q} q & 2 
\lambda^{3} q + i \lambda \partial_{x} q + \lambda \overline{q} 
q^{2} \\
2 \lambda^{3} \overline{q} - i \lambda \partial_{x} \overline{q} 
+ \lambda \overline{q}^{2} q & 2 i \lambda^{4} + i \lambda^{2} 
\overline{q} q 
\end{array} \! \right) \!,& 
\end{eqnarray}
with 
\begin{eqnarray}
q(x,t) \equiv Q(x,t)((\Psi^{-1}(x,t;0))_{11})^{2}, 
\end{eqnarray}
is the ``Kaup-Newell'' Lax pair for the derivative non-linear 
Schr\"{o}dinger equation, DNLSE {\rm \cite{a17}},
\begin{eqnarray}
i {\partial}_{t} q + {\partial}_{x}^{2} q - i {\partial}_{x} 
(\vert q \vert^{2} q) = 0.
\end{eqnarray}
\end{yergoo}

{\em Proof.\/} Differentiating $\Psi_{q} (x,t;\lambda) \equiv 
\Psi^{-1}(x,t;0) \Psi(x,t;\lambda)$ with respect to $x$ and 
$t$ and using the fact that $\Psi(x,t;0) = \exp \{- \frac{i 
\sigma_{3}}{2} \! \int_{x_{0}}^{x} \vert Q(\xi,t) \vert^{2} 
d \xi\}$, for some $x_{0} \in \Bbb R$, and $\Psi (x,t;
\lambda)$ satisfy (3) for $\lambda \! = \! 0$ and $\lambda 
\in \Bbb C \! \setminus \! \{0\}$, respectively, defining 
$q(x,t)$ as in (8), we get that $\Psi_{q}(x,t;\lambda)$ 
satisfies (5), where ${\cal U}_{q}(x,t;\lambda)$ and ${\cal 
V}_{q}(x,t;\lambda)$ are given by (6) and (7): (9) is the 
compatibility condition for (5). \ \ \rule{6pt}{6pt}   

\setcounter{one}{3}
\begin{one}
If $q(x,t)$ is a solution of the DNLSE (Eq.~(9)) such 
that $q(x,0) \in {\cal S}(\Bbb R)$, then $u(x,t) = 
\frac{1}{\sqrt{2s}} \exp \{\frac{i}{s}(x - \frac{t}{2 s})\} 
q(\frac{t}{s} - x,\frac{t}{2})$ satisfies the MNLSE (Eq.~(2)) 
with $u(x,0) \in {\cal S}(\Bbb R)$.
\end{one}

{\em Proof.\/} Direct substitution. \ \ \rule{6pt}{6pt}

\setcounter{tya}{0}
\begin{tya}
For $Q(x,t)$, as a function of $x$, $\in$ ${\cal S}(\Bbb 
R)$, define the vector functions $\underline{\psi}^{\pm}
(x,t;\linebreak[4] \lambda)$ and $\underline{\phi}^{\pm}
(x,t;\lambda)$ as the Jost solutions of the first equation 
of system~(3), namely, $(\partial_{x} - U(x,t;\lambda)) 
\underline{\psi}^{\pm}(x,t;\lambda) = 0$ and $(\partial_{x} 
- U(x,t;\lambda)) \underline{\phi}^{\pm}(x,t;\lambda) = 0$, 
with the following asymptotics,
\begin{eqnarray}
&\lim\limits_{x \rightarrow + \infty} \underline{\psi}^{+} 
(x,t;\lambda) = (0,1)^{{\rm T}} e^{i \lambda^{2} x + 2 i 
\lambda^{4} t}, \, \, \, \, \, \, \, \, \, \lim\limits_{x 
\rightarrow + \infty} \underline{\psi}^{-} (x,t;\lambda) = 
(1,0)^{{\rm T}} e^{-i \lambda^{2} x - 2 i \lambda^{4} t},& 
\nonumber \\
&\lim\limits_{x \rightarrow - \infty} \underline{\phi}^{+} 
(x,t;\lambda) = (1,0)^{{\rm T}} e^{-i \lambda^{2} x - 2 i 
\lambda^{4} t}, \, \, \, \, \, \, \, \, \, \lim\limits_{x 
\rightarrow - \infty} \underline{\phi}^{-} (x,t;\lambda) = 
(0,-1)^{{\rm T}} e^{i \lambda^{2} x + 2 i \lambda^{4} t},& 
\nonumber
\end{eqnarray}
where the superscripts $\pm$ mean $\Im(\lambda^{2}) {> 
\atop <} 0$, and ${\rm T}$ denotes transposition.
\end{tya}

\setcounter{pyt}{4}
\begin{pyt}
The Jost solutions (Definition~2.1) have the following 
properties: (i) 
\begin{eqnarray}
&\underline{\psi}^{\pm} (x,t;\lambda) = \mp a^{\pm}
(\lambda,t) \underline{\phi}^{\mp} (x,t;\lambda) + b^{\mp} 
(\lambda,t) \underline{\phi}^{\pm} (x,t;\lambda), \, \, \, 
\, \, \, \, \Im(\lambda^{2}) =0,& 
\nonumber 
\end{eqnarray}
where $a^{\pm}(\lambda,t) = W(\underline{\phi}^{\pm}(x,t;
\lambda),\underline{\psi}^{\pm} (x,t;\lambda))$, $b^{\pm}
(\lambda,t) = \mp W(\underline{\phi}^{\pm}(x,t;\lambda),
\underline{\psi}^{\mp}(x,t;\lambda))$, and $W(f,g)$ is the 
Wronskian of $f$ and $g$, $W(f,g) \equiv f_{1} g_{2} - f_{
2} g_{1}$; (ii) if $\underline{\psi}^{\pm} (x,t;\lambda)$ 
and $\underline{\phi}^{\pm}(x,t;\lambda)$ satisfy the second 
equation of system~(3) as well, i.e., $(\partial_{t} - V(x,
t;\lambda)) \underline{\psi}^{\pm}(x,t;\lambda) = 0$ and 
$(\partial_{t} - V(x,t;\lambda)) \underline{\phi}^{\pm}(x,
t;\lambda) = 0$, then $a^{\pm}(\lambda,t) = a^{\pm}(\lambda,
0) \equiv a^{\pm}(\lambda)$, and $b^{\pm}(\lambda,t) = b^{
\pm}(\lambda,0) e^{\pm 4 i \lambda^{4} t}$ $\equiv b^{\pm}
(\lambda) e^{\pm 4 i \lambda^{4} t}$; (iii) for $\Im(\lambda^{
2}) = 0$, $a^{+}(\lambda) a^{-}(\lambda) + b^{+}(\lambda) 
b^{-}(\lambda) = 1$, $a^{+}(\lambda,t) = \overline{a^{-} 
(\overline{\lambda},t)}$, $b^{+}(\lambda,t) = - \overline{
b^{-}(\overline{\lambda},t)}$, $a^{\pm}(-\lambda) = a^{\pm}
(\lambda)$, $b^{\pm}(\lambda) = - b^{\pm}(-\lambda)$, and 
$\frac{1}{2} \ln \! \left(\frac{a^{+}(\lambda)}{a^{-}(\lambda)} 
\right) =$ $- \int_{\widehat{\Gamma}} \frac{\mu \ln (1 + r^{+} 
(\mu) r^{-} (\mu))}{(\mu^{2} - \lambda^{2})} \frac{d \mu}{2 
\pi i}$, with $r^{\pm}(\lambda,t) = b^{\pm}(\lambda,t)/a^{\pm}
(\lambda,t) \equiv r^{\pm}(\lambda) e^{\pm 4 i \lambda^{4} t}$,
and $\widehat{\Gamma} = \{ \mathstrut \lambda;$ $\Im(\lambda^{
2}) = 0\}$ (oriented as in Fig.~1);
\begin{figure}[bht]
\begin{center}
\unitlength=1cm
\vspace{-1.0cm} 
\begin{picture}(6,6)(0,0) 
\thicklines 
\put(2.5,0){\vector(0,1){1.25}}
\put(2.25,1.25){\makebox(0,0){$\scriptstyle{}+$}}
\put(2.75,1.25){\makebox(0,0){$\scriptstyle{}-$}}
\put(2.5,1.25){\line(0,1){1.25}}
\put(2.5,5){\vector(0,-1){1.25}}
\put(2.5,2.5){\line(0,1){1.25}}
\put(2.25,3.75){\makebox(0,0){$\scriptstyle{}-$}}
\put(2.75,3.75){\makebox(0,0){$\scriptstyle{}+$}}
\put(2.5,2.5){\vector(-1,0){1.25}}
\put(3.75,2.75){\makebox(0,0){$\scriptstyle{}+$}}
\put(3.75,2.25){\makebox(0,0){$\scriptstyle{}-$}}
\put(1.25,2.5){\line(-1,0){1.25}}
\put(1.25,2.75){\makebox(0,0){$\scriptstyle{}-$}}
\put(1.25,2.25){\makebox(0,0){$\scriptstyle{}+$}}
\put(2.5,2.5){\vector(1,0){1.25}}
\put(3.75,2.5){\line(1,0){1.25}}
\put(4.0,3.75){\makebox(0,0){$\scriptstyle{}\Im 
(\lambda^{2}) \, > \, 0$}}
\put(4.0,1.25){\makebox(0,0){$\scriptstyle{}\Im 
(\lambda^{2}) \, < \, 0$}}
\put(1.0,1.25){\makebox(0,0){$\scriptstyle{}\Im 
(\lambda^{2}) \, > \, 0$}}
\put(1.0,3.75){\makebox(0,0){$\scriptstyle{}\Im 
(\lambda^{2}) \, < \, 0$}}
\put(4.75,2.75){\makebox(0,0){$\scriptstyle{}\Re 
(\lambda)$}}
\put(3.0,4.75){\makebox(0,0){$\scriptstyle{}\Im 
(\lambda)$}}
\end{picture}
\vspace{-0.725cm}
\end{center}
\caption{Continuous spectrum ${\widehat \Gamma}$.}
\end{figure}
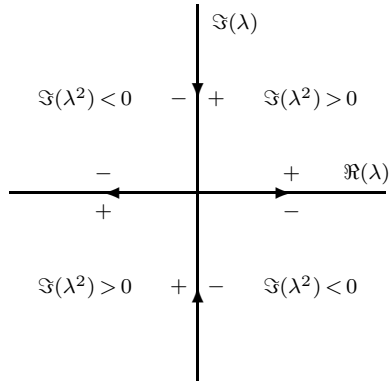
(iv) $a^{\pm}(\lambda)$ are analytically continuable to 
$\Im(\lambda^{2}) {> \atop <} 0$ with the following 
integral representation, $\pm \ln (a^{\pm}(\lambda)) = 
- \int_{\widehat{\Gamma}} \frac{\mu \ln (1 + r^{+}(\mu) 
r^{-} (\mu))}{(\mu^{2} - \lambda^{2})} \frac{d \mu}{2 
\pi i}$; and (v) for $\lambda \rightarrow \infty$, 
$\Im(\lambda^{2}) > 0$, $a^{+}(\lambda,t) = 1 + {\cal 
O}(\lambda^{-2})$, 
\begin{eqnarray}
&\underline{\psi}^{+}(x,t;\lambda) e^{-i \lambda^{2} x - 2 i 
\lambda^{4} t} = \left(\frac{Q(x,t)}{2 \lambda},1 \right)^{{
\rm T}} + {\cal O}(\lambda^{-2}),& \nonumber \\
&\underline{\phi}^{+}(x,t;\lambda) e^{i \lambda^{2} x + 2 i 
\lambda^{4} t} = \left(1,\frac{P(x,t)}{2 \lambda} \right)^{{
\rm T}} + {\cal O}(\lambda^{-2}),& \nonumber 
\end{eqnarray}
and, for $\lambda \rightarrow \infty$, $\Im(\lambda^{2}) < 0$, 
$a^{-}(\lambda,t) = 1 + {\cal O}(\lambda^{-2})$, 
\begin{eqnarray}
&\underline{\psi}^{-}(x,t;\lambda) e^{i \lambda^{2} x + 2 i 
\lambda^{4} t} = \left(1,\frac{P(x,t)}{2 \lambda} \right)^{{
\rm T}} + {\cal O}(\lambda^{-2}),& \nonumber \\
&\underline{\phi}^{-}(x,t;\lambda) e^{-i \lambda^{2} x - 2 i 
\lambda^{4} t} = - \left(\frac{Q(x,t)}{2 \lambda},1 \right)^{{
\rm T}} + {\cal O}(\lambda^{-2}).& \nonumber 
\end{eqnarray}
\end{pyt}

{\em Proof.\/} See \cite{a18}. \ \ \rule{6pt}{6pt}

{\em Remark 2.1.\/} We now adopt a convention which is 
adhered to {\em sensus strictu\/} throughout the paper: 
for each segment of an oriented contour, according to 
the given orientation, the ``$+$'' side is to the left 
and the ``$-$'' side is to the right as one traverses 
the contour in the direction of the orientation; hence, 
$(\cdot)_{+}$ and $(\cdot)_{-}$ denote, respectively, 
the non-tangential limits (boundary values) of $(\cdot)$ 
on an oriented contour {}from the ``$+$'' (left) side 
and ``$-$'' (right) side (this should \underline{{\bf 
not}} be confused with the $\pm$ superscripts).

As a result of Propositions 2.1--2.4, for the asymptotic 
analysis of the solution to the MNLSE, it is enough to 
investigate asymptotically the solutions of (4) and the 
corresponding function $\Psi(x,t;0)$ (Proposition~2.2): 
the main tool for this analysis is the following matrix 
RH factorisation problem.
\setcounter{three}{0}
\begin{three}
Let $Q(x,t)$, as a function of $x$, $\in \! {\cal S}(\Bbb 
R)$. Set $\Psi(x,t;\lambda) \! \equiv \! m(x,t;\lambda) 
\exp \{- i(\lambda^{2} x \linebreak[4] + 2 \lambda^{4} t) 
\sigma_{3}\}$. Define: (i) for $\Im (\lambda^{2}) > 0$,  
\begin{eqnarray}
&m(x,t;\lambda) \equiv
\left( \begin{array}{cc}
\frac{\underline{\phi}_{1}^{+}(x,t;\lambda)}{a^{+}
(\lambda,t)} & 
\underline{\psi}_{1}^{+}(x,t;\lambda) \\
\frac{\underline{\phi}_{2}^{+}(x,t;\lambda)}{a^{+}
(\lambda,t)} & 
\underline{\psi}_{2}^{+}(x,t;\lambda)
\end{array} \right) \! \exp \{i (\lambda^{2} x + 2 
\lambda^{4} t) \sigma_{3} \};& \nonumber
\end{eqnarray}
(ii) for $\Im (\lambda^{2}) < 0$, 
\begin{eqnarray}
&m(x,t;\lambda) \equiv
\left( \begin{array}{cc}
\underline{\psi}_{1}^{-} (x,t;\lambda) & - 
\frac{\underline{\phi}_{1}^{-} 
(x,t;\lambda)}{a^{-}(\lambda,t)} \\
\underline{\psi}_{2}^{-} (x,t;\lambda) & - 
\frac{\underline{\phi}_{2}^{-} 
(x,t;\lambda)}{a^{-}(\lambda,t)} 
\end{array} \right) \! \exp \{ i (\lambda^{2} x + 2 
\lambda^{4} t) \sigma_{3} \};& \nonumber
\end{eqnarray}
and (iii) $r(\lambda) \equiv r^{-}(\lambda)$. Then the $2 
\times 2$ matrix-valued function $m(x,t;\lambda)$ $(\det 
(m(x,t;\lambda)) \! = \! 1)$ solves the following RH 
problem: 
\begin{enumerate}
\item[a.] $m(x,t;\lambda)$ is holomorphic $\forall$ $\lambda 
\in \Bbb C \setminus \widehat{\Gamma};$ 
\item[b.] $m(x,t;\lambda)$ satisfies the following jump 
conditions,
\begin{eqnarray}
m_{+}(x,t;\lambda) = m_{-}(x,t;\lambda) \exp \{-i (\lambda^{2} 
x + 2 \lambda^{4} t) {\rm ad}(\sigma_{3})\} G(\lambda), \, \, 
\, \, \, \lambda \in \widehat{\Gamma}, \nonumber 
\end{eqnarray}
where
\begin{eqnarray}
G(\lambda) = 
\left(\! \begin{array}{cc}
1 - \overline{r(\overline{\lambda})} r(\lambda) & 
r(\lambda) \\
- \overline{r(\overline{\lambda})} & 1
\end{array} \! \right) \!, \nonumber
\end{eqnarray}
$r(\lambda) \in {\cal S}(\widehat{\Gamma})$, and 
$r(- \lambda) = - r(\lambda);$
\item[c.] as $\lambda \rightarrow \infty$, $\lambda \in 
\Bbb C \setminus \widehat{\Gamma}$,
\begin{eqnarray*}
m(x,t;\lambda) = {\rm I} + {\cal O}(\lambda^{-1}).
\end{eqnarray*}
\end{enumerate}
\end{three}

{\em Proof.\/} Follows {}from the definitions of $m(x,t;
\lambda)$ for $\Im(\lambda^{2}) {> \atop <} 0$ given in 
the Lemma and the properties of the Jost solutions and 
scattering data for~(3) given in Definition~2.1 and 
Proposition~2.5: for details, see \cite{a19}. 
\ \ \rule{6pt}{6pt}

\setcounter{eng}{1}
\begin{eng}
Let $\vert \vert r \vert \vert_{{\cal L}^{\infty}(\widehat{
\Gamma})} \equiv \sup\limits_{\lambda \in \widehat{\Gamma}} 
\vert r(\lambda) \vert < 1$. Then: (i) the RH problem 
formulated in Lemma~2.1 is uniquely solvable; (ii) $\Psi(x,t;
\lambda) \! \equiv \! m(x,t;\lambda) \exp \{- i(\lambda^{2} 
x \! + \! 2 \lambda^{4} t) \sigma_{3}\}$ is the solution of 
system~(3) with
\begin{eqnarray}
Q(x,t) \equiv 2 i \! \lim\limits_{\lambda \rightarrow \infty}
(\lambda m(x,t;\lambda))_{12} \, \, \, \, \, \, {\rm and} \, 
\, \, \, \, \, P(x,t) \equiv - 2 i \! \lim\limits_{\lambda 
\rightarrow \infty} (\lambda m(x,t;\lambda))_{21};
\end{eqnarray}
(iii) the functions $Q(x,t)$ and $P(x,t)$ defined by Eqs.~(10) 
satisfy Eqs.~(4), and
\begin{eqnarray}
q(x,t) \equiv Q(x,t)((m^{-1}(x,t;0))_{11})^{2}
\end{eqnarray}
satisfies the DNLSE (Eq.~(9)); and (iv) $m(x,t;\lambda)$ 
possesses the following symmetry reductions, $m(x,t;
- \lambda) = \sigma_{3} m(x,t;\lambda) \sigma_{3}$ 
and $m(x,t;\lambda) = \sigma_{1} \overline{m(x,t;
\overline{\lambda})} \sigma_{1}$.
\end{eng}

{\em Proof.\/} The solvability of the RH problem $\forall \, 
t$ follows from the condition $\vert \vert r \vert \vert_{
{\cal L}^{\infty}(\widehat{\Gamma})} \! < \! 1$ and the 
discussions in \cite{a18,a20} (see, also, Zhou's skew Schwarz 
reflection invariant symmetry arguments \cite{a21,a22}). The 
fact that $q(x,t)$ defined by (11) satisfies the DNLSE follows 
{}from Proposition~2.3. \ \ \rule{6pt}{6pt}
\section{Solution Procedure and Summary of Results}
Prior to describing the Deift and Zhou procedure \cite{a15} 
for obtaining the leading-order temporal asymptotics of (4), 
let us briefly review the Beals and Coifman formulation 
\cite{a14} for the solution of a RH problem on an oriented 
contour $\Xi$. The RH problem (in the absence of a discrete 
spectrum) on $\Xi$ consists of finding a $2 \times 2$ 
matrix-valued function $m(\lambda)$ such that: (i) $m
(\lambda)$ is holomorphic $\forall$ $\lambda \in \Bbb C 
\setminus \Xi$; (ii) $m_{+}(\lambda) = m_{-}(\lambda) v
(\lambda)$ $\forall \, \lambda \in \Xi$, for some jump 
matrix $v(\lambda) \colon \Xi \rightarrow {\rm Mat}(2,
\Bbb C)$; and (iii) as $\lambda \rightarrow \infty$, 
$\lambda \in \Bbb C \setminus \Xi$, $m(\lambda) 
\rightarrow {\rm I}$. Writing the jump matrix, 
$v(\lambda)$, in the following factorised form, $v(\lambda) 
\equiv ({\rm I} - w_{-}(\lambda))^{-1} ({\rm I} + w_{+}
(\lambda))$ $\forall$ $\lambda \in \Xi$, where $w_{\pm}
(\lambda) \in {\cal L}^{(2 \times 2)}_{2} (\Xi) \cap {\cal 
L}^{(2 \times 2)}_{\infty}(\Xi)$, with $\vert \vert w_{\pm} 
\vert \vert_{{\cal L}^{(2 \times 2)}_{2}(\Xi) \cap {\cal 
L}^{(2 \times 2)}_{\infty}(\Xi)}$ $\equiv$ $\vert \vert 
w_{\pm} \vert \vert_{{\cal L}^{(2 \times 2)}_{2}(\Xi)}$ $+$ 
$\vert \vert w_{\pm} \vert \vert_{{\cal L}^{(2 \times 2)}_{
\infty}(\Xi)}$, are off-diagonal upper/lower triangular 
matrices, denote $w = w_{+} + w_{-}$, and introduce the 
operator $C_{w}$ on ${\cal L}^{(2 \times 2)}_{2} (\Xi)$ as 
follows, 
\begin{eqnarray}
C_{w} f \equiv C_{+} (f \, w_{-}) + C_{-} (f \, w_{+}), 
\end{eqnarray}
where $f \! \in \! {\cal L}^{(2 \times 2)}_{2}(\Xi)$, 
and $C_{\pm} \colon {\cal L}^{(2 \times 2)}_{2}(\Xi) \! 
\rightarrow \! {\cal L}^{(2 \times 2)}_{2}(\Xi)$ denote 
the Cauchy operators (bounded monomorphisms),
\begin{eqnarray}
(C_{\pm} f)(\lambda) = \lim\limits_{{\lambda^{\prime} 
\rightarrow \lambda \atop \lambda^{\prime} \, \in \, \pm \, 
{\rm side} \, {\rm of} \, \Xi}} \int_{\Xi} \frac{f(\xi)}{
(\xi - \lambda^{\prime})} \frac{d \xi}{2 \pi i}. \nonumber
\end{eqnarray}
\setcounter{five}{0}
\begin{five}[\cite{a14}] 
If $\mu(\lambda)$ $(= m_{+}(\lambda)({\rm I} - w_{+}(\lambda)) 
= m_{-}(\lambda)({\rm I} + w_{-}(\lambda)))$ $\in$ ${\cal L}^{
(2 \times 2)}_{2}(\Xi)$ $+$ ${\cal L}^{(2 \times 2)}_{\infty}
(\Xi)$ solves the following linear singular integral equation,
\begin{eqnarray}
(\underline{{\bf Id}} - C_{w}) \mu = {\rm I}, 
\end{eqnarray}
where $\underline{{\bf Id}}$ is the identity operator on ${\cal 
L}^{(2 \times 2)}_{2}(\Xi) + {\cal L}^{(2 \times 2)}_{\infty} 
(\Xi)$, then the solution of the RH problem for $m(\lambda)$ is 
\begin{eqnarray}
m(\lambda) = {\rm I} + \int_{\Xi} \frac{\mu(\xi) w(\xi)}{
(\xi - \lambda)} \frac{d \xi}{2 \pi i}, \, \, \, \, \, \,
\lambda \in \Bbb C \setminus \Xi.
\end{eqnarray}
\end{five}

{}From Theorem~3.1 and (10), we obtain the following integral 
representation for $P(x,t)$ $(Q(x,t) = \overline{P(x,t)})$ in 
terms of the resolvent kernel of the linear singular integral 
equation (13), 
\begin{eqnarray}
P(x,t) = - i \! \left([ \sigma_{3}, \, \int_{\widehat{\Gamma}} 
((\underline{{\bf Id}} - C_{w_{x,t}})^{-1} {\rm I})(\xi) 
w_{x,t}(\xi)] \frac{d \xi}{2 \pi i} \right)_{21} \nonumber
\end{eqnarray}
(see Sec.~5), which can be written in the following, simpler 
form, 
\begin{eqnarray}
P(x,t) = \frac{1}{\pi} \int_{\widehat{\Gamma}} (\mu(x,t;
\xi))_{22} (w_{-}(\xi))_{21} e^{ 2 i t \theta (\xi) } d 
\xi, \, \, \, \, \, \, {\rm where} \, \, \, \, \, \theta
(\lambda) \equiv 2 \lambda^{2} (\lambda^{2} - 2 \lambda_{
0}^{2}),
\end{eqnarray}
with real, first-order (distinct) stationary phase points, 
$\{0,\pm \lambda_{0}\}=$ $\{0,\pm \frac{1}{2} \sqrt{- 
\frac{x}{t}}\}$ $(\{0,\pm \lambda_{0}\}$ \linebreak[4] $= 
\{\mathstrut \lambda^{\prime}; \, \partial_{\lambda} 
\theta(\lambda) \vert_{\lambda=\lambda^{\prime}}=0\})$.

At this point, the natural tendency would be to apply the 
method of steepest descents \cite{a23} to the integral in 
(15); however, even though $\theta(\lambda)$ is known and 
$w_{-}(\lambda)$ can be characterised completely, no {\em 
a priori\/} explicit information regarding the analytical 
properties of the resolvent kernel, $\mu(x,t;\lambda)$, 
is available. To eschew this manifest difficulty, Deift and 
Zhou \cite{a15}, based on the earlier, seminal work of Beals 
and Coifman \cite{a14}, introduced a novel non-linear analogue 
of the classical steepest-descent method and showed that, as 
$t \rightarrow + \infty$, via a sequence of transformations 
which convert the original (full) RH problem (Lemma~2.1) into 
an equivalent, auxiliary RH problem with jump matrix $v^{\prime}$ 
of the form $v^{\prime} \! = \! v_{{\rm model}} \! + \! v_{{\rm 
error}}$, where $v_{{\rm model}}$ denotes the jump matrix for 
an explicitly solvable model RH problem and $v_{{\rm error}}$ 
contains only terms which are decreasing as $t \! \rightarrow 
\! + \infty$, modulo some estimates, the solution of the 
original RH problem converges to the solution of the model RH 
problem. More precisely, the sequence of transformations which 
constitute the Deift and Zhou procedure \cite{a15} can be 
summarised as follows (even though the delineation of the 
procedure is general enough, the notation is specific to this 
paper): 
\begin{itemize}
\item[(i)] based on the classical method of steepest descents 
\cite{a23} for decomposing the complex plane of the spectral 
parameter $(\lambda)$ according to the signature of $\Re (i t 
\theta(\lambda))$, deform the original oscillatory RH problem 
to an auxiliary RH problem formulated on an (oriented) augmented 
contour $\Sigma$ in such a way that the respective jump matrices 
on $\widehat{\Gamma} \subset \Sigma$ and the finite ``triangular'' 
parts of $\Sigma \setminus \widehat{\Gamma}$, away {}from the 
neighbourhood of the real, first-order stationary phase points, 
$\{0,\pm \lambda_{0}\}$, converge (with respect to appropriately 
defined norms) to the identity $({\rm I})$, i.e., rewrite the 
original undulatory RH problem as an auxiliary RH problem on the 
paths of steepest descents with upper/lower triangular jump 
matrices whose structures depend on the signature of $\Re (i t 
\theta(\lambda))$ (see Sec.~4);
\item[(ii)] reduce the contour $\Sigma$ to a truncated contour 
$\Sigma^{\prime}$ which can be written as the disjoint union 
of three crosses, $\Sigma^{\prime} = \Sigma_{A^{\prime}} \cup 
\Sigma_{B^{\prime}} \cup \Sigma_{C^{\prime}}$ (according to 
the number of real, first-order stationary phase points: see 
Sec.~5);
\item[(iii)] analysing the higher order interaction between the 
three disjoint crosses, $\Sigma_{A^{\prime}}$, $\Sigma_{B^{
\prime}}$, and $\Sigma_{C^{\prime}}$, and noting that, as $t 
\rightarrow \pm \infty$, they are negligible (with respect 
to certain norms), separate out the contributions of the 
disjoint crosses in $\Sigma^{\prime}$ and show that the 
solution $P(x,t)$ ($Q(x,t) = \overline{P(x,t)})$ can be 
written as the linear superposition of the contributions of 
the various disjoint crosses (see Sec.~6); and 
\item[(iv)] localise the jump matrices of the most rapidly 
descented RH problems on the disjoint crosses to the 
neighbourhood of the real, first-order stationary phase 
points, $\{0,\pm \lambda_{0}\}$, and, under suitable 
scalings of the spectral parameter, reduce the respective 
RH problems to RH problems on $\Bbb R$ with constant jump 
matrices which can be solved explicitly (see Sec.~7).
\end{itemize}

In this paper, we prove that the application of the 
aforementioned procedure leads to the following results.

{\em Remark 3.1.\/} To facilitate the reading of the 
results stated in Theorems~3.2 and 3.3 (see below), 
as well as those which appear throughout the paper, 
the following preamble is necessary: (i) $M \! \in \! 
\Bbb R_{> 0}$ is a fixed constant; (ii) the ``symbols'' 
(``notations'') $\underline{c}$ and $c^{{\cal S}}$, 
respectively, which appear in the various error estimates 
are to be understood as follows, $\underline{c} \! \equiv 
\! \underline{c}(\lambda_{0}) \! \in \! {\cal L}^{\infty}
(\Bbb R_{> M})$, and $c^{{\cal S}} \! \equiv \! c^{{\cal 
S}}(\lambda_{0}) \! \in \! {\cal S}(\Bbb R_{> M})$; (iii) 
even though the ``symbols'' $\underline{c}$ and $c^{{\cal 
S}}$ appearing in the various error estimates are 
\underline{{\bf not\/}} equal and should properly be 
denoted as $\underline{c}_{1}(\lambda_{0})$, $\underline{
c}_{2}(\lambda_{0})$, etc., the simplified ``notations'' 
$\underline{c}$ and $c^{{\cal S}}$ are retained throughout 
since the main concern in this paper is not their explicit 
(functional) $\lambda_{0}$-dependence, but rather, the 
explicit class(es) to which they belong; and (iv) with the 
exception of Theorems~3.2 and 3.3 (see below), the arguments 
of the ``symbols'' appearing in the error estimations will 
be suppressed.

{\em Remark 3.2.\/} In Theorems~3.2 and 3.3 (see below), one 
should keep everywhere the upper signs as $t \! \rightarrow 
\! + \infty$ and the lower signs as $t \! \rightarrow \! - 
\infty$. 

\setcounter{six}{1}
\begin{six}
For $\vert \vert r \vert \vert_{{\cal L}^{\infty}(\widehat{
\Gamma})} \! < \! 1$, let $m(x,t;\lambda)$ be the solution 
of the RH problem formulated in Lemma~2.1 and $Q(x,t)$ and 
$q(x,t)$ be defined by Eqs.~(10) and (11), respectively. 
Then as $t \! \rightarrow \! \pm \infty$ and $x \! \rightarrow 
\! \mp \infty$ such that $\lambda_{0} \! \equiv \! \frac{1}{2} 
\sqrt{- \frac{x}{t}} \! > \! M$, 
\begin{eqnarray}
&Q(x,t) = \sqrt{\pm \frac{\nu(\lambda_{0})}{2 \lambda_{0}^{2} 
t}} \exp \! \left\{i \left(\theta^{\pm}(\lambda_{0}) + 
\widehat{\Phi}^{\pm}(\lambda_{0};t) \right) \right\} + {\cal 
O} \! \left( \frac{\underline{c}(\lambda_{0}) \ln \vert t 
\vert}{\lambda_{0} t} \right) \!,& \\  
&q(x,t) = \sqrt{\pm \frac{\nu(\lambda_{0})}{2 \lambda_{0}^{2} 
t}} \exp \! \left\{i \left(\theta^{\pm}(\lambda_{0}) + 
\widehat{\Phi}_{0}^{\pm}(\lambda_{0};t) \right) \right\} + {\cal 
O} \! \left( \frac{\underline{c}(\lambda_{0}) \ln \vert t 
\vert}{\lambda_{0} t} \right) \!,&
\end{eqnarray}
where 
\begin{eqnarray}
&\nu(z) = - \frac{1}{2 \pi} \ln(1 - \vert r (z) \vert^{2}),& \\ 
&\theta^{+}(z) = \frac{1}{\pi} \int_{0}^{z} \ln \! \vert \mu^{2} 
- z^{2} \vert \, d \ln(1 - \vert r (\mu) \vert^{2}) - \frac{1}{
\pi} \int_{0}^{\infty} \ln \! \vert \mu^{2} + z^{2} \vert \, d 
\ln(1 + \vert r (i \mu) \vert^{2}),& \\ 
&\theta^{-}(z) = \frac{1}{\pi} \int_{z}^{\infty} \ln \! \vert 
\mu^{2} - z^{2} \vert \, d \ln(1 - \vert r (\mu) \vert^{2}),& \\ 
&\widehat{\Phi}^{\pm}(\lambda_{0};t) = 4 \lambda_{0}^{4} t \mp 
\nu(\lambda_{0}) \ln \vert t \vert \mp \frac{3 \pi}{4} \pm \arg 
\Gamma (i \nu (\lambda_{0})) + \arg r(\lambda_{0}) \mp 3 \nu
(\lambda_{0}) \ln 2,& \\ 
&\widehat{\Phi}^{+}_{0}(\lambda_{0};t) = \widehat{\Phi}^{+} 
(\lambda_{0};t) + \frac{2}{\pi} \int_{0}^{\lambda_{0}} \frac{ 
\ln (1 - \vert r(\mu) \vert^{2})}{\mu} d \mu - \frac{2}{\pi} 
\int_{0}^{\infty} \frac{\ln(1 + \vert r(i \mu) \vert^{2})}{
\mu} d \mu,& \\ 
&\widehat{\Phi}^{-}_{0}(\lambda_{0};t) = \widehat{\Phi}^{-} 
(\lambda_{0};t) + \frac{2}{\pi} \int_{\lambda_{0}}^{\infty} 
\frac{\ln(1 - \vert r(\mu) \vert^{2})}{\mu} d \mu,&
\end{eqnarray}
and, as $t \! \rightarrow \! \pm \infty$ and $x \! \rightarrow 
\! \pm \infty$ such that $\lambda_{0}^{\prime} \! \equiv \! 
\frac{1}{2} \sqrt{\frac{x}{t}} \! > \! M$, 
\begin{eqnarray}
&Q(x,t) = \sqrt{\mp \frac{\nu(i \lambda_{0}^{\prime})}{2 
(\lambda_{0}^{\prime})^{2} t}} \exp \! \left\{i \left(
\phi^{\pm} (\lambda_{0}^{\prime}) + \widehat{\Theta}^{\pm}
(\lambda_{0}^{\prime};t) \right) \right\} + {\cal O} \! 
\left( \frac{\underline{c}(\lambda_{0}^{\prime}) \ln \vert 
t \vert}{\lambda_{0}^{\prime} t} \right) \!,& \\  
&q(x,t) = \sqrt{\mp \frac{\nu(i \lambda_{0}^{\prime})}{2 
(\lambda_{0}^{\prime})^{2} t}} \exp \! \left\{i \left(\phi^{
\pm}(\lambda_{0}^{\prime}) + \widehat{\Theta}_{0}^{\pm} 
(\lambda_{0}^{\prime};t) \right) \right\} + {\cal O} \! 
\left( \frac{\underline{c}(\lambda_{0}^{\prime}) \ln \vert 
t \vert}{\lambda_{0}^{\prime} t} \right) \!,&
\end{eqnarray}
where 
\begin{eqnarray}
&\nu(i z) = - \frac{1}{2 \pi} \ln (1 + \vert r (i z) \vert^{
2}),& \\ 
&\phi^{+}(z) = \frac{1}{\pi} \int_{0}^{z} \ln \! \vert \mu^{2} 
- z^{2} \vert \, d \ln (1 + \vert r (i \mu) \vert^{2}) - \frac{
1}{\pi} \int_{0}^{\infty} \ln \! \vert \mu^{2} + z^{2} \vert \, 
d \ln (1 - \vert r (\mu) \vert^{2}),& \\ 
&\phi^{-}(z) = \frac{1}{\pi} \int_{z}^{\infty} \ln \! \vert 
\mu^{2} - z^{2} \vert \, d \ln (1 + \vert r (i \mu) \vert^{
2}),& \\ 
&\widehat{\Theta}^{\pm} (\lambda_{0}^{\prime};t) = 4 
(\lambda_{0}^{\prime})^{4} t \mp \nu (i \lambda_{0}^{\prime}) 
\ln \vert t \vert + \widehat{\varepsilon}_{\pm} \pm \arg 
\Gamma (i \nu (i \lambda_{0}^{\prime}) ) + \arg r (i \lambda_{
0}^{\prime}) \mp 3 \nu (i \lambda_{0}^{\prime}) \ln 2,& \\ 
&\widehat{\Theta}^{+}_{0} (\lambda_{0}^{\prime};t) = 
\widehat{\Theta}^{+} (\lambda_{0}^{\prime};t) + \frac{2}{\pi} 
\int_{0}^{\lambda_{0}^{\prime}} \frac{\ln (1 + \vert r(i \mu) 
\vert^{2})}{\mu} d \mu - \frac{2}{\pi} \int_{0}^{\infty} 
\frac{\ln (1 - \vert r(\mu) \vert^{2})}{\mu} d \mu,& \\ 
&\widehat{\Theta}^{-}_{0} (\lambda_{0}^{\prime};t) = 
\widehat{\Theta}^{-} (\lambda_{0}^{\prime};t) + \frac{2}{\pi} 
\int_{\lambda_{0}^{\prime}}^{\infty} \frac{\ln(1 + \vert r(i 
\mu) \vert^{2})}{\mu} d \mu,&
\end{eqnarray}
$\widehat{\varepsilon}_{\pm} = (2 \pm 1) \pi /4$, and $\Gamma 
(\cdot)$ is the gamma function {\rm \cite{a24}.}
\end{six}

\setcounter{tree}{2}
\begin{tree}
For $\vert \vert r \vert \vert_{{\cal L}^{\infty}(\widehat{
\Gamma})} \! < \! 1$, let $m(x,t;\lambda)$ be the solution 
of the RH problem formulated in Lemma~2.1 and $u(x,t)$, the
solution of the MNLSE, be given as in Proposition~2.4 in 
terms of the function $q(x,t)$ given in Theorem~3.2. Then 
as $t \! \rightarrow \! \pm \infty$ and $x \! \rightarrow \! 
\pm \infty$ such that $\widehat{\lambda}_{0} \! \equiv \! 
\sqrt{\frac{1}{2}(\frac{x}{t} \! - \! \frac{1}{s})} \! > \! 
M$ and $\frac{x}{t} \! > \! \frac{1}{s}$, $s \! \in \! \Bbb 
R_{> 0}$,
\begin{eqnarray}
&u(x,t) = \sqrt{\pm \frac{\nu (\widehat{\lambda}_{0})}{2 
\widehat{\lambda}_{0}^{2} s t}} \exp \! \left\{i \left(
\theta^{\pm}(\widehat{\lambda}_{0}) + \widetilde{\Phi}^{
\pm}(\widehat{\lambda}_{0};t) \right) \right\} + {\cal O} 
\! \left( \frac{\underline{c}(\widehat{\lambda}_{0}) \ln 
\vert t \vert}{\widehat{\lambda}_{0} t} \right) \!,&
\end{eqnarray}
where 
\begin{eqnarray}
\widetilde{\Phi}^{+}(\widehat{\lambda}_{0};t) & = & 2 \left( 
\widehat{\lambda}_{0}^{2} + \frac{1}{2 s} \right)^{2} \! t - 
\nu (\widehat{\lambda}_{0}) \ln t - \frac{3 \pi}{4} + \arg 
\Gamma (i \nu (\widehat{\lambda}_{0})) + \arg r (\widehat{
\lambda}_{0}) \nonumber \\
 & - & 2 \nu ( \widehat{\lambda}_{0} ) \ln 2 + \frac{2}{\pi} 
\int_{0}^{\widehat{\lambda}_{0}} \frac{\ln (1 - \vert r(\mu) 
\vert^{2})}{\mu} d \mu -  \frac{2}{\pi} \int_{0}^{\infty} 
\frac{\ln (1 + \vert r(i \mu) \vert^{2})}{\mu} d \mu,
\end{eqnarray}
\begin{eqnarray}
\widetilde{\Phi}^{-}(\widehat{\lambda}_{0};t) & = & 2 \left(
\widehat{\lambda}_{0}^{2} + \frac{1}{2 s} \right)^{2} \! t + 
\nu (\widehat{\lambda}_{0}) \ln \vert t \vert + \frac{3 \pi}{4} 
- \arg \Gamma (i \nu (\widehat{\lambda}_{0})) + \arg r(\widehat{
\lambda}_{0})
\nonumber \\
 & + & 2 \nu (\widehat{\lambda}_{0}) \ln 2 + \frac{2}{\pi}
\int_{\widehat{\lambda}_{0}}^{\infty} \frac{\ln (1 - \vert 
r(\mu) \vert^{2})}{\mu} d \mu,
\end{eqnarray}
and, as $t \! \rightarrow \! \pm \infty$ and $x \! \rightarrow 
\! \mp \infty$ or $\pm \infty$ such that $\widehat{\lambda}_{
0}^{\prime} \! \equiv \! \sqrt{\frac{1}{2}(\frac{1}{s} \! - \! 
\frac{x}{t})} \! > \! M$ and $\frac{x}{t} \! < \! \frac{1}{s}$, 
$s \! \in \! \Bbb R_{> 0}$, 
\begin{eqnarray}
&u(x,t) = \sqrt{\mp \frac{ \nu (i \widehat{\lambda}_{0}^{\prime})
}{2 (\widehat{\lambda}_{0}^{\prime})^{2} s t}} \exp \! \left\{i 
\left(\phi^{\pm} (\widehat{\lambda}_{0}^{\prime}) + \widetilde{
\Theta}^{\pm}(\widehat{\lambda}_{0}^{\prime};t) \right) \right\} 
+ {\cal O} \! \left( \frac{\underline{c}(\widehat{\lambda}_{0}^{
\prime}) \ln \vert t \vert}{\widehat{\lambda}_{0}^{\prime} t} 
\right) \!,&
\end{eqnarray}
where 
\begin{eqnarray}
\widetilde{\Theta}^{+}(\widehat{\lambda}_{0}^{\prime};t) & = & 
2 \left( (\widehat{\lambda}_{0}^{\prime})^{2} - \frac{1}{2 s} 
\right)^{2} \! t - \nu (i \widehat{\lambda}_{0}^{\prime}) \ln t 
+ \frac{3 \pi}{4} + \arg \Gamma (i \nu (i \widehat{\lambda}_{0}^{
\prime})) + \arg r (i \widehat{\lambda}_{0}^{\prime}) \nonumber \\
 & - & 2 \nu (i \widehat{\lambda}_{0}^{\prime}) \ln 2 + \frac{2}{
\pi} \int_{0}^{\widehat{\lambda}_{0}^{\prime}} \frac{\ln (1 + 
\vert r(i \mu) \vert^{2})}{\mu} d \mu - \frac{2}{\pi} \int_{0}^{
\infty} \frac{\ln (1 - \vert r(\mu) \vert^{2})}{\mu} d \mu,
\end{eqnarray}
\begin{eqnarray}
\widetilde{\Theta}^{-} (\widehat{\lambda}_{0}^{\prime};t) & = & 
2 \left((\widehat{\lambda}_{0}^{\prime})^{2} - \frac{1}{2 s} 
\right)^{2} \! t + \nu (i \widehat{\lambda}_{0}^{\prime}) \ln 
\vert t \vert + \frac{\pi}{4} - \arg \Gamma (i \nu (i \widehat{
\lambda}_{0}^{\prime})) + \arg r (i \widehat{\lambda}_{0}^{
\prime}) \nonumber \\
 & + & 2 \nu ( i \widehat{\lambda}_{0}^{\prime} ) \ln 2 + 
\frac{2}{\pi} \int_{\widehat{\lambda}_{0}^{\prime}}^{\infty} 
\frac{\ln (1 + \vert r(i \mu) \vert^{2})}{\mu} d \mu,
\end{eqnarray}
and $\nu(\cdot)$, $\theta^{\pm}(\cdot)$, $\nu(i \cdot)$, and 
$\phi^{\pm}(\cdot)$ are given in Theorem~3.2, Eqs.~(18)--(20) 
and (26)--(28). 
\end{tree}

{\em Remark 3.3.\/} The results presented in Theorem~3.3 give 
the asymptotic expansion of $u(x,t)$ in the domain $x/t \neq 
1/s$. In the case $t \to \pm \infty$ and $x/t = 1/s$, $u(x,t)$ 
can be written as follows, 
$$
u(x,t) = \frac{u_{0}}{\sqrt{t}} + o \! \left( \frac{1}{\sqrt{
t}} \right) \!, 
$$
where $u_{0} \in \Bbb C \setminus \{0\}$ is some constant: 
the determination of $u_{0}$ requires special consideration 
and will be presented elsewhere. 

{\em Remark 3.4.\/} Some recent results concerning the 
asymptotic form of solutions as $t$ $\rightarrow \pm 
\infty$ of the DNLSE were written in terms of the so-called 
final states (the final-value problem) \cite{a25}; however, 
unlike the present article, the authors in \cite{a25} did 
{\bf not\/} treat the Cauchy initial-value problem.

{\em Remark 3.5.\/} In the proof of Theorems~3.2 and 3.3, 
according to the above-formulated scheme, we omit the 
details of intermediate estimations which are analogous 
to those in \cite{a15}: in corresponding places, we make 
exact references to the appropriate assertions in \cite{a15}, 
and \cite{a19}, where these details may be found; moreover, 
for $Q(x,t)$, $q(x,t)$, and $u(x,t)$, one should consider 
actually four different cases, depending on the quadrant in 
the $(x,t)$-plane. In the subsequent sections, we present 
only the proof for the function $Q(x,t)$ in the case $t \! 
\to \! + \infty$ and $x \! \to \! - \infty$ such that 
$\lambda_{0} \! \equiv \! \frac{1}{2} \sqrt{- \frac{x}{t}} 
\! > \! M$: the results for the remaining domains of the 
$(x,t)$-plane are obtained in an analogous manner. The 
corresponding results for the functions $q(x,t)$ and $u(x,t)$ 
are obtained {}from Propositions~2.3 and 2.4. 
\section{The Augmented RH Problem}
As explained in the previous section, we begin with the 
decomposition of the complex $\lambda$-plane according to the 
signature of the real part of the phase of the conjugating 
exponential of the oscillatory RH problem (Lemma 2.1), $\Re (i t 
\theta(\lambda) )$, where $\theta (\lambda) = 2 \lambda^{2} 
(\lambda^{2} - 2 \lambda_{0}^{2} )$, $\lambda_{0}^{2} = - 
\frac{x}{4 t}$ $> 0$, and $t > 0$ (see Fig.~2).
\begin{figure}[bht]
\vspace{-0.5cm}
\begin{center}
\unitlength=1cm
\begin{picture}(6,6)(0,0)
\thicklines
\put(3,3){\vector(1,0){0.75}}
\put(6,3){\vector(-1,0){0.85}}
\put(0,3){\vector(1,0){0.85}}
\put(3.75,3){\line(1,0){2.75}}
\put(3,3){\vector(-1,0){0.75}}
\put(2.25,3){\line(-1,0){2.75}}
\put(3,6){\vector(0,-1){2.25}}
\put(3,3.75){\line(0,-1){0.75}}
\put(3,0){\vector(0,1){2.25}}
\put(3,2.25){\line(0,1){0.75}}
\put(4.25,2.675){\makebox(0,0){$\scriptstyle{}\lambda_{0}$}}
\put(4.5,3){\makebox(0,0){$\bullet$}}
\put(2.875,2.675){\makebox(0,0){$\scriptstyle{}0$}}
\put(3,3){\makebox(0,0){$\bullet$}}
\put(1.825,2.675){\makebox(0,0){$\scriptstyle{}-\lambda_{0}$}}
\put(1.5,3){\makebox(0,0){$\bullet$}}
\put(3.95,5){\makebox(0,0){$\scriptstyle{}\Re(i t \theta 
(\lambda)) \, > \, 0$}}
\put(3.95,1){\makebox(0,0){$\scriptstyle{}\Re(i t \theta 
(\lambda)) \, < \, 0$}}
\put(2.05,1){\makebox(0,0){$\scriptstyle{}\Re(i t \theta 
(\lambda)) \, > \, 0$}}
\put(2.05,5){\makebox(0,0){$\scriptstyle{}\Re(i t \theta 
(\lambda)) \, < \, 0$}}
\put(5.7,3.4){\makebox(0,0){$\scriptstyle{}\Re(i t \theta 
(\lambda)) \, < \, 0$}}
\put(5.7,2.6){\makebox(0,0){$\scriptstyle{}\Re(i t \theta 
(\lambda)) \, > \, 0$}}
\put(0.3,2.6){\makebox(0,0){$\scriptstyle{}\Re(i t \theta 
(\lambda)) \, < \, 0$}}
\put(0.3,3.4){\makebox(0,0){$\scriptstyle{}\Re(i t \theta 
(\lambda)) \, > \, 0$}}
\qbezier(6,6)(3,3)(6,0)
\qbezier(0,6)(3,3)(0,0)
\qbezier[75](0,0)(3,3)(6,6)
\qbezier[75](0,6)(3,3)(6,0)
\end{picture}
\vspace{-0.725cm}
\end{center}
\caption{Signature graph of $\Re(i t \theta (\lambda))$.} 
\end{figure}
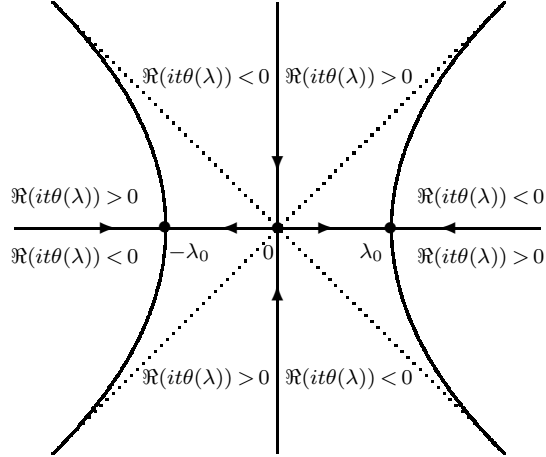

{\em Remark 4.1.\/} Note that, we have reoriented the 
contour $\widehat{\Gamma}$ in accordance with the 
signature of $\Re (i t \theta (\lambda))$ and the 
``sign'' convention of Remark~2.1.  

The main purpose of this section is to reformulate the 
original RH problem (Lemma 2.1) as an equivalent RH problem 
(see Lemma~4.1) on the augmented contour $\Sigma$ (see 
Fig.~3),
\begin{eqnarray}
\Sigma =  L \cup \overline{L} \cup \widehat{\Gamma}, 
\nonumber
\end{eqnarray}
where $L \equiv L^{0} \cup L^{1}$, and 
\begin{eqnarray}
&L^{0} = \{ \mathstrut \lambda; \mathstrut \lambda =
\lambda_{0} + \frac{ \lambda_{0} u }{2} e^{ \frac{3 \pi 
i}{4}}, \mathstrut u \in (- \infty, \sqrt{2}] \}
\cup \{ \mathstrut \lambda; \mathstrut \lambda = -
\lambda_{0} + \frac{ \lambda_{0} u}{2} e^{- \frac{i \pi}{
4}}, \mathstrut u \in (- \infty, \sqrt{2}] \},& \nonumber 
\\
&L^{1} = \{ \mathstrut \lambda; \mathstrut \lambda =
\frac{\lambda_{0} u}{2} e^{\frac{i \pi}{4}}, \mathstrut 
u \in \Bbb R \}.& \nonumber 
\end{eqnarray}

{\em Remark 4.2.\/} Actually, for the following analysis, 
the augmented contour $\Sigma$ can be chosen in different 
ways, not necessarily consisting of straight lines: its 
important characteristic is the position of the contour 
$L \cup \overline{L}$ with respect to the lines $\Re (i t 
\theta(\lambda)) = 0$.
\begin{figure}[bht]
\vspace{-0.5cm}
\begin{center}
\unitlength=1cm
\begin{picture}(7,7)(0,0)
\thicklines
\put(3.5,0){\vector(0,1){1.75}}   
\put(3.5,1.75){\line(0,1){3.5}}
\put(3.5,7){\vector(0,-1){1.75}}
\put(0,3.5){\vector(1,0){0.875}}
\put(0.875,3.5){\line(1,0){1.75}}
\put(3.5,3.5){\vector(-1,0){0.875}}
\put(3.5,3.5){\vector(1,0){0.875}}
\put(4.375,3.5){\line(1,0){1.75}}
\put(7,3.5){\vector(-1,0){0.875}}   
\put(3.3895,3.15){\makebox(0,0){$\scriptstyle{}0$}}
\put(3.5,3.5){\makebox(0,0){$\bullet$}}
\put(5.25,3.15){\makebox(0,0){$\scriptstyle{}\lambda_{0}$}}
\put(5.25,3.5){\makebox(0,0){$\bullet$}}
\put(1.75,3.15){\makebox(0,0){$\scriptstyle{}-\lambda_{0}$}}
\put(1.75,3.5){\makebox(0,0){$\bullet$}}
\put(5.25,3.5){\line(1,1){0.875}}
\put(7,5.25){\vector(-1,-1){0.875}}
\put(5.25,3.5){\line(1,-1){0.875}}
\put(7,1.75){\vector(-1,1){0.875}}
\put(5.25,3.5){\line(-1,1){0.4375}}
\put(4.375,4.375){\vector(1,-1){0.4375}}
\put(5.25,3.5){\line(-1,-1){0.4375}}
\put(4.375,2.625){\vector(1,1){0.4375}}
\put(3.5,3.5){\vector(1,1){0.4375}}
\put(3.9375,3.9375){\line(1,1){0.875}}
\put(6.5,6.5){\vector(-1,-1){1.6875}}
\put(3.5,3.5){\vector(1,-1){0.4375}}
\put(3.9375,3.0625){\line(1,-1){0.875}}
\put(6.5,0.5){\vector(-1,1){1.6875}}   
\put(3.5,3.5){\vector(-1,1){0.4375}}
\put(3.0625,3.9375){\line(-1,1){0.875}} 
\put(0.5,6.5){\vector(1,-1){1.6875}}
\put(3.5,3.5){\vector(-1,-1){0.4375}}   
\put(3.0625,3.0625){\line(-1,-1){0.875}}
\put(0.5,0.5){\vector(1,1){1.6875}}
\put(1.75,3.5){\line(1,1){0.4375}}
\put(2.625,4.375){\vector(-1,-1){0.4375}}
\put(1.75,3.5){\line(1,-1){0.4375}}
\put(2.625,2.625){\vector(-1,1){0.4375}}
\put(1.75,3.5){\line(-1,1){0.875}}  
\put(0,5.25){\vector(1,-1){0.875}}
\put(1.75,3.5){\line(-1,-1){0.875}}
\put(0,1.75){\vector(1,1){0.875}}
\put(6.125,3.75){\makebox(0,0){$\scriptstyle{}\Omega_{5}$}}
\put(6.125,3.25){\makebox(0,0){$\scriptstyle{}\Omega_{11}$}}
\put(0.875,3.75){\makebox(0,0){$\scriptstyle{}\Omega_{16}$}}
\put(0.875,3.25){\makebox(0,0){$\scriptstyle{}\Omega_{10}$}}
\put(5.25,4.25){\makebox(0,0){$\scriptstyle{}\Omega_{1}$}}
\put(5.25,2.75){\makebox(0,0){$\scriptstyle{}\Omega_{4}$}}
\put(4.375,3.75){\makebox(0,0){$\scriptstyle{}\Omega_{12}$}}
\put(4.375,3.25){\makebox(0,0){$\scriptstyle{}\Omega_{6}$}}
\put(4.125,4.75){\makebox(0,0){$\scriptstyle{}\Omega_{13}$}}
\put(4.125,2.25){\makebox(0,0){$\scriptstyle{}\Omega_{7}$}}
\put(2.875,4.75){\makebox(0,0){$\scriptstyle{}\Omega_{8}$}}
\put(2.875,2.25){\makebox(0,0){$\scriptstyle{}\Omega_{14}$}}
\put(2.625,3.75){\makebox(0,0){$\scriptstyle{}\Omega_{9}$}}
\put(2.625,3.25){\makebox(0,0){$\scriptstyle{}\Omega_{15}$}}
\put(1.75,4.25){\makebox(0,0){$\scriptstyle{}\Omega_{3}$}}
\put(1.75,2.75){\makebox(0,0){$\scriptstyle{}\Omega_{2}$}}
\end{picture}
\vspace{-0.725cm}
\end{center}
\caption{Augmented contour $\Sigma$.}
\end{figure}
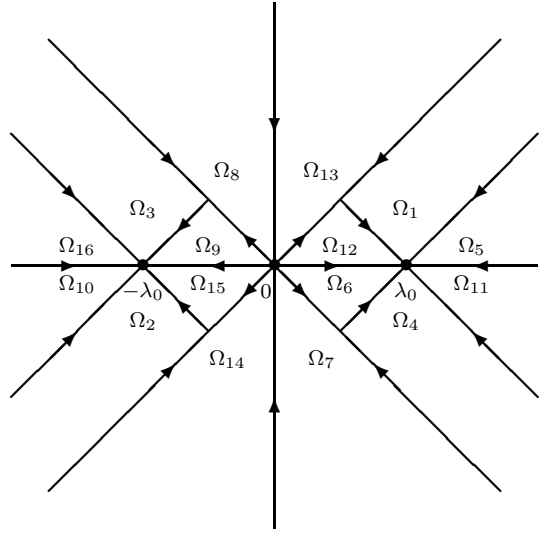

In order to define the conjugation matrices on $\Sigma$ and 
exploit the analyses in \cite{a14,a15}, we need to formulate 
two technical propositions: the first concerns the triangular 
factorisation of the conjugation matrices of the original RH 
problem (Lemma 2.1), and the second pertains to a special 
decomposition for the reflection coefficient, $r(\lambda)$. 
Write the jump matrices of the original RH problem (Lemma~2.1) 
in upper/lower triangular form for $\lambda \in (- \infty,- 
\lambda_{0}) \cup (+\lambda_{0},+\infty)$ as  
\begin{eqnarray}
&e^{-i t \theta (\lambda) {\rm ad} (\sigma_{3})} G(\lambda) = 
e^{-i t \theta (\lambda) {\rm ad} (\sigma_{3})} 
\! \left( \! \begin{array}{cc}
         1 & r(\lambda) \\
         0 & 1
       \end{array} \! \right)
\! \! \left( \! \begin{array}{cc}
         1 & 0 \\
         - \overline{r(\overline{\lambda})} & 1 
       \end{array} \! \right) \!,& 
\end{eqnarray}
and, in lower/upper triangular form for $\lambda \in 
(- \lambda_{0},+ \lambda_{0}) \cup (-i \infty,+i \infty)$ 
as 
\begin{eqnarray}
&e^{-i t \theta (\lambda) {\rm ad} (\sigma_{3})} G(\lambda) 
= e^{-i t \theta (\lambda) {\rm ad} (\sigma_{3})} 
\! \left( \! \begin{array}{cc}
1 & 0 \\
- \overline{r(\overline{\lambda})} (1 - \overline{r 
(\overline{\lambda})} r(\lambda))^{-1} & 1
\end{array} \! \right)& \nonumber \\ 
&\times \left( \! \begin{array}{cc}
(1 - \overline{r(\overline{\lambda})} r(\lambda)) & 0 \\
0 & (1 - \overline{r(\overline{\lambda})} r(\lambda))^{-1} 
\end{array} \! \right) \! \!
\left( \! \begin{array}{cc} 
1 & r(\lambda) (1 - \overline{r(\overline{\lambda})} r 
(\lambda))^{-1} \\
0 & 1 
\end{array} \! \right) \!.&  
\end{eqnarray}
To eliminate the diagonal matrix between the lower/upper 
triangular factors in (39), we, according to the scheme in 
\cite{a15}, introduce the auxiliary function $\delta(\lambda)$ 
which solves the following discontinuous scalar RH problem: 
\begin{eqnarray}
&\delta_{+} (\lambda) = \left\{ \begin{array}{l}
\delta_{-} (\lambda) = \delta(\lambda), \, \, \, \, \, \lambda 
\in (-\infty,-\lambda_{0}) \cup (+\lambda_{0},+\infty), \\
\delta_{-}(\lambda) (1 - \overline{r(\overline{\lambda})} r 
(\lambda)), \, \, \, \, \lambda \in (-\lambda_{0},+\lambda_{0}) 
\cup (- i \infty, + i \infty), 
\end{array} \right. \\
&\delta(\lambda) \rightarrow 1 \, \, \, {\rm as} \, \, \, 
\lambda \rightarrow \infty.& \nonumber 
\end{eqnarray}

\setcounter{yereg}{0}
\begin{yereg}
The unique solution of the above RH problem (40) can be 
written as 
\begin{eqnarray}
&\delta(\lambda) = \left(\! \left(\frac{\lambda - \lambda_{
0}}{\lambda} \right) \! \! \left(\frac{\lambda + \lambda_{0}
}{\lambda} \right) \! \right)^{i \nu} \! e^{\chi_{+}(\lambda)} 
e^{\chi_{-}(\lambda)} e^{\widehat{\chi}_{+}(\lambda)} 
e^{\widehat{\chi}_{-}(\lambda)},& 
\end{eqnarray}
where $\nu \equiv \nu(\lambda_{0})$ is given in Theorem~3.2, 
Eq.~(18), and
\begin{eqnarray}
&\chi_{\pm}(\lambda) = \frac{1}{2 \pi i} \int_{0}^{\pm 
\lambda_{0}} \ln \! \left(\! \frac{1 - \vert r(\mu) \vert^{
2}}{1 - \vert r(\lambda_{0}) \vert^{2}} \! \right) \! 
\frac{d \mu}{(\mu - \lambda)}, \, \, \, \, \, \, \, 
\widehat{\chi}_{\pm}(\lambda) = \int_{\pm i \infty}^{i 0} 
\frac{\ln (1 - \overline{r(\overline{\mu})} r(\mu))}{(\mu - 
\lambda)} \frac{d \mu}{2 \pi i} \!:& 
\end{eqnarray} 
in Eq.~(41), $(\lambda - \mu )^{\pm i \nu} \equiv \exp \{\pm 
i \nu \ln (\lambda - \mu) \}$, where the principal branch of 
the logarithmic functions, $\ln (\lambda - \mu)$, $\mu \in \{0,
\pm \lambda_{0}\}$, has been chosen, with the branch cuts along 
$(-\infty,\mu)$, $\ln (\lambda - \mu) \equiv \ln \! \vert 
\lambda - \mu \vert + i \arg (\lambda - \mu)$, $\arg (\lambda - 
\mu) \in (- \pi,\pi)$; moreover, the function $\delta(\lambda)$ 
possesses the following properties, 
\begin{eqnarray}
&\delta(\lambda) = \left(\overline{\delta(\overline{\lambda})} 
\right)^{-1} = \left(\overline{\delta (- \overline{\lambda})} 
\right)^{-1} \Rightarrow \delta(\lambda) = \delta(- \lambda ),& 
\\ 
&\vert \delta_{\pm} (\lambda) \vert^{2} \leq (1 - \sup\limits_{
\mu \in \widehat{\Gamma}} \overline{r(\overline{\mu})} r(\mu))^{
\pm 1} < \infty \, \, \, \, \, \, \forall \, \, \, \, \lambda 
\in \widehat{\Gamma},& 
\end{eqnarray}
and $\vert \vert (\delta(\cdot))^{\pm 1} \vert \vert_{{\cal 
L}^{\infty}(\Bbb C)} \equiv \sup\limits_{\lambda \in \Bbb C} 
\vert (\delta(\lambda))^{\pm 1} \vert < \infty$.
\end{yereg}

{\em Proof.\/} It is well known that the scalar RH problem 
for $\delta(\lambda)$ stated in (40) can be solved explicitly 
(see, for example, \cite{a26,a27}): in our case, the solution 
is given in (41) and (42). As a consequence of (41) and (42), 
one proves the symmetry properties in (43): {}from (40), 
using (43), one deduces inequalities~(44); therefore, {}from 
the maximum modulus principle and the fact that $\delta
(\lambda)$ has no zeros for $\lambda \in \Bbb C$, one gets 
that $\delta(\lambda) \in {\cal L}^{\infty}(\Bbb C)$. 
\ \ \rule{6pt}{6pt}

The conjugation matrices for the RH problem on the augmented 
contour $\Sigma$ should, of course, be written in terms of 
the matrix elements of the original RH problem on $\widehat{
\Gamma}$ (Lemma~2.1); but, since the reflection coefficient, 
$r(\lambda)$, does not, in general, have analytical continuation 
off $\widehat{\Gamma}$, we, following \cite{a15}, decompose it 
as the sum of an analytically continuable part and a negligible 
non-analytic remainder. To formulate an exact result which we 
use later, let us define: (i)  
\begin{eqnarray}
\rho(\lambda) \equiv \left\{ \begin{array}{l}
r(\lambda) (1 - \overline{r(\overline{\lambda})} r(\lambda))^{
-1}, \, \, \, \, \, \lambda \in (- \lambda_{0},+ \lambda_{0}) 
\cup (- i \infty,+ i \infty), \\
- r(\lambda), \, \, \, \, \, \lambda \in (- \infty,- \lambda_{0}) 
\cup (+ \lambda_{0},+ \infty);
 \end{array} \right.
\end{eqnarray}
and (ii) the contour $L_{\widehat{\delta}} \equiv L_{\widehat{
\delta}}^{0} \cup L_{\widehat{\delta}}^{1}$, where 
\begin{eqnarray}
&L_{\widehat{\delta}}^{0} = \{\mathstrut \lambda; \mathstrut
\lambda = \lambda_{0} + \frac{\lambda_{0} u}{2} e^{\frac{3  
\pi i}{4}}, \mathstrut u \in (\widehat{\delta},\sqrt{2}]\}
\cup \{\mathstrut \lambda; \mathstrut \lambda = - \lambda_{0} 
+ \frac{\lambda_{0} u}{2} e^{- \frac{i \pi}{4}}, \mathstrut u 
\in (\widehat{\delta}, \sqrt{2}]\},& \nonumber \\
&L_{\widehat{\delta}}^{1} = \{\mathstrut \lambda; \mathstrut
\lambda = \frac{\lambda_{0} u}{2} e^{\frac{i \pi}{4}}, 
\mathstrut u \in [- \sqrt{2},- \widehat{\delta}) \cup 
(\widehat{\delta},\sqrt{2}]\}.& \nonumber
\end{eqnarray}

\setcounter{seventh}{1}  
\begin{seventh}
For each $l \in \Bbb Z_{\geq 1}$, there exists a decomposition 
of the function $\rho(\lambda)$ in Eq.~(45), 
\begin{eqnarray}
\rho(\lambda) = h_{I}(\lambda) + R(\lambda) + h_{II}(\lambda),
\, \, \, \, \, \, \lambda \in \widehat{\Gamma},
\end{eqnarray}
such that $h_{I}(\lambda)$ is analytic on $\widehat{\Gamma}$ 
(generally, it has no analytic continuation off $\widehat{
\Gamma})$, $R(\lambda)$ is a piecewise-rational function such 
that $\left. \frac{d^{j} \rho(\lambda)}{d \lambda^{j}} 
\right\vert_{\lambda \in \mho} = \left. \frac{d^{j} R(\lambda)}{
d \lambda^{j}} \right\vert_{\lambda \in \mho}$, $0 \leq j \leq 
12 l + 1$, where $\mho$ $\equiv \{0,\pm \lambda_{0}\}$, and 
$h_{II}(\lambda)$ has an analytic continuation to $L$; moreover, 
in the domain $\lambda_{0} \! > \! M$, the following estimates 
are valid as $t \! \rightarrow \! + \infty$,
\begin{eqnarray*}
&\vert e^{- 2 i t \theta(\lambda)} h_{I}(\lambda) \vert 
\leq \frac{\underline{c}}{(1 + \vert \lambda \vert^{3})
(\lambda_{0}^{2} t)^{l}}, \, \, \, \, \lambda \in \widehat{
\Gamma}, \, \, \, \, \, \, \, \, \vert e^{- 2 i t 
\theta(\lambda)} h_{II}(\lambda) \vert \leq \frac{\underline{
c}}{(1 + \vert \lambda \vert^{3})(\lambda_{0}^{2} t)^{l}}, \, 
\, \, \, \lambda \in L,& \\
&\vert e^{- 2 i t \theta(\lambda)} R(\lambda) \vert \leq 
\underline{c} \exp \{- 2 \lambda_{0}^{4} \widehat{\delta}^{
2} t\}, \, \, \, \, \lambda \in L_{\widehat{\delta}},& 
\end{eqnarray*}
and $\widehat{\delta} \! \in \! \Bbb R_{> 0}$ is sufficiently 
small.
\end{seventh}
   
{\em Proof.\/} Rather technical: proceed analogously as in 
the proof of Proposition~1.92 in \cite{a15} by expanding 
$\rho(\lambda)$ in terms of a rational polynomial approximation 
in the neighbourhood of the real, first-order stationary phase 
points, $\{0,\pm \lambda_{0}\}$, and show that, for $l \! \in 
\! \Bbb Z_{\geq 1}$, $\vert e^{- 2 i t \theta(\lambda)} h_{I}
(\lambda) \vert \! \leq \! \frac{\underline{c}}{(1 + \vert 
\lambda \vert^{3}) \vert x \vert^{l}}$, $\lambda \! \in \! 
\widehat{\Gamma}$, $\vert e^{- 2 i t \theta(\lambda)} h_{II}
(\lambda) \vert \! \leq \! \frac{\underline{c}}{(1 + \vert 
\lambda \vert^{3}) \vert x \vert^{l}}$, $\lambda \! \in \! 
L$, and $\vert e^{- 2 i t \theta(\lambda)} R(\lambda) \vert 
\! \leq \! \underline{c} \exp \{- \frac{1}{2} \lambda_{0}^{2} 
\widehat{\delta}^{2} \vert x \vert\}$, $\lambda \! \in \! 
L_{\widehat{\delta}}$. Using the relation $\vert x \vert \! 
= \! 4 \lambda_{0}^{2} t$, one obtains the result stated in 
the Proposition: for details, see \cite{a19}. 
\ \ \rule{6pt}{6pt}

\setcounter{chorss}{0}
\begin{chorss}
Let $m(x,t;\lambda)$ be the solution of the RH problem 
formulated in Lemma~2.1. Set $m^{\Delta}(x,t;\lambda) 
\! \equiv \! m(x,t;\lambda)(\Delta(\lambda))^{-1}$, where
\begin{eqnarray}
\Delta(\lambda) \equiv (\delta(\lambda))^{\sigma_{3}},
\end{eqnarray}
and $\delta(\lambda)$ is given in Proposition~4.1. Define
\begin{eqnarray*}
m^{\sharp}(x,t;\lambda) \equiv
\left\{ \begin{array}{l}
m^{\Delta}(x,t;\lambda), \, \, \, \, \, \lambda \in 
\Omega_{1} \cup \Omega_{2} \cup \Omega_{3} \cup 
\Omega_{4}, \\
m^{\Delta}(x,t;\lambda) ({\rm I} - (w_{-}^{a})_{x,t,\delta}
)^{-1}, \, \, \, \, \, \lambda \in \Omega_{5} \cup \Omega_{6} 
\cup \Omega_{7} \cup \Omega_{8} \cup \Omega_{9} \cup \Omega_{
10}, \\
m^{\Delta}(x,t;\lambda) ({\rm I} + (w_{+}^{a})_{x,t,\delta})^{
-1}, \, \, \, \, \, \lambda \in \Omega_{11} \cup \Omega_{12} 
\cup \Omega_{13} \cup \Omega_{14} \cup \Omega_{15} \cup 
\Omega_{16}. 
\end{array} \right.
\end{eqnarray*}
Then $m^{\sharp}(x,t;\lambda)$ solves the following (augmented) 
RH problem on $\Sigma$,
\begin{eqnarray}
&\left. \begin{array}{c}
m_{+}^{\sharp}(x,t;\lambda) = m_{-}^{\sharp}(x,t;\lambda)
v_{x,t,\delta}^{\sharp}(\lambda), \, \, \, \, \, \lambda 
\in \Sigma, \\
m^{\sharp}(x,t;\lambda) \rightarrow {\rm I} \, \, \, \, 
{\rm as} \, \, \, \, \lambda \rightarrow \infty, 
\end{array} \right.&
\end{eqnarray}
where
\begin{eqnarray}
&v_{x,t,\delta}^{\sharp}(\lambda) \! \equiv \! ({\rm I} \! 
- \! (w_{-}^{\sharp})_{x,t,\delta})^{-1}({\rm I} \! + \! 
(w_{+}^{\sharp})_{x,t,\delta}) \! = \! 
\left\{ \! \begin{array}{l}
({\rm I} \! - \! (w_{-}^{0})_{x,t,\delta})^{-1} 
({\rm I} \! + \! (w_{+}^{0})_{x,t,\delta}), \, \, \, \,
\lambda \! \in \! \widehat{\Gamma}, \\
({\rm I} \! + \! (w_{+}^{a})_{x,t,\delta}), \, \, \, \, 
\lambda \! \in \! L, \\
({\rm I} \! - \! (w_{-}^{a})_{x,t,\delta})^{-1}, \, \, \, 
\, \lambda \! \in \! \overline{L}, 
\end{array} \right.&
\end{eqnarray}
and
\begin{eqnarray}
&(w^{(0,a)}_{\pm})_{x,t,\delta} = (\delta_{\pm}(\lambda))^{{
\rm ad}(\sigma_{3})} \exp \left\{- i t \theta(\lambda) {\rm 
ad}(\sigma_{3}) \right\} w^{(0,a)}_{\pm},& \\
&w_{+}^{0} =  h_{I}(\lambda) \sigma_{+}, \, \, \, \, \, \, 
\, \, w_{+}^{a} = (h_{II}(\lambda) + R(\lambda)) \sigma_{+},& 
\\    
&w_{-}^{0} = - \overline{h_{I}(\lambda)} \sigma_{-}, \, \, 
\, \, \, \, \, \, w_{-}^{a} = - (\overline{h_{II}(\overline{
\lambda})} + \overline{R(\overline{\lambda})}) \sigma_{-}.&
\end{eqnarray}
\end{chorss}
 
{\em Proof.\/} In terms of the function $m^{\Delta}(x,t;
\lambda)$ defined in the Lemma, the original oscillatory RH 
problem (Lemma~2.1) can be rewritten in the following form,
\begin{eqnarray}
\left. \begin{array}{c}
m_{+}^{\Delta}(x,t;\lambda) = m_{-}^{\Delta}(x,t;\lambda) 
({\rm I} - (w_{-})_{x,t,\delta})^{-1} ({\rm I} + 
(w_{+})_{x,t,\delta}), \, \, \, \, \lambda \in \widehat{
\Gamma}, \\ 
m^{\Delta}(x,t;\lambda) \rightarrow {\rm I} \, \, \, \, 
{\rm as} \, \, \, \, \lambda \rightarrow \infty,
\end{array} \right.
\end{eqnarray}
where
\begin{eqnarray}
(w_{\pm})_{x,t,\delta} = (\delta_{\pm}(\lambda))^{{\rm ad} 
(\sigma_{3})} \exp \left\{- i t \theta(\lambda) {\rm ad} 
(\sigma_{3}) \right\} w_{\pm}, 
\end{eqnarray}
and
\begin{eqnarray}
w_{+} = \rho (\lambda) \sigma_{+}, \, \, \, \, \, \, \, \,
w_{-} = - \overline{\rho(\overline{\lambda})} \sigma_{-}. 
\end{eqnarray}
Defining $m^{\sharp}(x,t;\lambda)$ as in the Lemma, one 
arrives, as a consequence of Proposition~4.2, (47), and 
(50)--(55), to the RH problem stated in the Lemma. 
\ \ \rule{6pt}{6pt}
\section{RH Problem on the Truncated Contour}
The goal of this section is to get rid (asymptotically) 
of the contribution of the functions $h_{I}(\lambda)$ 
and $h_{II}(\lambda)$ (Proposition~4.2) to the conjugation 
matrices of the augmented RH problem: simultaneously, it 
is proved that, as $t \! \rightarrow \! + \infty$, under 
the condition $\lambda_{0} \! > \! M$, the contribution 
to the asymptotics of the function $P(x,t)$ $(Q(x,t) \! 
= \! \overline{P(x,t)})$ coming {}from $\widehat{\Gamma}$ 
and the finite ``triangular'' segments $L_{\widehat{\delta}}$ 
and $\overline{L_{\widehat{\delta}}}$, respectively, are 
negligible; hence, we derive the RH problem on the truncated 
contour $\Sigma^{\prime}$ 
(see Fig.~4),
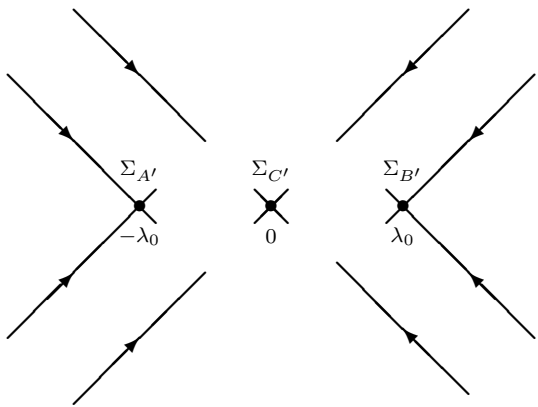
\begin{figure}[bht]
\vspace{-0.25cm}
\begin{center}
\unitlength=1cm
\begin{picture}(6,6)(0,0)
\thicklines
\put(3.5,3.10){\makebox(0,0){$\scriptstyle{}0$}}
\put(3.5,3.5){\makebox(0,0){$\bullet$}}
\put(5.25,3.10){\makebox(0,0){$\scriptstyle{}\lambda_{0}$}}
\put(5.25,3.5){\makebox(0,0){$\bullet$}}
\put(1.75,3.10){\makebox(0,0){$\scriptstyle{}-\lambda_{0}$}}  
\put(1.75,3.5){\makebox(0,0){$\bullet$}}
\put(5.03125,3.28125){\line(1,1){1.09375}}
\put(7,5.25){\vector(-1,-1){0.875}}
\put(5.03125,3.71875){\line(1,-1){1.09375}}
\put(7,1.75){\vector(-1,1){0.875}}
\put(4.375,4.375){\line(1,1){0.875}}
\put(6.125,6.125){\vector(-1,-1){0.875}}
\put(4.375,2.75){\line(1,-1){0.875}}
\put(6.125,0.9975){\vector(-1,1){0.875}}   
\put(3.71875,3.71875){\line(-1,-1){0.4375}}
\put(3.71875,3.28125){\line(-1,1){0.4375}}
\put(1.96875,3.71875){\line(-1,-1){1.09375}}
\put(0,1.75){\vector(1,1){0.875}}
\put(1.96875,3.28125){\line(-1,1){1.09375}}
\put(0,5.25){\vector(1,-1){0.875}}
\put(2.625,4.375){\line(-1,1){0.875}}
\put(0.875,6.125){\vector(1,-1){0.875}}
\put(2.625,2.625){\line(-1,-1){0.875}}
\put(0.875,0.875){\vector(1,1){0.875}}
\put(3.5,3.975){\makebox(0,0){$\scriptstyle{}\Sigma_{
C^{\prime}}$}}
\put(5.25,3.975){\makebox(0,0){$\scriptstyle{}\Sigma_{
B^{\prime}}$}}
\put(1.75,3.975){\makebox(0,0){$\scriptstyle{}\Sigma_{
A^{\prime}}$}}
\end{picture}
\vspace{-1.125cm}
\end{center}
\caption{Truncated contour $\Sigma^{\prime}$.}
\end{figure}
\begin{eqnarray}
\Sigma^{\prime} = \Sigma \setminus (L_{\widehat{\delta}} 
\cup \overline{L_{\widehat{\delta}}} \cup {\widehat \Gamma}). 
\nonumber 
\end{eqnarray}
Instead of (10), {}from Lemma~4.1, one can write 
\begin{eqnarray}
P(x,t) = i \! \lim\limits_{{\lambda \rightarrow \infty \atop 
\lambda \in \Omega_{1}}} \! (\lambda [\sigma_{3},m^{\sharp} 
(x,t; \lambda)])_{21}, \, \, \, \, \, \, \, \, Q(x,t) = 
\overline{P(x,t)};
\end{eqnarray}
whence, following the Beals and Coifman formulation 
(Theorem~3.1), by setting
\begin{eqnarray}
w^{\sharp}_{x,t,\delta} = (w_{+}^{\sharp})_{x,t,\delta} 
+ (w_{-}^{\sharp})_{x,t,\delta}, \nonumber
\end{eqnarray}
where $(w_{\pm}^{\sharp})_{x,t,\delta}$ are given 
in~(49)--(52), one gets the following integral representation 
for $P(x,t)$,
\begin{eqnarray}
P(x,t) = - i \left(\int_{\Sigma} [\sigma_{3},((\underline{{\bf
Id}} - C_{w_{x,t,\delta}^{\sharp}})^{-1} {\rm I})(\xi) 
w^{\sharp}_{x,t,\delta}(\xi)] \frac{d \xi}{2 \pi i} \right)_{21},
\end{eqnarray}
where $(\underline{{\bf Id}} - C_{w_{x,t,\delta}^{\sharp}} )$ 
is invertible as an operator in ${\cal L}^{(2 \times 2)}_{2} 
(\Sigma) + {\cal L}^{(2 \times 2)}_{\infty}(\Sigma)$ \cite{a14}.
To reduce the RH problem on $\Sigma$ to the one on the contour 
$\Sigma^{\prime}$, we need to estimate corrections which arise 
as a result of the corresponding reduction of the integration 
contour in (57). To do this, we decompose $w_{x,t,\delta}^{
\sharp}$ as
\begin{eqnarray}
w_{x,t,\delta}^{\sharp} = w^{e} + w^{\prime}, \, \, \, \, \, 
\, \, \, w^{e} \equiv w^{a} + w^{b} + w^{c}:
\end{eqnarray}
(i) $w^{a} = w_{x,t,\delta}^{\sharp} \! \! \mid_{\widehat{
\Gamma}}$ is the restriction of $w_{x,t,\delta}^{\sharp}$ 
to ${\widehat \Gamma}$ and consists of the contribution to 
$w_{x,t,\delta}^{\sharp}$ {}from $h_{I}(\lambda)$ and 
$\overline{h_{I}(\lambda)}$; (ii) $w^{b}$ has support on 
$L \cup \overline{L}$ and consists of the contribution 
to $w_{x,t,\delta}^{\sharp}$ {}from $h_{II}(\lambda)$ and 
$\overline{h_{II}(\overline{\lambda})}$; and (iii) $w^{c}$ 
has support on $L_{\widehat{\delta}} \cup \overline{L_{
\widehat{\delta}}}$ and consists of the contribution 
to $w_{x,t,\delta}^{\sharp}$ {}from $R(\lambda)$ and 
$\overline{R(\overline{\lambda})}$. The main idea of the 
estimation of the integral in (57) will be (based on 
Proposition~4.2) to show that, as $t \! \rightarrow \! + 
\infty$ $(x / t \! \sim \! {\cal O}(1))$, $w^{\prime} 
\vert_{\Sigma \setminus \Sigma^{\prime}} \! \to \! 0$ 
and $w^{e} \! \to \! 0$, in the sense of appropriately 
defined operator norms, and, therefore, to ``lump'' the 
contribution to $w_{x,t,\delta}^{\sharp}$ {}from 
$R(\lambda)$ and $\overline{R(\overline{\lambda})}$ into 
the factor $w^{\prime}$, which is supported on 
$\Sigma^{\prime}$, and show that it encapsulates the 
leading order asymptotics. 
\setcounter{nine}{0}
\begin{nine}
For arbitrary $l \! \in \! \Bbb Z_{\geq 1}$ and sufficiently 
small $\widehat{\delta} \! \in \! \Bbb R_{> 0}$, as $t \! 
\rightarrow \! + \infty$ such that $\lambda_{0} \! > \! M$,
\begin{eqnarray*}
&\vert \vert w^{a} \vert \vert_{{\cal L}^{(2 \times 2)}_{k}
({\widehat \Gamma})} \! \leq \! \frac{\underline{c}}{
(\lambda_{0}^{2} t)^{l}}, \, \, \, \vert \vert w^{b} \vert 
\vert_{{\cal L}^{(2 \times 2)}_{k}(L \cup \overline{L})} \! 
\leq \! \frac{\underline{c}}{(\lambda_{0}^{2} t)^{l}}, \, \, 
\, \vert \vert w^{c} \vert \vert_{{\cal L}^{(2 \times 2)}_{k}
(L_{\widehat{\delta}} \cup \overline{L_{\widehat{\delta}}})} 
\! \leq \! \underline{c} e^{- 2 \lambda_{0}^{4} \widehat{
\delta}^{2} t},& \\
&\vert \vert w^{\prime} \vert \vert_{{\cal L}^{(2 \times 
2)}_{2}(\Sigma)} \! \leq \! \frac{\underline{c}}{(\lambda_{
0}^{2} t)^{1/4}}, \, \, \, \, \, \vert \vert w^{\prime} \vert 
\vert_{{\cal L}^{(2 \times 2)}_{1}(\Sigma)} \! \leq \! 
\frac{\underline{c}}{\sqrt{\lambda_{0}^{2} t}},& 
\end{eqnarray*}
where $k \! \in \! \{1,2,\infty\}$. 
\end{nine}

{\em Proof.\/} Consequence of Proposition~4.2, the following 
estimates,  
$$
(w_{\pm}^{\sharp})_{x,t,\delta}, \, \, w^{\sharp}_{x,t,\delta} 
\in {\cal L}_{k}^{(2 \times 2)}(\Sigma) \cap {\cal L}_{
\infty}^{(2 \times 2)}(\Sigma), \, \, \, \, \, k \in \{1,2\},
$$ 
and analogous calculations as in Lemma~2.13 of \cite{a15}: for 
details, see \cite{a19}. \ \ \rule{6pt}{6pt}

\setcounter{rov}{0}
\begin{rov}
Denote by ${\cal N}(\cdot)$ the space of bounded linear 
operators acting in ${\cal L}^{(2 \times 2)}_{2}(\cdot)$. 
\end{rov}
       
\setcounter{ten}{1}
\begin{ten}
As $t \! \rightarrow \! + \infty$ such that $\lambda_{0} 
\! > \! M$, $(\underline{{\bf Id}} - C_{w^{\sharp}_{x,t,
\delta}})^{-1} \in {\cal N}(\Sigma) \Leftrightarrow 
(\underline{{\bf Id}} - C_{w^{\prime}})^{-1} \in {\cal 
N}(\Sigma)$.  
\end{ten}

{\em Proof.\/} Consequence of the following inequality, 
$\vert \vert C_{w_{x,t,\delta}^{\sharp}} \! - C_{w^{
\prime}} \vert \vert_{{\cal N}(\Sigma)} \! \leq \! 
\underline{c} \vert \vert w^{e} \vert \vert_{{\cal L}^{
(2 \times 2)}_{\infty}(\Sigma)}$, the fact that $\vert 
\vert w^{e} \vert \vert_{{\cal L}^{(2 \times 2)}_{\infty}
(\Sigma)} (\leq \underline{c} \vert x \vert^{-l}) \! 
\leq \! \underline{c}(\lambda_{0}^{2} t)^{-l} \! \leq 
\! \underline{c}$ ((58) and Lemma~5.1), and the second 
resolvent identity. \ \ \rule{6pt}{6pt}

{\em Remark 5.1.\/} Actually, the operator $(\underline{{\bf 
Id}} - C_{w^{\prime}})^{-1}$ acts in ${\cal L}^{(2 \times 
2)}_{2}(\Sigma) + {\cal L}^{(2 \times 2)}_{\infty}(\Sigma)$,
and the operator $C_{w_{x,t,\delta}^{\sharp}}$ acts {}from 
${\cal L}^{(2 \times 2)}_{2}(\Sigma) + {\cal L}^{(2 \times 
2)}_{\infty}(\Sigma)$ into ${\cal L}^{(2 \times 2)}_{2}
(\Sigma)$: we consider their restrictions to ${\cal L}^{(2 
\times 2)}_{2}(\Sigma)$. 

\setcounter{twelve}{0}
\begin{twelve}
If $(\underline{{\bf Id}} - C_{w^{\prime}})^{-1} \! \in \! 
{\cal N}(\Sigma)$, then for arbitrary $l \! \in \! \Bbb Z_{
\geq 1}$, as $t \! \rightarrow \! + \infty$ such that 
$\lambda_{0} \! > \! M$,
\begin{eqnarray}
P(x,t) = - i \left(\int_{\Sigma} [\sigma_{3},((\underline{{
\bf Id}} - C_{w^{\prime}})^{-1} {\rm I})(\xi) w^{\prime}
(\xi)] \frac{d \xi}{2 \pi i} \right)_{21} + {\cal O} \! 
\left( \frac{\underline{c}}{(\lambda_{0}^{2} t)^{l}} 
\right) \!.
\end{eqnarray}
\end{twelve}

{\em Proof.\/} {}From the second resolvent identity, one can 
derive the following expression (see Eq.~(2.27) in \cite{a15}):
\begin{eqnarray*}
&\int_{\Sigma} ((\underline{{\bf Id}} - C_{w_{x,t,\delta}^{
\sharp}})^{-1} {\rm I}) (\xi) w_{x,t,\delta}^{\sharp}(\xi) 
\frac{d \xi}{2 \pi i} \! = \! \int_{\Sigma} ((\underline{{\bf 
Id}} - C_{w^{\prime}})^{-1} {\rm I}) (\xi) w^{\prime}(\xi) 
\frac{d \xi}{2 \pi i} \! + \! I \! + \! II \! + \! III \! + 
\! IV,&
\end{eqnarray*}
where
\begin{eqnarray*}
&\left. \begin{array}{l}
I \! \equiv \! \int_{\Sigma} w^{e} (\xi) \frac{d \xi}{2 \pi 
i}, IV \! \equiv \! \int_{\Sigma} (((\underline{{\bf Id}} \! 
- \! C_{w^{\prime}})^{-1} C_{w^{e}} (\underline{{\bf Id}} \! 
- \! C_{w_{x,t,\delta}^{\sharp}})^{-1})(C_{w_{x,t,\delta}^{
\sharp}} {\rm I}))(\xi) w_{x,t,\delta}^{\sharp} (\xi) \frac{
d \xi}{2 \pi i}, \\
II \! \equiv \! \int_{\Sigma}((\underline{{\bf Id}} \! - \! 
C_{w^{\prime}})^{-1}(C_{w^{e}}{\rm I}))(\xi) w_{x,t,\delta
}^{\sharp}(\xi) \frac{d \xi}{2 \pi i}, III \! \equiv \! 
\int_{\Sigma}((\underline{{\bf Id}} \! - \! C_{w^{\prime}}
)^{-1} (C_{w^{\prime}}{\rm I}))(\xi) w^{e}(\xi) \frac{d 
\xi}{2 \pi i}.
\end{array} \right. 
\end{eqnarray*}

The terms $I$, $II$, $III$, and $IV$ have, respectively, 
for $l \! \in \! \Bbb Z_{\geq 1}$, the following estimate, 
${\cal O}(\underline{c} \vert x \vert^{-l})$ (hence, using 
the relation $\vert x \vert \! = \! 4 \lambda_{0}^{2} t$, 
${\cal O}(\underline{c}(\lambda_{0}^{2} t)^{-l}))$. Let us 
prove, for example, the last estimate: 
\begin{eqnarray*}
&\vert IV \vert \! \leq \! \vert \vert (\underline{{\bf Id}} 
\! - \! C_{w^{\prime}})^{-1} C_{w^{e}}(\underline{{\bf Id}} \! 
- \! C_{w_{x,t,\delta}^{\sharp}})^{-1} (C_{w_{x,t,\delta}^{
\sharp}} {\rm I}) \vert \vert_{{\cal L}^{(2\times2)}_{2}(\Sigma)}
\vert \vert w_{x,t,\delta}^{\sharp} \vert \vert_{{\cal L}^{(2 
\times 2)}_{2}(\Sigma)}& \\
&\! \! \! \leq \! \vert \vert (\underline{{\bf Id}} \! - \! 
C_{w^{\prime}})^{-1} \vert \vert_{{\cal N}(\Sigma)} \vert \vert 
C_{w^{e}} \vert \vert_{{\cal N}(\Sigma)} \vert \vert 
(\underline{{\bf Id}} \! - \! C_{w_{x,t,\delta}^{\sharp}})^{-1} 
\vert \vert_{{\cal N}(\Sigma)} \vert \vert C_{w_{x,t,\delta}^{
\sharp}} \! {\rm I} \vert \vert_{{\cal L}^{(2 \times 2)}_{2} 
(\Sigma)} \vert \vert w_{x,t,\delta}^{\sharp} \vert \vert_{{\cal 
L}^{(2 \times 2)}_{2}(\Sigma)}& \\ 
&\! \leq \! \underline{c} \vert \vert C_{w^{e}} \vert \vert_{{
\cal N}(\Sigma)} \vert \vert C_{w_{x,t,\delta}^{\sharp}} {\rm 
I} \vert \vert_{{\cal L}^{(2 \times 2)}_{2}(\Sigma)} \vert \vert 
w_{x,t,\delta}^{\sharp} \vert \vert_{{\cal L}^{(2\times2)}_{2}
(\Sigma)} \! \leq \! \underline{c} \vert \vert w^{e} \vert 
\vert_{{\cal L}^{(2 \times 2)}_{\infty}(\Sigma)} (\vert \vert 
w_{x,t,\delta}^{\sharp} \vert \vert_{{\cal L}^{(2 \times 2)}_{
2}(\Sigma)})^{2}.&
\end{eqnarray*}
{}From Lemma~5.1, the arithmetic mean inequality, and the 
Cauchy-Schwarz inequality, one shows that,
\begin{eqnarray*}
(\vert \vert w_{x,t,\delta}^{\sharp} \vert \vert_{{\cal 
L}^{(2 \times 2)}_{2}(\Sigma)})^{2} & \leq & 4 (\vert 
\vert w^{a} \vert \vert_{{\cal L}^{(2 \times 2)}_{2}({
\widehat \Gamma})})^{2} + 4 (\vert \vert w^{b} \vert 
\vert_{{\cal L}^{(2 \times 2)}_{2} (L \cup \overline{
L})})^{2} \\
 & + & 4(\vert \vert w^{c} \vert \vert_{{\cal L}^{(2 
\times 2)}_{2} (L_{\widehat{\delta}} \cup \overline{L_{
\widehat{\delta}}})})^{2} + 4(\vert \vert w^{\prime} 
\vert \vert_{{\cal L}^{(2 \times 2)}_{2}(\Sigma)})^{2} 
\leq \underline{c};
\end{eqnarray*}
whence, estimate $\vert IV \vert$ follows {}from the fact 
that $\vert \vert w^{e} \vert \vert_{{\cal L}^{(2 \times 
2)}_{\infty}(\Sigma)} \! \leq \! \underline{c}(\lambda_{
0}^{2} t)^{-l}$ ((58) and Lemma 5.1). \ \ \rule{6pt}{6pt}

Let us now show that, in the sense of appropriately defined 
operator norms, one may always choose to delete (or add) a 
portion of a contour(s) on which the jump is ${\rm I} \! + 
\! {\cal O}(\frac{\underline{c}}{(\lambda_{0}^{2} t)^{l}})$, 
$l \! \in \! \Bbb Z_{\geq 1}$ and arbitrarily large, and 
without altering the RH problem in the operator sense.

\setcounter{eeh}{1}
\begin{eeh}[{\rm \cite{a15}}] 
Let: (i) $R_{\Sigma^{\prime}} \colon {\cal L}^{(2 \times 2)}_{
2}(\Sigma) \rightarrow {\cal L}^{(2 \times 2)}_{2} (\Sigma^{
\prime})$ denote the restriction map; (ii) $\underline{{\bf 
I}}_{\Sigma^{\prime} \rightarrow \Sigma} \colon {\cal L}^{(2 
\times 2)}_{2}(\Sigma^{\prime}) \rightarrow {\cal L}^{(2 \times 
2)}_{2}(\Sigma)$ denote the embedding; (iii) $C_{w^{\prime}}^{
\Sigma} \colon {\cal L}^{(2 \times 2)}_{2}(\Sigma) \rightarrow 
{\cal L}^{(2 \times 2)}_{2}(\Sigma)$ denote the operator in 
Eq.~(12) with $w \leftrightarrow w^{\prime}$; (iv) $C_{w^{
\prime}}^{\Sigma^{\prime}} \colon {\cal L}^{(2 \times 2)}_{2}
(\Sigma^{\prime}) \rightarrow {\cal L}^{(2 \times 2)}_{2}
(\Sigma^{\prime})$ denote the operator in Eq.~(12) with $w 
\leftrightarrow w^{\prime} \vert_{ \Sigma^{\prime}}$; (v) 
$C_{w^{\prime}}^{E} \colon {\cal L}^{(2 \times 2)}_{2}
(\Sigma^{\prime}) \rightarrow {\cal L}^{(2 \times 2)}_{2}
(\Sigma)$ denote the restriction of $C_{w^{\prime}}^{\Sigma}$ 
to ${\cal L}^{(2 \times 2)}_{2}(\Sigma^{\prime})$; and (vi) 
$\underline{{\bf Id}}_{\Sigma^{\prime}}$ and $\underline{{\bf 
Id}}_{\Sigma}$ denote, respectively, the identity operators 
on ${\cal L}^{(2 \times 2)}_{2} (\Sigma^{\prime})$ and ${\cal 
L}^{(2 \times 2)}_{2}(\Sigma)$.
\end{eeh}
 
\setcounter{thirteen}{2}
\begin{thirteen}
\begin{eqnarray}
&\left. \begin{array}{c}
C_{w^{\prime}}^{\Sigma} C_{w^{\prime}}^{E} = C_{w^{\prime}}^{
E} C_{w^{\prime}}^{\Sigma^{\prime}}, \\
(\underline{{\bf Id}}_{\Sigma^{\prime}} - C_{w^{\prime}}^{
\Sigma^{\prime}})^{-1} = R_{\Sigma^{\prime}} (\underline{{\bf 
Id}}_{\Sigma} - C_{w^{\prime}}^{\Sigma})^{-1} \underline{{\bf 
I}}_{\Sigma^{\prime} \rightarrow \Sigma}, \\ 
( \underline{{\bf Id}}_{\Sigma} - C_{w^{\prime}}^{\Sigma})^{
-1} = \underline{{\bf Id}}_{\Sigma} + C_{w^{\prime}}^{E} 
(\underline{{\bf Id}}_{\Sigma^{\prime}} - C_{w^{\prime}}^{
\Sigma^{\prime}})^{-1} R_{\Sigma^{\prime}}.
\end{array} \right.
\end{eqnarray}
\end{thirteen}

{\em Proof.\/} See Lemma~2.56 in \cite{a15}. \ \ \rule{6pt}{6pt}
 
\setcounter{fourteen}{1}
\begin{fourteen}
If $(\underline{{\bf Id}}_{\Sigma} - C_{w^{\prime}}^{\Sigma}) 
\in {\cal N}(\Sigma)$, then for arbitrary $l \! \in \! \Bbb 
Z_{\geq 1}$, as $t \! \rightarrow \! + \infty$ such that 
$\lambda_{0} \! > \! M$, 
\begin{eqnarray}
P(x,t) = - i \left(\int_{\Sigma^{\prime}} [\sigma_{3},((
\underline{{\bf Id}}_{\Sigma^{\prime}} - C_{w^{\prime}}^{
\Sigma^{\prime}})^{-1} I)(\xi) w^{\prime} (\xi)] \frac{d 
\xi}{2 \pi i} \right)_{21} + {\cal O} \! \left(\frac{
\underline{c}}{(\lambda_{0}^{2} t)^{l}} \right) \!.
\end{eqnarray}
\end{fourteen}

{\em Proof.\/} The boundedness of $\vert \vert (\underline{
{\bf Id}}_{\Sigma^{\prime}} - C_{w^{\prime}}^{\Sigma^{\prime}
})^{-1} \vert \vert_{{\cal N}(\Sigma^{\prime})}$ follows 
{}from the assertion of the Lemma and identity~(60): the 
remainder is a consequence of Proposition~5.1. 
\ \ \rule{6pt}{6pt}

{}From Proposition~5.2, it is clear that the asymptotic 
expansion (to ${\cal O}(\frac{\underline{c}}{(\lambda_{0}^{2} 
t)^{l}})$, $l \! \in \! \Bbb Z_{\geq 1})$ can be constructed 
by means of the following RH problem on the contour 
$\Sigma^{\prime}$,
\begin{eqnarray}
\left. \begin{array}{c}
m_{+}^{\Sigma^{\prime}}(x,t;\lambda) = m_{-}^{\Sigma^{\prime}} 
(x,t;\lambda) v_{x,t,\delta}^{\Sigma^{\prime}}(\lambda), \, \, 
\, \, \, \lambda \in \Sigma^{\prime}, \\  
m^{\Sigma^{\prime}}(x,t;\lambda) \rightarrow {\rm I} \, \, \, \, 
\, {\rm as} \, \, \, \, \, \lambda \rightarrow \infty, \, \, \, 
\, \, \lambda \in \Bbb C \setminus \Sigma^{\prime},
\end{array} \right.
\end{eqnarray}
where $v^{\Sigma^{\prime}}_{x,t,\delta}(\lambda) = ({\rm I} 
- (w^{\Sigma^{\prime}}_{-})_{x,t,\delta})^{-1} ({\rm I} +
(w^{\Sigma^{\prime}}_{+})_{x,t,\delta})$, with 
\begin{eqnarray}
&(w^{\Sigma^{\prime}}_{+})_{x,t,\delta} = (\delta (\lambda))^{
{\rm ad}(\sigma_{3})} e^{- i t \theta(\lambda) {\rm ad}(\sigma_{
3})} R(\lambda) \sigma_{+}, \, \, \, \, \, \, (w^{\Sigma^{
\prime}}_{-})_{x,t,\delta} = 0, \, \, \, \, \, \lambda \in L \! 
\setminus \! L_{\widehat{\delta}},& \\
&(w^{\Sigma^{\prime}}_{+})_{x,t,\delta} = 0, \, \, \, \, \, \,
(w^{\Sigma^{\prime}}_{-})_{x,t,\delta} = - (\delta(\lambda))^{{
\rm ad} (\sigma_{3})} e^{- i t \theta(\lambda) {\rm ad} (\sigma_{
3})} \overline{R(\lambda)} \sigma_{-}, \, \, \, \, \, \lambda \in 
\overline{L} \! \setminus \! \overline{L_{\widehat{\delta}}}.& 
\end{eqnarray}
Denote $(w^{\Sigma^{\prime}})_{x,t,\delta} = (w^{\Sigma^{
\prime}}_{+})_{x,t,\delta} + (w^{\Sigma^{\prime}}_{-})_{x,t,
\delta}$, so that $(w^{\Sigma^{\prime}})_{x,t,\delta} = w^{
\prime} \vert_{\Sigma^{\prime}}$; then, according to Theorem~3.1, 
the solution of RH problem~(62) has the following integral 
representation,  
\begin{eqnarray}
m^{\Sigma^{\prime}}(x,t;\lambda) = {\rm I} + \int_{\Sigma^{\prime}}
\frac{\mu^{\Sigma^{\prime}}(x,t;\xi) (w^{\Sigma^{\prime}}
(\xi))_{x,t,\delta}}{(\xi - \lambda)} \frac{d \xi}{2 \pi i},
\, \, \, \, \, \, \lambda \in \Bbb C \setminus \Sigma^{\prime}, 
\nonumber
\end{eqnarray}
where $\mu^{\Sigma^{\prime}}(x,t;\lambda) \equiv (\underline{{
\bf Id}}_{\Sigma^{\prime}} - C_{w^{\prime}}^{\Sigma^{\prime}})^{
-1} {\rm I}$.

{\em Remark 5.2.\/} In (64), $\overline{R(\lambda)}$ is the 
same piecewise-rational function $R(\lambda)$ appearing in (63), 
except with the complex conjugated coefficients.  

\section{RH Problem on the Disjoint Crosses} 
In this section, we make a further simplification of the RH 
problem on the truncated contour $\Sigma^{\prime}$ by reducing 
it to the one which is stated on the three disjoint crosses, 
$\Sigma_{A^{\prime}}$, $\Sigma_{B^{\prime}}$, and $\Sigma_{C^{
\prime}}$, and prove that the leading term of the asymptotic 
expansion for $P(x,t)$ (Proposition~5.2, (61)) can be written 
as the sum of three terms corresponding to the solutions of 
three auxiliary RH problems, each of which is set on one of 
the crosses; moreover, the solution of the latter RH problem 
can be presented in terms of an exactly solvable model matrix 
RH problem, which is studied in the next section. At the end 
of this section, we also prove the basic bound on 
$(\underline{{\bf Id}}_{\Sigma^{\prime}} - C_{w^{\prime}}^{
\Sigma^{\prime}})^{-1}$ (Proposition~5.2).

Let us prepare the notations which are needed for exact 
formulations. Write $\Sigma^{\prime}$ as the disjoint union 
of the three crosses, $\Sigma_{A^{\prime}}$, $\Sigma_{B^{
\prime}}$, and $\Sigma_{C^{\prime}}$, extend the contours 
$\Sigma_{A^{\prime}}$, $\Sigma_{B^{\prime}}$, and $\Sigma_{
C^{\prime}}$ (with orientations unchanged) to the following 
ones, 
\begin{eqnarray}
&\widehat{\Sigma}_{A^{\prime}} = \{\mathstrut \lambda; \, 
\lambda = - \lambda_{0} + \frac{\lambda_{0} u}{2} e^{\pm 
\frac{i \pi}{4}}, \, u \in \Bbb R \}, \, \, \, \, \, \, \, 
\widehat{\Sigma}_{B^{\prime}} = \{\mathstrut \lambda; \, 
\lambda = \lambda_{0} + \frac{\lambda_{0} u}{2} e^{\pm 
\frac{3 \pi i}{4}}, \, u \in \Bbb R \},& \nonumber \\
&\widehat{\Sigma}_{C^{\prime}} = \{\mathstrut \lambda; \, 
\lambda = \frac{\lambda_{0} u}{2} e^{\pm \frac{i \pi}{4}}, 
\, u \in \Bbb R \},& \nonumber
\end{eqnarray}
and define by $\Sigma_{A}$, $\Sigma_{B}$, and $\Sigma_{C}$, 
respectively, the contours $\{ \mathstrut \lambda; \, \lambda 
= \frac{\lambda_{0} u}{2} e^{\pm \frac{i \pi}{4}}, \, u \in 
\Bbb R \}$ oriented inward as in $\Sigma_{A^{\prime}}$ and 
$\widehat{\Sigma}_{A^{\prime}}$, inward as in $\Sigma_{B^{
\prime}}$ and $\widehat{\Sigma}_{B^{\prime}}$, and inward/outward 
as in $\Sigma_{C^{\prime}}$ and $\widehat{\Sigma}_{C^{\prime}}$. 

For $k \in \{A,B,C\}$, introduce the following operators, 
\begin{eqnarray}
&N_{k} \colon {\cal L}^{2} (\widehat{\Sigma}_{k^{\prime}}) 
\rightarrow {\cal L}^{2} (\Sigma_{k}), \, \, \, \, \, f 
(\lambda) \mapsto (N_{k} f)(\lambda) = f (\lambda_{k} + 
\varepsilon_{k}),& 
\end{eqnarray}
where
\begin{eqnarray}
&\lambda_{A} = - \lambda_{0}, \, \, \, \, \, \lambda_{B} = 
\lambda_{0}, \, \, \, \, \, \lambda_{C} = 0, \, \, \, \, \, 
\varepsilon_{A} = \varepsilon_{B} = \lambda (16 \lambda_{
0}^{2} t)^{-1/2}, \, \, \, \, \, \varepsilon_{C} = \lambda 
(8 \lambda_{0}^{2} t)^{-1/2}.&
\end{eqnarray}
Considering the action of the operators $N_{k}$ on $\delta
(\lambda) e^{-i t \theta (\lambda)}$, we find that, for $k 
\in \{A,B,C\}$, $I_{A} \equiv (- \infty,\lambda_{A})$, $I_{B} 
\equiv (\lambda_{B},+ \infty)$, and $I_{C} \equiv (- 
\lambda_{0},+ \lambda_{0})$,
\begin{eqnarray}
&N_{k} \{\delta(\lambda) e^{-i t \theta (\lambda)}\} = 
\delta_{k}^{0} \delta_{k}^{1}(\lambda), \, \, \, \, \, 
\, \Re (\lambda) \in I_{k},& \nonumber 
\end{eqnarray}
where, as a result of the second of (15), (41), and (42), 
\begin{eqnarray}
&\delta_{l}^{0} \! = \! (16 \lambda_{0}^{4} t)^{- \frac{i 
\nu}{2}} \exp \left \{2 i \lambda_{0}^{4} t \! + \! 
\sum\limits_{m \in {\cal M}} \chi_{m}(\lambda_{l}) \right\}, 
\, \, l \! \in \! \{A,B\}, \, \, \delta_{C}^{0} \! = \! 
\exp \left\{\sum\limits_{m \in {\cal M}} \chi_{m}(0) \right
\} \!,& \\
&\delta_{l}^{1}(\lambda) = \frac{(\lambda \lambda_{0})^{i \nu} 
(\varepsilon_{l} + 2 \lambda_{l})^{i \nu}}{(\varepsilon_{l} + 
\lambda_{l})^{2 i \nu}} \exp \left\{- \frac{i \lambda^{2}}{2} 
(1 + \frac{\varepsilon_{l}}{2 \lambda_{0}})^{2} + \sum\limits_{
m \in {\cal M}} (\chi_{m}(\lambda_{l} + \varepsilon_{l}) - 
\chi_{m} (\lambda_{l})) \right\} \!,& \\ 
&\delta_{C}^{1}(\lambda) = (\varepsilon_{C}^{2} - \lambda_{
0}^{2})^{i\nu} \exp \left\{\frac{i \lambda^{2}}{2} (1 - 
\frac{4 \varepsilon_{C}^{2}}{\lambda_{0}^{2}}) + \sum\limits_{
m \in {\cal M}} (\chi_{m}(\varepsilon_{C}) - \chi_{m} (0)) 
\right \} \!,&
\end{eqnarray}
with ${\cal M} \equiv \{A,B,+,-\}$, and
\begin{eqnarray}
&\chi_{k}(\lambda_{l}) \! = \! \frac{1}{2 \pi i} \int_{0}^{
\lambda_{k}} \! \ln \! \left(\! \frac{1 - \vert r(\mu) 
\vert^{2}}{1 - \vert r(\lambda_{0}) \vert^{2}} \! \right) \! 
\frac{d \mu}{(\mu - \lambda_{l})}, \, \, \, \chi_{\pm}
(\lambda_{l}) \! = \! \frac{i}{2 \pi} \! \int_{\pm \infty}^{0} 
\! \ln \! \vert i \mu - \lambda_{l} \vert d \ln (1 + \vert r(i 
\mu) \vert^{2}),& \\  
&\chi_{k} (0) = \frac{i}{2 \pi} \int_{0}^{\lambda_{k}} \ln \! 
\vert \mu \vert \, d \ln (1 - \vert r(\mu) \vert^{2}), \, \, 
\, \, \chi_{\pm}(0) = \int_{\pm \infty}^{0} \frac{\ln (1 + 
\vert r(i \mu) \vert^{2})}{\mu} \frac{d \mu}{2 \pi i}.& 
\nonumber
\end{eqnarray}

Set
\begin{eqnarray}
&\Delta_{k}^{0} = (\delta_{k}^{0} )^{\sigma_{3}},& 
\end{eqnarray}
and let $\widetilde{\Delta_{k}^{0}}$ denote right 
multiplication by $\Delta_{k}^{0}$: 
\begin{eqnarray}
&\widetilde{\Delta_{k}^{0}} \phi \equiv \phi \Delta_{k}^{0}.& 
\end{eqnarray}

Denote
\begin{eqnarray}
&w^{k^{\prime}} (\lambda) = 
\left\{ \begin{array}{l}
(w^{\Sigma^{\prime}})_{x,t,\delta}, \, \, \, \, \, \lambda 
\in \Sigma_{k^{\prime}}, \\
0, \, \, \, \, \, \lambda \in \Sigma^{\prime} \setminus 
\Sigma_{k^{\prime}},
\end{array} \right. \, \, \, {\rm and} \, \, \, \, \, \, 
\widehat{w}^{k^{\prime}}(\lambda) = 
\left\{ \begin{array}{l}
w^{k^{\prime}}, \, \, \, \, \, \lambda \in \Sigma_{k^{
\prime}}, \\ 0, \, \, \, \, \, \lambda \in \widehat{\Sigma}_{
k^{\prime}} \setminus \Sigma_{k^{\prime}}.
\end{array} \right.&
\end{eqnarray}
According to this,
\begin{eqnarray} 
&(w^{\Sigma^{\prime}})_{x,t,\delta} = \sum\limits_{k \in 
\{A,B,C\}} w^{k^{\prime}}, \, \, \, \, \, 
C^{\Sigma^{\prime}}_{(w^{\Sigma^{\prime}})_{x,t,\delta}}
\equiv C^{\Sigma^{\prime}}_{w^{\prime}} =
\sum\limits_{k \in \{A,B,C\}} \! C_{w^{k^{\prime}}}^{
\Sigma^{\prime}} \equiv \sum\limits_{k \in \{A,B,C\}} 
\! C_{w^{k^{\prime}}}^{\Sigma_{k^{\prime}}}:&
\end{eqnarray} 
hereafter, we do not introduce a special notation for 
$w^{k^{\prime}} \vert_{\Sigma_{k^{\prime}}^{\prime}}$. Let 
us prove some technical results concerning the operators
$C_{w^{k^{\prime}}}^{\Sigma_{k^{\prime}}}$ and $C_{\widehat{
w}^{k^{\prime}}}^{\widehat{\Sigma}_{k^{\prime}}}$.
\setcounter{seep}{0}
\begin{seep}
For $k \in \{A,B,C\}$, 
\begin{eqnarray}
&C_{\widehat{w}^{k^{\prime}}}^{\widehat{\Sigma}_{k^{\prime}}} 
= (N_{k})^{-1}(\widetilde{\Delta_{k}^{0}})^{-1} C_{w^{k}}^{
\Sigma_{k}}(\widetilde{\Delta_{k}^{0}}) N_{k}, \, \, \, \, \, 
\, \, w^{k} \equiv (\Delta_{k}^{0})^{-1} (N_{k} \widehat{w}^{
k^{\prime}})(\Delta_{k}^{0}),&
\end{eqnarray}
where
\begin{eqnarray}
&\left. \begin{array}{c}
C_{w^{k}}^{\Sigma_{k}} \vert_{{\cal L}^{(2 \times 2)}_{2} 
(\overline{L_{k}})} = - C_{+} (\cdot (\overline{R 
(\varepsilon_{k} + \lambda_{k})} (\delta_{k}^{1} (\lambda))^{
-2} \sigma_{-})), \\ 
C_{w^{k}}^{\Sigma_{k}} \vert_{{\cal L}^{(2 \times 2)}_{2}
(L_{k})} = C_{-} (\cdot (R(\varepsilon_{k} + \lambda_{k})
(\delta_{k}^{1} (\lambda))^{2} \sigma_{+})).
\end{array} \right.&
\end{eqnarray}
Here, the rays $L_{k}$ are defined as follows,
\begin{eqnarray} 
&L_{l} \equiv \{ \mathstrut \lambda; \, \lambda = \frac{
\lambda_{l} u}{2} (16 \lambda_{0}^{2} t)^{1/2} e^{- 
\frac{i \pi}{4}}, \, u \in (- \varepsilon,+ \infty) \}, \, 
\, \, \, \, l \in \{A,B\},& 
\nonumber \\
&L_{C} \equiv \{ \mathstrut \lambda; \, \lambda = \frac{ 
\lambda_{0} u}{2} (8 \lambda_{0}^{2} t)^{1/2} e^{\frac{i 
\pi}{4}}, \, u \in \Bbb R \},& \nonumber
\end{eqnarray} 
so that $\Sigma_{k^{\prime}} = L_{k} \cup \overline{L_{k}}$. 
\end{seep}

{\em Proof.\/} We consider the case $k \! = \! B$: the cases 
$k \! = \! A$ and $C$ follow in an analogous manner. Since, 
{}from the first of (67), $\vert \delta_{B}^{0} \vert = 1$, 
it follows {}from the definition of the operator $\widetilde{
\Delta_{B}^{0}}$ in (72) that $(\widetilde{\Delta_{B}^{0}})^{
\dagger} = (\widetilde{\Delta_{B}^{0}})^{-1}$, where $\dagger$ 
denotes Hermitian conjugation $(\widetilde{\Delta_{B}^{0}}$ is 
a unitary operator). Changing variables ((65)), recalling (12), 
using (71) and (72), as well as the unitarity of $\widetilde{
\Delta_{B}^{0}}$, one obtains (75), where $C_{w^{B}}^{\Sigma_{
B}} = C_{(\Delta_{B}^{0})^{-1}(N_{B} \widehat{w}^{B^{\prime}})
(\Delta_{B}^{0})}$ $= C_{+} (\cdot (\Delta_{B}^{0})^{-1} (N_{B} 
\widehat{w}_{-}^{B^{\prime}}) (\Delta_{B}^{0})) + C_{-} (\cdot 
(\Delta_{B}^{0})^{-1} (N_{B} \widehat{w}_{+}^{B^{\prime}}) 
(\Delta_{B}^{0}))$. Using (63) and (64), one shows that $((
\Delta_{B}^{0})^{-1} (N_{B} \widehat{w}_{+}^{B^{\prime}}) 
(\Delta_{B}^{0}))(\lambda) = 0$ on $\overline{L_{B}}$ and 
$((\Delta_{B}^{0})^{-1} (N_{B} \widehat{w}_{-}^{B^{\prime}}) 
(\Delta_{B}^{0})) (\lambda) = 0$ on $L_{B}$, so that (76) are 
valid. \ \ \rule{6pt}{6pt}

\setcounter{sixteen}{0}
\begin{sixteen}
Let $\kappa \! \in \! (0,1)$, $p(B) \! = \! - p(A) \! = \! 1$, 
$p(C) \! = \! - \lambda_{0}^{2}/(2 \lambda)$, ${\rm sgn\/}(A) \! 
= \! {\rm sgn\/}(B) \! = \! - {\rm sgn\/}(C) \! = \! 1$. Then 
$\forall \, \lambda \! \in \! \overline{L_{k}} \! \subset \! 
\Sigma_{k}$, as $t \! \rightarrow \! + \infty$ such that 
$\lambda_{0} \! > \! M$,
\begin{eqnarray}
&\vert \overline{R(\varepsilon_{k} + \lambda_{k})} 
(\delta_{k}^{1} (\lambda))^{-2} - \overline{R(\lambda_{
k}^{\pm})} (2 \lambda p(k))^{- 2 i \nu} e^{i {\rm sgn}(k) 
\lambda^{2}} \vert \leq \frac{c(k) \ln t}{\sqrt{\lambda_{
0}^{2} t}} e^{- 2 \kappa \lambda_{0}^{4} t u^{2}},&  
\end{eqnarray}
and $\forall \, \lambda \! \in \! L_{k} \! \subset \! 
\Sigma_{k}$,
\begin{eqnarray}
&\vert R (\varepsilon_{k} + \lambda_{k}) (\delta_{k}^{1} 
(\lambda))^{2} - R(\lambda_{k}^{\pm})(2 \lambda p(k))^{2 i 
\nu} e^{- i {\rm sgn}(k) \lambda^{2}} \vert \leq \frac{c(k) 
\ln t}{\sqrt{\lambda_{0}^{2} t}} e^{- 2 \kappa \lambda_{
0}^{4} t u^{2}},& 
\end{eqnarray}  
where $L_{k}$ (resp. $\overline{L_{k}})$, $k \in \{A,B,C\}$, 
are defined in Proposition~6.1, $u \in (-\varepsilon,+ 
\infty)$, with $0$ $< \varepsilon < \sqrt{2}$, $R(\lambda_{
0}^{+}) = \lim\limits_{\Re (\lambda) \downarrow \lambda_{0}} 
\! R(\lambda) = - r(\lambda_{0})$, $R(\lambda_{0}^{-}) = 
\lim\limits_{\Re (\lambda) \uparrow \lambda_{0}} \! R(\lambda) 
= r(\lambda_{0})(1 - \vert r(\lambda_{0}) \vert^{2})^{-1}$, 
$R(- \lambda_{0}^{+}) = \lim\limits_{\Re (\lambda) \downarrow 
- \lambda_{0}} \! R(\lambda) = - r(\lambda_{0}) (1 - \vert 
r(\lambda_{0}) \vert^{2})^{-1}$, $R(- \lambda_{0}^{-}) = 
\lim\limits_{\Re (\lambda) \uparrow - \lambda_{0}} \! R 
(\lambda) = r( \lambda_{0})$, $R(0^{\pm}) = 0$ since $r(0) \! 
= \! r(i0) \! = \! 0$, $c(B) \! = \! c^{{\cal S}}$, $c(A) \! 
= \! c^{{\cal S}}$, and $c(C) \! = \! \underline{c}$.
\end{sixteen}

{\em Proof.\/} We prove inequality~(77) for $k \! = \! B$: 
the other results follow in an analogous manner. By using 
(68), for $\lambda \in \overline{L_{B}} \subset \Sigma_{B}$ 
and $\kappa \in (0,1)$, write
\begin{eqnarray}
\overline{R( \lambda_{0} + \varepsilon_{B} )} 
(\delta_{B}^{1} (\lambda))^{-2} - \overline{R(\lambda_{0}^{
\pm})} (2 \lambda)^{- 2 i \nu} e^{i \lambda^{2} } = e^{\frac{
i \kappa \lambda^{2}}{2}} (I + II + III), \nonumber
\end{eqnarray}
where
\begin{eqnarray}
I&=&(2 \lambda)^{-2 i \nu} e^{i(1-\frac{\kappa}{2}) 
\lambda^{2}} (\overline{R(\lambda_{0} + \varepsilon_{B})} 
- \overline{R( \lambda_{0}^{\pm})}), \nonumber \\
II&=&(2 \lambda)^{-2 i \nu} e^{i \left(1 - \frac{\kappa}{2} 
\right) \lambda^{2}} \overline{R(\lambda_{0} + \varepsilon_{
B})} \left[\left(\frac{\lambda_0(\lambda_0+\varepsilon_B/2)}
{(\lambda_0+\varepsilon_B)^2}\right)^{-2i\nu}-1\right], 
\nonumber \\
III&=&(2 \lambda)^{-2 i \nu} e^{i \left(1 - \frac{\kappa}{2} 
\right) \lambda^{2}} \overline{R(\lambda_{0} + \varepsilon_{
B})} \left( \frac{\lambda_{0} (\lambda_0+\varepsilon_B/2)}
{(\lambda_0+\varepsilon_B)^2}\right)^{-2i\nu} \! \! \left(
e^{{\cal Z}} -1 \right), \nonumber 
\end{eqnarray}
and, in the last formula, ${\cal Z} \equiv i \lambda^{2} 
\left(\frac{\varepsilon_{B}^{2}}{ 4 \lambda_{0}^{2}} + 
\frac{\varepsilon_{B}}{\lambda_{0}} \right)- 2 \sum\limits_{
m \in {\cal M}} (\chi_{m} (\lambda_{0} + \varepsilon_{B}) - 
\chi_{m} (\lambda_{0}))$. Note that $\vert \exp \{ \frac{i 
\kappa \lambda^{2}}{2} \} \vert$ $=$ $\exp \{- 2 \kappa 
\lambda_{0}^{4} t u^{2} \}$, which gives the exponential 
factor in (77). The terms $I$, $II$, and $III$ can be 
estimated in the following way.
$$
\vert I \vert \leq \vert (2 \lambda)^{-2 i \nu} \vert 
\left\vert e^{i (1 - \frac{\kappa}{2}) \lambda^{2} } 
\right\vert \vert \varepsilon_{B} \vert
\sup\limits_{\lambda \in \overline{L_{B}}} 
\left\vert \partial_{\lambda} \overline{R (\lambda)} 
\right\vert
\leq e^{\frac{\pi \nu}{2}}
\frac{\vert \lambda \vert e^{- (1 - \frac{\kappa}{2}) 
\vert \lambda \vert^{2}}}{\sqrt{16 \lambda_{0}^{2} t}}
\left\vert \left\vert \overline{R^{\prime}} \right\vert 
\right\vert_{{\cal L}^{\infty} \left(\overline{L_{B}} 
\right)} \leq \frac{\underline{c}}{\sqrt{\lambda_{0}^{
2} t}}, \nonumber  
$$ 
where $\underline{c}$ is independent of $\lambda$. 
$$
\vert II \vert \leq e^{\frac{\pi \nu}{2}} 
e^{-(1-\frac{\kappa}{2})
\vert \lambda \vert^{2} } \left\vert \left\vert 
\overline{R} \right\vert \right\vert_{ {\cal L}^{\infty} 
\left( \overline{L_{B}} \right) } \vert II_{1} + II_{2} 
\vert, 
$$
where
\begin{eqnarray}
&II_{1} = \left (\frac{\lambda_0(\lambda_0+\varepsilon_B/2)}
{(\lambda_0+\varepsilon_B)^2}\right)^{-2i\nu}
\left\{ 1 - \left( \frac{ \lambda_{0} }{
\lambda_{0} + \varepsilon_{B}/2 } \right)^{-2 i \nu} \right\}, 
\, \, \, \, \, II_{2} = \left\{\left(\frac{\lambda_{0}}{
\lambda_{0} + \varepsilon_{B}} \right)^{-4 i \nu} - 1 
\right \} \!.& \nonumber
\end{eqnarray}
To estimate $\vert II_{2} \vert$, one proceeds as follows:
\begin{eqnarray*}
&e^{ - (1-\frac{\kappa}{2}) \vert \lambda \vert^{2} } \vert 
II_{2} \vert \leq e^{ - (1-\frac{\kappa}{2}) \vert \lambda 
\vert^{2} } \left\vert \int_{1}^{ 1 + \varepsilon_B/ 
\lambda_0} \xi^{4 i \nu -1} (4 i \nu ) d \xi \right\vert& \\
&\leq 4 \nu e^{ - (1-\frac{\kappa}{2}) \vert \lambda 
\vert^{2}} \sup \{\mathstrut \vert \xi^{2 i \nu -1} \vert; 
\mathstrut \xi = 1 + \frac{s}{\lambda_{0}} \varepsilon_{B}, 
\mathstrut s \in [0,1] \} \frac{\vert \varepsilon_{B} 
\vert}{\lambda_{0}} \leq \frac{\underline{c}}{\lambda_{0}^{
2}} t^{-1/2},& 
\end{eqnarray*}
since $\vert \xi^{2 i \nu -1} \vert \leq \sqrt{2} \exp \{- 2 
\nu \arg (\xi) \}$ and $-\frac{3 \pi}{4} < \arg(\xi) < \frac{
\pi}{4}$: in this estimation, one uses the fact that $\vert 
\xi \vert \geq 1 / \sqrt{2}$ and $0 < \nu \leq \nu_{\max} 
\equiv - \frac{1}{2 \pi} \ln (1 - \sup\limits_{\mu \in \widehat{
\Gamma}} \overline{r(\overline{\mu})} r(\mu)) < \infty$. Since 
the first term on the right-hand side of the equation for 
$II_{1}$ is bounded, namely, $$\left\vert \left 
(\frac{\lambda_0(\lambda_0 + \varepsilon_B/2)}{(\lambda_0
+ \varepsilon_B)^2}\right)^{-2i\nu} \right\vert \leq e^{\frac{
\pi \nu}{2}},$$ one gets an analogous estimate for $\vert 
II_{1} \vert$.
\begin{eqnarray}
\vert III \vert & \leq & e^{\pi \nu} \left\vert 
\left\vert \overline{R} \right\vert \right\vert_{ 
{\cal L}^{\infty} (\overline{L_{B}})} e^{-(1-\frac{
\kappa}{2}) \vert \lambda \vert^{2}} \sup\limits_{0 
\leq s \leq 1} \left\vert \frac{d}{d s} e^{s {\cal 
Z}} \right\vert \leq \underline{c} e^{-(1-\frac{
\kappa}{2}) \vert \lambda \vert^{2}} \vert {\cal Z} 
\vert \sup\limits_{0 \leq s \leq 1} \vert e^{s {\cal 
Z}} \vert \nonumber \\
 & \leq & \underline{c} e^{- (1 - \frac{\kappa}{2} - 
\frac{\varepsilon s}{2 \sqrt{2}}) \vert \lambda 
\vert^{2}} \vert {\cal Z} \vert \leq \underline{c} e^{- 
(1 - \frac{\kappa}{2} - \frac{\varepsilon s}{2 \sqrt{2}}) 
\vert \lambda \vert^{2}} \left(\frac{\lambda^{4}}{64 
\lambda_{0}^{4} t} + \frac{\lambda^{3}}{\sqrt{16 
\lambda_{0}^{4} t}} + {\cal O} \! \left(\frac{c^{{\cal 
S}} \ln t}{\sqrt{\lambda_{0}^{2} t}} \right) \! \right) 
\!, \nonumber 
\end{eqnarray}
where one uses the boundedness of $\exp \{ \sum\limits_{m 
\in {\cal M}}(\chi_{m}(\lambda_{0} + \varepsilon_{B}) - 
\chi_{m}(\lambda_{0})) \},$ which follows {}from (44), (67), 
and (68). The term ${\cal O}(c^{{\cal S}} (\lambda_{0}^{2} 
t)^{-1/2} \ln t)$ appears in $\vert III \vert$ due to the 
estimation $\chi_{l}(\lambda_{0} + \varepsilon_{B}) - 
\chi_{l}(\lambda_{0})$, $l \! \in \! \{A,B\}$, which can 
be obtained by using the Lipschitz property of the function 
$\ln \! \left(\! \frac{1 - \vert r(\mu) \vert^{2}}{1 - \vert 
r(\lambda_{0}) \vert^{2}} \! \right)$, $\vert \mu \vert \! 
< \! \lambda_{0}$, and integrating by parts the first of (70): 
analogous estimations for $\chi_{\pm}(\lambda_{0} + 
\varepsilon_{B}) - \chi_{\pm}(\lambda_{0})$ are ${\cal O} 
(\underline{c} (\lambda_{0}^{2} t)^{-1/2})$. 
\ \ \rule{6pt}{6pt}

\setcounter{nineteen}{1}
\begin{nineteen}
For general operators $C_{w^{k^{\prime}} }^{\Sigma^{\prime}}$, 
$k \in \{1,2,\ldots,N\}$, if $(\underline{{\bf Id}}_{\Sigma^{
\prime}} - C_{w^{k^{\prime}}}^{\Sigma^{\prime}})^{-1}$ exist, 
then
\begin{eqnarray}  
&(\underline{{\bf Id}}_{\Sigma^{\prime}} + 
\sum\limits_{1 \leq \alpha \leq N} 
C_{w^{\alpha^{\prime}}}^{\Sigma^{\prime}} 
( \underline{{\bf Id}}_{\Sigma^{\prime}} - 
C_{w^{\alpha^{\prime}}}^{\Sigma^{\prime}})^{-1}) 
(\underline{{\bf Id}}_{\Sigma^{\prime}} - 
\sum\limits_{1 \leq \beta \leq N} C_{w^{
\beta^{\prime}}}^{\Sigma^{\prime}}) =& 
\nonumber \\ 
&\underline{{\bf Id}}_{\Sigma^{\prime}} 
- \sum\limits_{1 \leq \alpha \leq N} 
\sum\limits_{1 \leq \beta \leq N} 
( 1 - \delta_{\alpha \beta} ) ( \underline{{\bf 
Id}}_{\Sigma^{\prime}} - 
C_{w^{\alpha^{\prime}}}^{\Sigma^{\prime}} )^{-1} 
C_{w^{\alpha^{\prime}}}^{\Sigma^{\prime}} 
C_{w^{\beta^{\prime}}}^{\Sigma^{\prime}},& \nonumber
\end{eqnarray}
and
\begin{eqnarray}  
&( \underline{{\bf Id}}_{\Sigma^{\prime}} - 
\sum\limits_{1 \leq \beta \leq N} 
C_{w^{\beta^{\prime}}}^{\Sigma^{\prime}} 
) ( \underline{{\bf Id}}_{\Sigma^{\prime}} 
+ \sum\limits_{1 \leq \alpha \leq N} 
C_{w^{\alpha^{\prime}}}^{\Sigma^{\prime}} 
( \underline{{\bf Id}}_{\Sigma^{\prime}} - 
C_{w^{\alpha^{\prime}}}^{\Sigma^{\prime}})^{-1}) 
=& \nonumber \\ 
&\underline{{\bf Id}}_{\Sigma^{\prime}}  
- \sum\limits_{1 \leq \alpha \leq N} 
\sum\limits_{1 \leq \beta \leq N} ( 1 - 
\delta_{\alpha \beta} ) 
C_{w^{\alpha^{\prime}}}^{\Sigma^{\prime}} 
C_{w^{\beta^{\prime}}}^{\Sigma^{\prime}} ( 
\underline{{\bf Id}}_{\Sigma^{\prime}} - 
C_{w^{\beta^{\prime}}}^{\Sigma^{\prime}} )^{-1},& 
\nonumber
\end{eqnarray}
where $\delta_{\alpha \beta}$ is the Kronecker delta.
\end{nineteen}

{\em Proof.\/} Assumption of existence of general operators
$(\underline{{\bf Id}}_{\Sigma^{\prime}} - C_{w^{k^{\prime}
}}^{\Sigma^{\prime}})^{-1}$, $k \in \{1,2,\ldots,N\}$, 
induction, and a straightforward application of the second 
resolvent identity. \ \ \rule{6pt}{6pt}

\setcounter{fifteen}{1}
\begin{fifteen}
For $\alpha \not= \beta \in \{A^{\prime},B^{\prime}, 
C^{\prime}\}$, as $t \! \rightarrow \! + \infty$ such 
that $\lambda_{0} \! > \! M$,
\begin{eqnarray*}
&\vert \vert C_{w^{\alpha}}^{\Sigma^{\prime}} C_{w^{
\beta}}^{\Sigma^{\prime}} \vert \vert_{{\cal N}(\Sigma^{
\prime})} \leq \frac{\underline{c}}{\lambda_{0} \sqrt{t}}, 
\, \, \, \, \, \, \, \vert \vert C_{w^{\alpha}}^{
\Sigma^{\prime}} C_{w^{\beta}}^{\Sigma^{\prime}} 
\vert \vert_{{\cal L}^{(2 \times 2)}_{\infty}
(\Sigma^{\prime}) \rightarrow {\cal L}^{(2 \times 2)}_{2}
(\Sigma^{\prime})} \leq \frac{\underline{c}}{(\lambda_{
0}^{2} t)^{1/4} \sqrt{\lambda_{0}^{4} t}}.& 
\end{eqnarray*}
\end{fifteen}

{\em Proof.\/} Analogous to Lemma~3.5 in \cite{a15}. \ \ 
\rule{6pt}{6pt}

\setcounter{twenty}{2}
\begin{twenty}
If, for $k \in \{A,B,C\}$, $(\underline{{\bf Id}}_{\Sigma_{
k^{\prime}}} - C_{w^{k^{\prime}}}^{\Sigma_{k^{\prime}}})^{-1} 
\in {\cal N} (\Sigma_{k^{\prime}})$, then as $t \! \rightarrow 
\! + \infty$ such that $\lambda_{0} \! > \! M$,
\begin{eqnarray}
&P(x,t) = - i \sum\limits_{k \in \{A,B,C\}} \left(
\int_{\Sigma_{k^{\prime}}} [\sigma_{3},((\underline{{\bf 
Id}}_{\Sigma_{k^{\prime}}} - C_{w^{k^{\prime}}}^{\Sigma_{
k^{\prime}}})^{-1} {\rm I})(\xi) w^{k^{\prime}}(\xi)] 
\frac{d \xi}{2 \pi i} \right)_{21} + {\cal O}(\frac{
\underline{c}}{\lambda_{0} t}).&
\end{eqnarray}
\end{twenty}

{\em Proof.\/} Using Proposition~6.2 and the second resolvent 
identity, one writes 
\begin{eqnarray}
&(\underline{{\bf Id}}_{\Sigma^{\prime}} - 
\sum\limits_{k \in \{A,B,C\}} C_{w^{k^{\prime}}}^{\Sigma^{
\prime}})^{-1} = \underline{{\bf D}}_{\Sigma^{\prime}}
+ \underline{{\bf D}}_{\Sigma^{\prime}} (\underline{{\bf 
Id}}_{\Sigma^{\prime}} - \underline{{\bf E}}_{\Sigma^{\prime}}
)^{-1} \underline{{\bf E}}_{\Sigma^{\prime}},&
\end{eqnarray}
where
\begin{eqnarray}
&\underline{{\bf D}}_{\Sigma^{\prime}}\equiv
\underline{{\bf Id}}_{\Sigma^{\prime}} +
\sum\limits_{k \in \{A,B,C\}} 
C_{w^{k^{\prime}}}^{\Sigma^{\prime}}  
(\underline{{\bf Id}}_{\Sigma^{\prime}} - 
C_{w^{k^{\prime}}}^{\Sigma^{\prime}})^{-1},& \nonumber \\
&\underline{{\bf E}}_{\Sigma^{\prime}} \equiv \sum\limits_{
\alpha,\beta \in \{A,B,C\}} (1 - \delta_{\alpha \beta}) 
C_{w^{\alpha^{\prime}}}^{\Sigma^{\prime}} 
C_{w^{\beta^{\prime}}}^{\Sigma^{\prime}} 
( \underline{{\bf Id}}_{\Sigma^{\prime}} 
- C_{w^{\beta^{\prime}}}^{\Sigma^{\prime}})^{-1}.& 
\nonumber
\end{eqnarray}
Since, {}from Lemma~5.1, $\vert \vert C_{w^{k^{\prime}}}^{
\Sigma^{\prime}} \vert \vert_{{\cal N}(\Sigma^{\prime})} 
\leq \vert \vert w^{k^{\prime}} \vert \vert_{{\cal L}^{(2 
\times 2)}_{2}(\Sigma^{\prime})} \leq \frac{\underline{
c}}{(\lambda_{0}^{2} t)^{1/4}} \leq \underline{c}$, and, 
by assumption $\vert \vert (\underline{{\bf Id}}_{\Sigma^{
\prime}} - C_{w^{k^{\prime}}}^{\Sigma^{\prime}})^{-1} \vert 
\vert_{{\cal N} (\Sigma^{\prime})} < \infty$, via Lemma~6.2, 
one finds that, as $t \! \rightarrow \! + \infty$ such that 
$\lambda_{0} \! > \! M$,
$$
\vert \vert \underline{{\bf D}}_{\Sigma^{\prime}} \vert 
\vert_{{\cal N}(\Sigma^{\prime})} \leq \underline{c}, \, 
\, \, \, \, {\rm and} \, \, \, \, \, \vert \vert 
(\underline{{\bf Id}}_{\Sigma^{\prime}} - \underline{{\bf 
E}}_{\Sigma^{\prime}})^{-1} \vert \vert_{{\cal N} 
(\Sigma^{\prime})} \leq \underline{c}.
$$
Taking account of the second resolvent identity, one also 
finds that, 
\begin{eqnarray}
&\vert \vert \underline{{\bf E}}_{\Sigma^{\prime}}
{\rm I} \vert \vert_{ {\cal L}^{(2 \times 2)}_{2} 
(\Sigma^{\prime})}
\leq \sum\limits_{\alpha, \beta \in \{A,B,C\}} (1 - \delta_{
\alpha \beta}) \vert \vert C_{w^{\alpha^{\prime}}}^{\Sigma^{
\prime}} C_{w^{\beta^{\prime}}}^{\Sigma^{\prime}} {\rm I} \vert 
\vert_{{\cal L}^{(2 \times 2)}_{2} (\Sigma^{\prime})}& 
\nonumber \\
&+ \sum\limits_{\alpha, \beta \in \{A,B,C\}} (1 - \delta_{
\alpha \beta}) \vert \vert C_{w^{\alpha^{\prime}}}^{\Sigma^{
\prime}} C_{w^{\beta^{\prime}}}^{\Sigma^{\prime}} \vert 
\vert_{ {\cal N} (\Sigma^{\prime}) } \vert \vert (\underline{
{\bf Id}}_{\Sigma^{\prime}} - C_{w^{\beta^{\prime}}}^{\Sigma^{
\prime}})^{-1} \vert \vert_{{\cal N} (\Sigma^{\prime})} \vert 
\vert C_{w^{\beta^{\prime}}}^{\Sigma^{\prime}} {\rm I} \vert 
\vert_{{\cal L}^{(2 \times 2)}_{2} (\Sigma^{\prime})}.& 
\nonumber 
\end{eqnarray}
Noting {}from Lemma~5.1 that, 
\begin{eqnarray*}
\vert \vert C_{w^{k^{\prime}}}^{\Sigma^{\prime}} {\rm I} 
\vert \vert_{{\cal L}^{(2 \times 2)}_{2}(\Sigma^{\prime})} 
\leq \vert \vert w^{k^{\prime}} \vert \vert_{{\cal L}^{(2 
\times 2)}_{2}(\Sigma^{\prime})} \leq \frac{\underline{
c}}{(\lambda_{0}^{2} t)^{1/4}}, 
\end{eqnarray*}
one finds, using Lemma~6.2, and applying once more the second 
resolvent identity,
\begin{eqnarray*}
&\vert \vert \underline{{\bf E}}_{\Sigma^{\prime}} {\rm 
I} \vert \vert_{{\cal L}^{(2 \times 2)}_{2} (\Sigma^{
\prime})} \leq \frac{\underline{c}}{\lambda_{0} t^{3/4}}.& 
\end{eqnarray*}
Now, {}from the Cauchy-Schwarz inequality and Lemma~5.1, 
\begin{eqnarray*}
&\vert \vert \underline{{\bf E}}_{\Sigma^{\prime}} w^{
\Sigma^{\prime}} \vert \vert_{ {\cal L}^{(2 \times 2)}_{1} 
(\Sigma^{\prime})} \leq \vert \vert \underline{{\bf E}}_{
\Sigma^{\prime}} {\rm I} \vert \vert_{{\cal L}^{(2 \times 
2)}_{2} (\Sigma^{\prime})} \vert \vert w^{\Sigma^{\prime}} 
\vert \vert_{{\cal L}^{(2 \times 2)}_{2}(\Sigma^{\prime})}
\leq \frac{\underline{c}}{\lambda_{0} t^{3/4}} \frac{
\underline{c}}{(\lambda_{0}^{2} t)^{1/4}} \leq \frac{
\underline{c}}{\lambda_{0} t};&
\end{eqnarray*}
hence, recalling~(80), $(\underline{{\bf Id}}_{\Sigma^{\prime}} 
- \! \sum\limits_{k \in \{A,B,C\}} C_{w^{k^{\prime}}}^{\Sigma^{
\prime}})^{-1} \in {\cal N}(\Sigma^{\prime})$. {}From 
Proposition~5.2 and the above estimates as $t \! \rightarrow \! 
+ \infty$ such that $\lambda_{0} \! > \! M$,
\begin{eqnarray}
&\int_{\Sigma^{\prime}} ((\underline{{\bf Id}}_{\Sigma^{
\prime}} - C_{w^{\prime}}^{\Sigma^{\prime}})^{-1} {\rm I})
(\xi) w^{\Sigma^{\prime}}(\xi) d \xi = \int_{\Sigma^{\prime}}
(\underline{{\bf D}}_{\Sigma^{\prime}} {\rm I}) w^{\Sigma^{
\prime}}(\xi) d \xi + {\cal O}(\frac{\underline{c}}{\lambda_{0} 
t}) + {\cal O} \! \left(\frac{\underline{c}}{(\lambda_{0}^{2} 
t)^{l}} \right) \!,&
\end{eqnarray}
for arbitrary $l \! \in \! \Bbb Z_{\geq 1}$. Recalling that 
$w^{\Sigma^{\prime}} \! = \! \sum\limits_{k \in \{A,B,C\}} 
w^{k^{\prime}}$, the integral on the right-hand side of~(81) 
can be written as follows:
\begin{equation}
\int_{\Sigma^{\prime}} (\underline{{\bf D}}_{\Sigma^{\prime}} 
{\rm I}) w^{\Sigma^{\prime}} (\xi) d \xi = \int_{\Sigma^{\prime}} 
(\underline{{\bf Id}}_{\Sigma^{\prime}} \! \! \sum\limits_{k \in 
\{A,B,C\}} \! w^{k^{\prime}} \! + \! \sum\limits_{\alpha, \beta 
\in \{A,B,C\}} \! \! C_{w^{\alpha^{\prime}}}^{\Sigma^{\prime}} 
(( \underline{{\bf Id}}_{\Sigma^{\prime}} - 
C_{w^{\alpha^{\prime}}}^{\Sigma^{\prime}})^{-1} 
{\rm I}) w^{\beta^{\prime}}) (\xi) d \xi.
\end{equation}
To estimate the right-hand side of (82), consider, say, the 
following integral, $\int_{\Sigma^{\prime}} (C_{w^{A^{
\prime}}}^{\Sigma^{\prime}}$ \linebreak[4] $\cdot (\underline{
{\bf Id}}_{\Sigma^{\prime}} - C_{w^{A^{\prime}}}^{\Sigma^{
\prime}})^{-1} {\rm I}) (\xi) w^{B^{\prime}} (\xi) d \xi$: 
\begin{eqnarray}
&\vert \int_{\Sigma^{\prime}} (C_{w^{A^{\prime}}}^{\Sigma^{
\prime}} (\underline{{\bf Id}}_{\Sigma^{\prime}} - C_{w^{
A^{\prime}}}^{\Sigma^{\prime}})^{-1} {\rm I}) (\xi) w^{B^{
\prime}}(\xi) d \xi \vert \leq \vert \int_{\Sigma^{\prime}} 
(C_{w^{A^{\prime}}}^{\Sigma^{\prime}} {\rm I}) (\xi) w^{B^{
\prime}} (\xi) d \xi \vert& \nonumber \\ 
&+ \vert \int_{\Sigma^{\prime}} (C_{w^{A^{\prime}}}^{\Sigma^{
\prime}} (\underline{{\bf Id}}_{\Sigma^{\prime}} - C_{w^{A^{
\prime}}}^{\Sigma^{\prime}})^{-1} C_{w^{A^{\prime}}}^{\Sigma^{
\prime}} {\rm I}) (\xi) w^{B^{\prime}} (\xi) d \xi \vert \leq 
\vert \int\limits_{\Sigma_{B^{\prime}}} (\int\limits_{\Sigma_{
A^{\prime}}} \frac{ w^{A^{\prime}} (\eta)}{(\eta - \xi)} 
\frac{d \eta}{2 \pi i}) w^{B^{\prime}} (\xi) d \xi \vert& 
\nonumber \\
&+ \vert \int\limits_{\Sigma_{B^{\prime}}} (\int\limits_{
\Sigma_{A^{\prime}}} \frac{((\underline{{\bf Id}}_{\Sigma^{
\prime}} - C_{w^{A^{\prime}}}^{\Sigma^{\prime}})^{-1} C^{
\Sigma^{\prime}}_{w^{A^{\prime}}} {\rm I}) (\eta) w^{A^{
\prime}} (\eta)}{(\eta - \xi)} \frac{d \eta}{2 \pi i}) w^{B^{
\prime}} (\xi) d \xi \vert \leq \frac{ \vert \vert w^{A^{
\prime}} \vert \vert_{{\cal L}^{(2 \times 2)}_{1} (\Sigma_{
A^{\prime}})} \vert \vert w^{B^{\prime}} \vert \vert_{{\cal 
L}^{(2 \times 2)}_{1} (\Sigma_{B^{\prime}})}}{2 \pi \, {\rm 
dist} (\Sigma_{A^{\prime}},\Sigma_{B^{\prime}} )}& \nonumber \\
&+ \frac{ \vert \vert ( \underline{{\bf Id}}_{\Sigma^{\prime}} 
- C_{w^{A^{\prime}}}^{\Sigma^{\prime}})^{-1} C_{w^{A^{\prime}}
}^{\Sigma^{\prime}} {\rm I} \vert \vert_{{\cal L}^{(2 \times 
2)}_{2} (\Sigma_{A^{\prime}})} \vert \vert w^{A^{\prime}} 
\vert \vert_{{\cal L}^{(2\times2)}_{2} (\Sigma_{A^{\prime}})} 
\vert \vert w^{B^{\prime}} \vert \vert_{{\cal L}^{(2 \times 
2)}_{1} (\Sigma_{B^{\prime}})}}{2 \pi \, {\rm dist} (\Sigma_{
A^{\prime}},\Sigma_{B^{\prime}})} &\nonumber \\   
&\leq \frac{\underline{c}}{\lambda_{0}} \vert \vert w^{A^{
\prime}} \vert \vert_{{\cal L}^{(2\times2)}_{1} (\Sigma_{
A^{\prime}})} \vert \vert w^{B^{\prime}} \vert \vert_{{\cal 
L}^{(2\times2)}_{1} (\Sigma_{B^{\prime}})}& \nonumber \\
&+ \frac{\underline{c}}{\lambda_{0}} \vert \vert 
(\underline{{\bf Id}}_{\Sigma^{\prime}} - C_{w^{A^{
\prime}}}^{\Sigma^{\prime}})^{-1} \vert \vert_{{\cal N} 
(\Sigma_{A^{\prime}})} \vert \vert C_{w^{A^{\prime}}}^{
\Sigma^{\prime}} {\rm I} \vert \vert_{{\cal L}^{(2 \times 
2)}_{2} (\Sigma_{A^{\prime}})} \vert \vert w^{A^{\prime}} 
\vert \vert_{{\cal L}^{(2\times2)}_{2} (\Sigma_{A^{\prime}})} 
\vert \vert w^{B^{\prime}} \vert \vert_{{\cal L}^{(2 \times 
2)}_{1} (\Sigma_{B^{\prime}})},& \nonumber 
\end{eqnarray}
since $2 \pi \, {\rm dist}(\Sigma_{A^{\prime}}, \Sigma_{
B^{\prime}}) \! \geq \! \lambda_{0} / \underline{c}$. 
Hence, by Lemma~6.2 and the assumption that $\vert \vert 
(\underline{{\bf Id}}_{\Sigma^{\prime}} - C_{w^{k^{
\prime}}}^{\Sigma^{\prime}})^{-1} \vert \vert_{{\cal N}
(\Sigma^{\prime})} < \infty$, one gets, for $\alpha 
\not= \beta \in \{A,B,C\}$, the following estimation: 
\begin{eqnarray}
&\vert \int_{\Sigma^{\prime}}  
(C_{w^{\alpha^{\prime}}}^{\Sigma^{\prime}} (\underline{
{\bf Id}}_{\Sigma^{\prime}} - C_{w^{\alpha^{\prime}}}^{
\Sigma^{\prime}})^{-1} {\rm I}) (\xi) w^{\beta^{\prime}} 
(\xi) d \xi \vert \leq \frac{\underline{c}}{\lambda_{0}} 
(\vert \vert w^{\prime} \vert \vert_{{\cal L}^{(2 \times 
2)}_{1} (\Sigma^{\prime})})^{2}& 
\nonumber \\
&+ \frac{\underline{c}}{\lambda_{0}} \vert \vert 
(\underline{{\bf Id}}_{\Sigma^{\prime}} - C_{w^{\alpha^{
\prime}}}^{\Sigma^{\prime}})^{-1} \vert \vert_{{\cal N} 
(\Sigma_{\alpha^{\prime}})} (\vert \vert w^{\alpha^{\prime}} 
\vert \vert_{{\cal L}^{(2 \times 2)}_{2} (\Sigma_{\alpha^{
\prime}})})^{2} \vert \vert w^{\beta^{\prime}} \vert 
\vert_{{\cal L}^{(2 \times 2)}_{1} (\Sigma_{\beta^{\prime}})} 
\leq \frac{\underline{c}}{\lambda_{0} t};& 
\nonumber
\end{eqnarray}
therefore, applying the second resolvent identity to the 
right-hand side of (82), one finds that,
\begin{eqnarray}
\int_{\Sigma^{\prime}}((\underline{{\bf Id}}_{\Sigma^{
\prime}} - C_{w^{\prime}}^{\Sigma^{\prime}})^{-1} {\rm 
I})(\xi) w^{\Sigma^{\prime}}(\xi) d \xi & = & \sum\limits_{
k \in \{A,B,C\}} \int_{\Sigma^{\prime}}((\underline{{\bf 
Id}}_{\Sigma^{\prime}} - C_{w^{k^{\prime}}}^{\Sigma^{\prime}
})^{-1} {\rm I})(\xi) w^{k^{\prime}}(\xi) d \xi \nonumber \\
 & + & {\cal O} \! \left(\frac{\underline{c}}{\lambda_{0} 
t} \right) + {\cal O} \! \left(\frac{\underline{c}}{
(\lambda_{0}^{2} t)^{l}} \right) \!,
\end{eqnarray}
for arbitrary $l \! \in \! \Bbb Z_{\geq 1}$. Now, {}from 
Definition~5.2 and Lemma~5.3 (identity~(60)), 
\begin{eqnarray}
(\underline{{\bf Id}}_{\Sigma_{k^{\prime}}} - C_{w^{k^{
\prime}}}^{\Sigma_{k^{\prime}}} )^{-1} = R_{\Sigma_{k^{
\prime}}} (\underline{{\bf Id}}_{\Sigma^{\prime}} - C_{
w^{k^{\prime}}}^{\Sigma^{\prime}})^{-1} \underline{{\bf 
I}}_{\Sigma^{\prime}_{k^{\prime}} \rightarrow \Sigma}; 
\end{eqnarray}
hence, substituting identity~(84) into (83), and recalling 
(80) and (81), the proof is complete. \ \ \rule{6pt}{6pt}

\setcounter{hme}{3}
\begin{hme}
For $k \in \{A,B,C\}$, $(\underline{{\bf Id}}_{\Sigma_{
k^{\prime}}} - C_{w^{k^{\prime}}}^{\Sigma_{k^{\prime}}})^{
-1} \in {\cal N} (\Sigma_{k^{\prime}})$.
\end{hme}

{\em Remark 6.1.\/} This result was proved in \cite{a15}: 
here, we briefly reproduce this proof since we need a model 
RH problem which arises in it in order to obtain the explicit 
asymptotic formulae presented in Theorem~3.2. 

{\em Proof.\/} Consider, say, the case $k \! = \! B$: the 
cases $k \! = \! A$ and $C$ follow in an analogous manner. 
Define the function (see Fig.~5(a))
\begin{eqnarray*}
&w_{\pm}^{B^{0}} (\lambda) \equiv ((\Delta_{B}^{0})^{-1} 
(N_{B} \widehat{w}_{\pm}^{B^{\prime}}) (\Delta_{B}^{0})) 
(\lambda) = \pm \widehat{R}^{\pm} (\lambda_{0}) (2 
\lambda)^{\pm 2 i \nu} e^{\mp i \lambda^{2}} \sigma_{
\pm},& \nonumber 
\end{eqnarray*}
where $\widehat{w}^{B^{\prime}}_{\pm}(\lambda)$ is given 
in (73), and
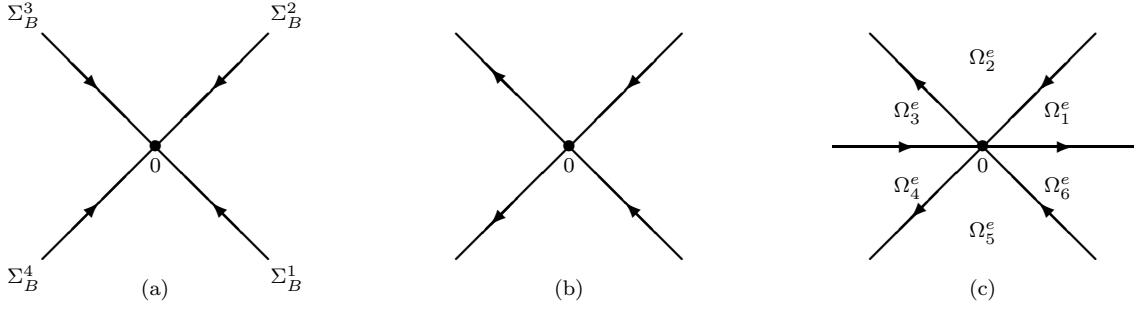
\begin{figure}[bht]
\begin{center}
\unitlength=1cm
\vspace{-0.50cm}
\begin{picture}(15,4)(-0.325,0)
\thicklines
\put(2,2){\line(1,1){0.75}}
\put(3.5,3.5){\vector(-1,-1){0.75}}
\put(2,2){\line(-1,1){0.75}}
\put(0.5,3.5){\vector(1,-1){0.75}}
\put(2,2){\line(1,-1){0.75}}
\put(3.5,0.5){\vector(-1,1){0.75}}
\put(2,2){\line(-1,-1){0.75}}
\put(0.5,0.5){\vector(1,1){0.75}}
\put(3.75,3.75){\makebox(0,0){$\scriptstyle{}\Sigma_{B}^{2}$}}
\put(3.75,0.25){\makebox(0,0){$\scriptstyle{}\Sigma_{B}^{1}$}}
\put(0.25,3.75){\makebox(0,0){$\scriptstyle{}\Sigma_{B}^{3}$}}
\put(0.25,0.25){\makebox(0,0){$\scriptstyle{}\Sigma_{B}^{4}$}}
\put(2,1.75){\makebox(0,0){$\scriptstyle{}0$}}
\put(2,2){\makebox(0,0){$\bullet$}}
\put(2,0.10){\makebox(0,0){$\scriptstyle{}{\rm (a)}$}}
\put(7.5,2){\line(1,1){0.75}}
\put(9.0,3.5){\vector(-1,-1){0.75}}
\put(7.5,2){\line(1,-1){0.75}}
\put(9.0,0.5){\vector(-1,1){0.75}}
\put(7.5,2){\vector(-1,1){1.0425}}
\put(6,3.5){\line(1,-1){0.75}}
\put(7.5,2){\vector(-1,-1){1.0425}}
\put(6,0.5){\line(1,1){0.75}}
\put(7.5,2.0){\makebox(0,0){$\bullet$}}
\put(7.5,1.75){\makebox(0,0){$\scriptstyle{}0$}}
\put(7.5,0.10){\makebox(0,0){$\scriptstyle{}{\rm (b)}$}}
\put(11,2){\vector(1,0){1.1}}
\put(12,2){\line(1,0){1.0}}
\put(13,2){\vector(1,0){1.2}}
\put(14,2){\line(1,0){1}}
\put(13,2){\line(1,1){0.75}}
\put(14.5,3.5){\vector(-1,-1){0.75}}
\put(13,2){\line(1,-1){0.75}}
\put(14.5,0.5){\vector(-1,1){0.75}}
\put(13,2){\vector(-1,1){0.951}}
\put(11.5,3.5){\line(1,-1){0.75}}
\put(13,2){\vector(-1,-1){0.951}}
\put(11.5,0.5){\line(1,1){0.75}}
\put(13,1.75){\makebox(0,0){$\scriptstyle{}0$}}
\put(13,2){\makebox(0,0){$\bullet$}}
\put(14,2.45){\makebox(0,0){$\scriptstyle{}\Omega_{1}^{e}$}}
\put(13,3.15){\makebox(0,0){$\scriptstyle{}\Omega_{2}^{e}$}}
\put(12,2.45){\makebox(0,0){$\scriptstyle{}\Omega_{3}^{e}$}}
\put(12,1.45){\makebox(0,0){$\scriptstyle{}\Omega_{4}^{e}$}}
\put(13,0.85){\makebox(0,0){$\scriptstyle{}\Omega_{5}^{e}$}}
\put(14,1.45){\makebox(0,0){$\scriptstyle{}\Omega_{6}^{e}$}}
\put(13,0.10){\makebox(0,0){$\scriptstyle{}{\rm (c)}$}}
\end{picture}
\vspace{-0.65cm}
\end{center}
\caption{(a) $\Sigma_{B}$; (b) $\Sigma_{B,r}$; and (c) 
$\Sigma_{ex} \equiv \Sigma_{B,r} \cup \Bbb R$.}
\end{figure}
\begin{eqnarray}
&\widehat{R}^{+} (\lambda_{0})
= \left\{ \begin{array}{l}
R(\lambda_{0}^{+}), \, \, \, \, \, \lambda \in 
\Sigma_{B}^{1}, \\
R(\lambda_{0}^{-}), \, \, \, \, \, \lambda \in 
\Sigma_{B}^{3}, 
\end{array} \right. \, \, \, 
{\rm and} \, \, \, \, \, \, 
\widehat{R}^{-}(\lambda_{0})
= \left\{ \begin{array}{l}
\overline{R(\lambda_{0}^{+})}, \, \, \, \, \, 
\lambda \in \Sigma_{B}^{2}, \\
\overline{R(\lambda_{0}^{-})}, \, \, \, \, \, 
\lambda \in \Sigma_{B}^{4}. 
\end{array} \right.& \nonumber
\end{eqnarray}
Now, defining as usual $w^{B^{0}} = w^{B^{0}}_{+} + w^{B^{
0}}_{-}$, and using Lemma~6.1, one finds that,
\begin{eqnarray}
&\vert \vert w^{B} - w^{B^{0}} \vert \vert_{\aleph} = \vert 
\vert (\Delta_{B}^{0})^{-1} (N_{B} \widehat{w}^{B^{\prime}}) 
(\Delta_{B}^{0}) - (\Delta_{B}^{0})^{-1} (N_{B} \widehat{w}^{
B^{0}}) (\Delta_{B}^{0}) \vert \vert_{\aleph} \leq \frac{
c^{{\cal S}} \ln t}{\sqrt{\lambda_{0}^{2} t}},& 
\end{eqnarray}
where $\aleph \! \equiv \! {\cal L}^{(2 \times 2)}_{p}
(\Sigma_{B})$, $p \! \in \! \{1,2,\infty\}$. Hence, as $t \! 
\rightarrow \! + \infty$ such that $\lambda_{0} \! > \! M$,
\begin{eqnarray}
&\vert \vert C_{w^{B}}^{\Sigma_{B}} - C_{w^{B^{0}}}^{
\Sigma_{B}} \vert \vert_{{\cal L}^{(2 \times 2)}_{2}
(\Sigma_{B})} \leq \frac{c^{{\cal S}} \ln t}{\sqrt{
\lambda_{0}^{2} t}},& 
\end{eqnarray}
and, consequently ({}from the second resolvent identity), 
one sees that, for sufficiently large $t$, $(\underline{
{\bf Id}}_{\Sigma_{B}} - C_{w^{B^{0}}}^{\Sigma_{B}} )^{-1} 
\in {\cal N}(\Sigma_{B}) \Rightarrow (\underline{{\bf 
Id}}_{\Sigma_{B}} - C_{w^{B}}^{\Sigma_{B}} )^{-1} \in 
{\cal N} (\Sigma_{B})$. Reorient $\Sigma_{B}$ as in 
Fig.~5(b) and define $w^{B,r}(\lambda) = w_{+}^{B,r}
(\lambda) + w_{-}^{B,r}(\lambda)$, 
where 
\begin{eqnarray}
&w_{\pm}^{B,r}(\lambda) = w_{\pm}^{B^{0}}(\lambda), \, \, \, \,
\, \Re (\lambda) > 0, \, \, \, \, \, \, {\rm and} \, \, \, \, \,  
\, w_{\pm}^{B,r}(\lambda) = - w_{\mp}^{B^{0}}(\lambda), \, \, \, 
\, \, \Re (\lambda) < 0,& \nonumber
\end{eqnarray}
so that one can consider the operator $C_{w^{B,r}}^{\Sigma_{
B,r}} = C_{+}(\cdot w_{-}^{B,r}) + C_{-} (\cdot w_{+}^{B,r})$, 
where the Cauchy operators are now taken with respect to 
$\Sigma_{B,r}$. Extend $\Sigma_{B,r} \rightarrow \Sigma_{ex} 
\equiv \Sigma_{B,r} \cup \Bbb R$, with the orientation in 
Fig.~5(c), and set 
\begin{eqnarray}
w^{ex} (\lambda) \equiv \left\{ \begin{array}{l}
w^{B,r} (\lambda), \, \, \, \, \, \lambda \in \Sigma_{B,r}, 
\\
0, \, \, \, \, \, \lambda \in \Sigma_{ex} \setminus 
\Sigma_{B,r}.
\end{array} \right. \nonumber
\end{eqnarray}
{}From Lemma~5.3, it follows that $(\underline{{\bf Id}}_{
\Sigma_{ex}} - C_{w^{ex}}^{\Sigma_{ex}} )^{-1} \in {\cal N} 
(\Sigma_{ex}) \Rightarrow (\underline{{\bf Id}}_{\Sigma_{B,r}} 
- C_{w^{B,r}}^{\Sigma_{B,r}})^{-1}$ $\in$ ${\cal N} (\Sigma_{
B,r})$ $\Rightarrow (\underline{{\bf Id}}_{\Sigma_{B}} - C_{
w^{B^{0}}}^{\Sigma_{B}})^{-1} \in {\cal N}(\Sigma_{B})$. Define 
a piecewise analytic $2 \times 2$ matrix-valued function 
$\phi(\lambda)$ as follows,
\begin{eqnarray}
&\phi(\lambda) \equiv \left\{ \begin{array}{l}
(2 \lambda)^{- i \nu \sigma_{3} }, \, \, \, \, \, \lambda 
\in \Omega_{2}^{e} \cup \Omega_{5}^{e}, \\
(2 \lambda)^{- i \nu \sigma_{3} } ({\rm I} - w^{ex}_{-} 
(\lambda))^{-1}, \, \, \, \, \, \lambda \in \Omega_{1}^{e} 
\cup \Omega_{4}^{e}, \\
(2 \lambda)^{- i \nu \sigma_{3}} ({\rm I} + w^{ex}_{+} 
(\lambda))^{-1}, \, \, \, \, \, \lambda \in \Omega_{3}^{e} 
\cup \Omega_{6}^{e}, 
\end{array} \right.&
\end{eqnarray}
where ${\rm I} \pm w_{\pm}^{ex}(\lambda)$ are, respectively, 
the analytic continuation of ${\rm I} \pm w^{ex}(\lambda)$ 
into $\Omega_{1}^{e}$, $\Omega_{3}^{e}$, $\Omega_{4}^{e}$, 
and $\Omega_{6}^{e}$. {}From the estimates above for $w^{B^{
0}}_{\pm}(\lambda)$, one finds that,
\begin{eqnarray}
&w^{ex}_{-}(\lambda) \rightarrow 0 \, \, \, \, \, {\rm as} 
\, \, \, \, \, \lambda \rightarrow \infty, \, \, \, \, \, \, 
\lambda \in \Omega_{1}^{e} \cup \Omega_{4}^{e},& \nonumber \\
&w_{+}^{ex}(\lambda) \rightarrow 0 \, \, \, \, \, {\rm as} \, 
\, \, \, \, \lambda \rightarrow \infty, \, \, \, \, \, \lambda 
\in \Omega_{3}^{e} \cup \Omega_{6}^{e}.& \nonumber 
\end{eqnarray}
For $\lambda \in \Sigma_{ex}$, set $V^{e,\phi} (\lambda) = 
\phi_{-} (\lambda) ({\rm I} - w_{-}^{ex})^{-1} 
({\rm I} + w_{+}^{ex}) \phi_{+}^{-1}(\lambda) \equiv 
( {\rm I} - w_{-}^{ex,\phi} )^{-1} ({\rm I} + w_{+}^{ex,
\phi})$, where $\phi_{\pm}(\lambda)$ are the non-tangential 
limits of $\phi(\lambda)$ as $\lambda$ approaches $\Sigma_{
ex}$ {}from the ``$\pm$'' side, respectively (Remark~2.1). For 
$\lambda \in \Sigma_{B,r}$, $V^{e,\phi} (\lambda) = {\rm I}$, 
and, for $\lambda \in \Sigma_{ex} \setminus \Sigma_{B,r}$, 
$V^{e,\phi}(\lambda) = \phi_{-}(\lambda) \phi_{+}^{-1}
(\lambda)$, whence, one gets that,
\begin{eqnarray}
&V^{e,\phi}(\lambda) = \exp \left\{- \frac{i \lambda^{2}}{2} 
{\rm ad} (\sigma_{3}) \right\} v(\lambda_{0}), \, \, \, \, \, 
\, \Re (\lambda) < 0,& \\
&V^{e,\phi}(\lambda) = \exp \left\{- \frac{i \lambda^{2}}{2} 
{\rm ad} (\sigma_{3}) \right\} v(\lambda_{0}), \, \, \, \, \, 
\, \Re (\lambda) > 0,& 
\end{eqnarray} 
where
\begin{eqnarray}
&v(\lambda_{0}) = \left(\! \begin{array}{cc}
1 - \vert r(\lambda_{0}) \vert^{2} & r(\lambda_{0}) \\
- \overline{r(\lambda_{0})} & 1
\end{array} \! \right) \!, \, \, \, \, \, \, 
\det (v(\lambda_{0})) = 1.&
\end{eqnarray}
Hence, the jump matrix, $V^{e,\phi}(\lambda)$, can be 
characterised as follows:
\begin{eqnarray}
&V^{e,\phi}(\lambda) = \left\{ \begin{array}{l}
{\rm I}, \, \, \, \, \, \lambda \in \Sigma_{B,r}, \\
\exp \{- \frac{i \lambda^{2}}{2} {\rm ad} (\sigma_{3}) \} 
v(\lambda_{0}), \, \, \, \, \, \lambda \in \Sigma_{ex} 
\setminus \Sigma_{B,r}.
\end{array} \right.& \nonumber
\end{eqnarray}
On $\Bbb R$, one has that,
\begin{eqnarray}
&V^{e,\phi}(\lambda) = ({\rm I} - w_{-}^{ex,\phi})^{-1} ({\rm 
I} + w_{+}^{ex,\phi}) = ({\rm I} - \overline{r(\lambda_{0})} 
e^{i \lambda^{2}} \sigma_{-}) ({\rm I} + r(\lambda_{0}) e^{- 
i \lambda^{2}} \sigma_{+}).& \nonumber  
\end{eqnarray}
Set $C_{e,\phi} = C_{+}(\cdot w_{-}^{ex,\phi}) + C_{-}(\cdot 
w_{+}^{ex,\phi})$ as the associated operator on $\Sigma_{ex}$, 
with $w^{ex,\phi} = w_{+}^{ex,\phi} + w_{-}^{ex,\phi}$. By 
Lemma~5.3, the boundedness of $C_{e,\phi}$, i.e., $\vert \vert 
(\underline{{\bf Id}}_{\Sigma_{ex}} - C_{w^{ex,\phi}}^{\Sigma_{
ex}})^{-1} \vert \vert_{{\cal N}(\Sigma_{ex})}$, follows {}from
the boundedness of the operator $C_{w^{ex,\phi} \vert_{\Bbb R}} 
\colon {\cal L}^{(2 \times 2)}_{2}(\Bbb R) \rightarrow {\cal 
L}^{(2 \times 2)}_{2} (\Bbb R)$ associated with the restriction 
of $w^{ex,\phi}$ to $\Bbb R$; but $\vert \vert C_{w^{ex,\phi}
\vert_{\Bbb R}} \vert \vert_{{\cal N}(\Bbb R)} \leq 
\sup\limits_{\lambda \in \widehat{\Gamma}} \vert r(\lambda_{0}) 
\exp \{- i \lambda^{2}\} \vert \leq \vert \vert r \vert 
\vert_{{\cal L}^{\infty}(\widehat{\Gamma})} < 1$, and hence, 
by the second resolvent identity,
\begin{eqnarray}
&\vert \vert (\underline{{\bf Id}}_{\Bbb R} - C_{w^{ex,\phi} 
\vert_{\Bbb R}})^{-1} \vert \vert_{{\cal N} (\Bbb R)} \leq (1 
- \vert \vert r \vert \vert_{{\cal L}^{\infty} (\widehat{
\Gamma})})^{-1} \, < \, \infty \, \Rightarrow& 
\nonumber \\
&(\underline{{\bf Id}}_{\Bbb R} - C_{w^{ex,\phi} \vert_{\Bbb 
R}}^{\Sigma_{ex}})^{-1} \in {\cal N} (\Bbb R) \Rightarrow 
(\underline{{\bf Id}}_{\Sigma_{ex}} - C_{w^{ex,\phi}}^{
\Sigma_{ex}})^{-1} \in {\cal N}(\Sigma_{ex}):& \nonumber
\end{eqnarray}
this completes the proof. \ \ \rule{6pt}{6pt}
\section{Model RH Problem}
In this section, we reduce the evaluation of the integrals 
in Lemma~6.3 to three RH problems on $\Bbb R$ which can be 
solved explicitly. 

For $k \in \{A,B,C\}$, define
\begin{eqnarray*}
m^{k^{0}}(\lambda) = {\rm I} + \int_{\Sigma_{k}} \frac{((
\underline{{\bf Id}}_{\Sigma_{k}} - C_{w^{k^{0}}}^{\Sigma_{k}}
)^{-1} {\rm I}) (\xi) w^{k^{0}} (\xi)}{(\xi - \lambda)} \frac{
d \xi}{2 \pi i}, \, \, \, \, \, \lambda \in \Bbb C \setminus 
\Sigma_{k}. 
\end{eqnarray*}
Now, {}from Theorem~3.1, we find that $m^{k^{0}}(\lambda)$ solves 
the following RH problem,
\begin{eqnarray}
&\left. \begin{array}{c}
m_{+}^{k^{0}}(\lambda) = m_{-}^{k^{0}}(\lambda) ({\rm I} 
- w_{-}^{k^{0}} )^{-1} ( {\rm I} + w_{+}^{k^{0}}), 
\, \, \, \, \, \, \lambda \in \Sigma_{k}, \\
m^{k^{0}} (\lambda) = {\rm I} - \frac{m_{1}^{k^{0}}}{\lambda} 
+ {\cal O}(\lambda^{-2}) \, \, \, \, \, {\rm as} \, \, \, \, 
\, \lambda \rightarrow \infty, \, \, \, \, \, \, \lambda \in 
\Bbb C \setminus \Sigma_{k}.
\end{array} \right.& 
\nonumber
\end{eqnarray}
Substituting into (79) of Lemma~6.3 inequalities~(85) and 
(86) (and their analogues for $\Sigma_{A}$ and $\Sigma_{C}$), 
we obtain
\begin{eqnarray}
&P(x,t) \! = \! \frac{i}{\sqrt{4 \lambda_{0}^{2} t}} 
\{(\delta_{A}^{0})^{-2} (m_{1}^{A^{0}})_{21} \! + \! 
(\delta_{B}^{0})^{-2} (m_{1}^{B^{0}})_{21} \! + \! 
\sqrt{2} (\delta_{C}^{0})^{-2} (m_{1}^{C^{0}})_{21}\} 
\! + \! {\cal O} \! \left( \frac{\underline{c} \ln 
t}{\lambda_{0} t} \right) \!.&
\end{eqnarray}
We consider in detail only case $B$. Let us introduce the 
function
$$
{\cal D}(\lambda) = m^{B^{0}}(\lambda) (\phi (\lambda))^{-1}
e^{- \frac{i \lambda^{2}}{2} \sigma_{3}}, \, \, \, \, \, \,
\lambda \in \Bbb C \setminus \Sigma_{ex},
$$
where $\phi(\lambda)$ is defined in (87), and notice that it 
is holomorphic in $\Bbb C \setminus \Bbb R$; in particular, 
it has no jumps across $\Sigma_{ex} \setminus \Bbb R$. 
Across $\Bbb R$, oriented {}from $- \infty$ to $+ \infty$, using 
(87)--(90), one finds that ${\cal D} (\lambda)$ solves the 
following RH problem: 
\begin{eqnarray}
&{\cal D}_{+}(\lambda) = {\cal D}_{-}(\lambda) v(\lambda_{0}), 
\, \, \, \, \, \lambda \in \Bbb R,& \\
&{\cal D}(\lambda) = (({\rm I} - \frac{ m_{1}^{B^{0}}}{\lambda}) 
+ {\cal O} (\lambda^{-2})) (2 \lambda)^{i \nu \sigma_{3}} e^{- 
\frac{i \lambda^{2}}{2} \sigma_{3}} \, \, \, \, \, {\rm as} \, 
\, \, \, \, \lambda \rightarrow \infty, \, \, \, \, \lambda \in 
\Bbb C \setminus \Bbb R. 
\end{eqnarray}
Since, according to (92) and (93), $\det({\cal D}(\lambda))$ is a 
holomorphic, bounded in $\Bbb C$ function, it is, by Liouville's
theorem, a constant. {}From (92), one finds that, $\det({\cal D} 
(\lambda)) = 1$. By differentiating (93), we see that $\partial_{
\lambda} {\cal D}(\lambda)$ also solves (92), so that $\partial_{
\lambda} {\cal D}(\lambda) ({\cal D}(\lambda))^{-1}$ is an entire 
function of $\lambda$. Using (93), 
\begin{eqnarray}
\partial_{\lambda} {\cal D}(\lambda)({\cal D}(\lambda))^{-1}
= - i \lambda \sigma_{3} - i [\sigma_{3},m_{1}^{B^{0}}] + 
{\cal O}(\lambda^{-1}). \nonumber
\end{eqnarray}
Applying Liouville's theorem to the left-hand side of the last 
equation, we arrive at the following ODE for ${\cal D}(\lambda)$,
\begin{eqnarray}
&(\partial_{\lambda} {\cal D} + i \lambda \sigma_{3} {\cal D}) = 
\beta {\cal D},& 
\end{eqnarray}
where
\begin{eqnarray}
\beta = - i [\sigma_{3},m_{1}^{B^{0}}] = \beta_{21} \sigma_{-} 
+ \beta_{12} \sigma_{+}; \nonumber
\end{eqnarray}
hence, 
\begin{eqnarray}
&(m_{1}^{B^{0}})_{21} = - \frac{i}{2} \beta_{21}.& \nonumber
\end{eqnarray}
It is well known \cite{a9,a26,a27} that the fundamental solution 
of (94) can be written in terms of the parabolic-cylinder 
function, $D_{a}(\cdot)$ \cite{a24}; in particular,  
\begin{eqnarray}
&\left. \begin{array}{c}
{\cal D}_{+11}(\lambda) = 2^{\frac{i \nu}{2}}
e^{- \frac{ 3 \pi \nu}{4}} D_{i \nu} (\sqrt{2} \lambda 
e^{- \frac{3 \pi i}{4}}), \, \, \, \, \, {\cal D}_{+22}
(\lambda) = 2^{-\frac{i \nu}{2}} e^{ \frac{\pi \nu}{4}} 
D_{- i \nu} (\sqrt{2} \lambda e^{- \frac{i \pi}{4}}), \\ 
{\cal D}_{+12}(\lambda) = (\beta_{21})^{-1} 2^{-\frac{i 
\nu}{2}} e^{\frac{ \pi \nu}{4}} \{\partial_{\lambda} D_{-i 
\nu} (\sqrt{2} \lambda e^{- \frac{i \pi}{4}}) - i \lambda 
D_{-i \nu} (\sqrt{2} \lambda e^{- \frac{i \pi}{4}}) \}, \\
{\cal D}_{+21} (\lambda) = (\beta_{12})^{-1} 2^{\frac{i 
\nu}{2}} e^{- \frac{3 \pi \nu}{4}} \{ \partial_{
\lambda} D_{i \nu} (\sqrt{2} \lambda e^{- \frac{3 \pi 
i}{4}}) + i \lambda D_{i \nu} (\sqrt{2} \lambda e^{- \frac{3 
\pi i}{4}}) \},
\end{array} \right.&
\end{eqnarray}
and
\begin{eqnarray}
&\left. \begin{array}{c}
{\cal D}_{-11} (\lambda) = 2^{\frac{i \nu}{2}}
e^{ \frac{ \pi \nu }{ 4 } } D_{i \nu} ( \sqrt{2} 
\lambda e^{ \frac{ i \pi }{4} } ), \, \, \, \, \, 
{\cal D}_{-22} (\lambda) = 2^{-\frac{i \nu}{2}} 
e^{- \frac{ 3 \pi \nu }{ 4 } } D_{- i \nu} ( \sqrt{2} 
\lambda e^{ \frac{3 \pi i}{4}}), \\
{\cal D}_{-12} (\lambda) = (\beta_{21})^{-1} 2^{-\frac{
i \nu}{2}} e^{ - \frac{ 3 \pi \nu }{ 4 } } 
\{ \partial_{\lambda} D_{-i \nu} ( \sqrt{2} \lambda 
e^{ \frac{ 3 \pi i }{4} } ) - i \lambda D_{-i \nu} 
(\sqrt{2} \lambda e^{ \frac{ 3 \pi i }{4} } ) \}, \\
{\cal D}_{-21} (\lambda) = (\beta_{12})^{-1} 2^{\frac{i 
\nu}{2}} e^{ \frac{ \pi \nu }{ 4 } } 
\{ \partial_{\lambda} D_{i \nu} ( \sqrt{2} \lambda e^{ 
\frac{ i \pi }{4} } ) + i \lambda D_{i \nu} (\sqrt{2} 
\lambda e^{ \frac{ i \pi }{4} } ) \},
\end{array} \right.&
\end{eqnarray}
where $\beta_{12} \beta_{21} = 2 \nu$. Now, substituting 
(95) and (96) into (92), we find that  
\begin{eqnarray}
&- \beta_{12} \overline{r(\lambda_{0})} = 2^{i \nu} 
e^{ - \frac{\pi \nu}{2}} W(D_{ i \nu} (\sqrt{2} \lambda 
e^{ \frac{i \pi }{4}}),D_{i \nu} (\sqrt{2} \lambda e^{- 
\frac{3 \pi i}{4}})),& \nonumber
\end{eqnarray}
where $W(f,g)$ $(\equiv f(\lambda) \frac{d g(\lambda)}{
d \lambda} - g(\lambda) \frac{d f(\lambda)}{d \lambda})$ 
is the Wronskian of $f \! = \! D_{i \nu}(\sqrt{2} \lambda 
e^{\frac{i \pi}{4}})$ and $g \! = \! D_{i \nu}(\sqrt{2} 
\lambda e^{- \frac{3 \pi i}{4}})$,
\begin{eqnarray} 
&W(D_{i \nu} (\sqrt{2} \lambda e^{\frac{i \pi}{4}}),D_{i 
\nu} (\sqrt{2} \lambda e^{- \frac{3 \pi i}{4}})) = \frac{2 
\sqrt{\pi} e^{\frac{i \pi}{4}}}{\Gamma(- i \nu)},& 
\nonumber
\end{eqnarray}
and $\Gamma(\cdot)$ is the gamma function \cite{a24}; hence, 
\begin{eqnarray}
(m_{1}^{B^{0}})_{21} = \frac{2^{- i \nu} \sqrt{\pi} 
e^{- \frac{\pi \nu}{2}} e^{\frac{i \pi}{4}}}{\Gamma (i \nu) 
r(\lambda_{0})}.
\end{eqnarray}
{}From the $\sigma_{3}$ symmetry reduction for $m(\lambda)$, 
i.e., $m(-\lambda) = \sigma_{3} 
m(\lambda) \sigma_{3}$, we have that,
\begin{eqnarray}
(m_{1}^{A^{0}})_{21} = (m_{1}^{B^{0}})_{21}.
\end{eqnarray}
In the case $C$, $r(0) = r(i 0) = 0$, so that the corresponding
model RH problem~(92)--(93) $(\lambda_{0} = 0)$ has identity 
conjugation matrix, $v(0) = {\rm I}$, whence,
\begin{eqnarray} 
&(m_{1}^{C^{0}})_{21} = 0.& 
\end{eqnarray}
Hence, substituting (71), (97), (98), and (99) into (91), 
and recalling that $Q(x,t)$ $=$ $\overline{P(x,t)}$, we 
obtain (16), (18), (19), and (21) in Theorem~3.2.
\section{Asymptotic Evaluation of $m(x,t;0)$}
In order to complete the proofs for $q(x,t)$ and $u(x,t)$ 
given in Theorems~3.2 and 3.3, we need to evaluate $m(x,t;
0)$ asymptotically (Lemma~2.1, (11)). 
\setcounter{dva}{0}
\begin{dva}
If $Q(x,0) \in {\cal S} (\Bbb R)$, then $m(x,t;0)$ $=$ $\exp 
\{- \frac{i \sigma_{3}}{2} \int_{+ \infty}^{x} \vert Q(\xi,t) 
\vert^{2} d \xi\}$. 
\end{dva}

{\em Proof.\/} {}From the proof of Proposition~2.2, it follows 
that, $m(x,t;0) = \Psi (x,t;0) =$ $\exp \{- \frac{i 
\sigma_{3}}{2} \int_{x_{0}}^{x} \vert Q(\xi,t) \vert^{2} d \xi\}$, 
for some $x_{0} \in \Bbb R$. Since $Q(x,0) \in {\cal S}(\Bbb R)$, 
it follows {}from Definition~2.1, Proposition~2.5, and the 
definitions of $m(x,t;\lambda)$ for $\Im(\lambda^{2}) {> \atop 
<} 0$ given in Lemma~2.1 that, $\lim\limits_{x \rightarrow + 
\infty} \! m(x,t;0)$ $= {\rm I}$: note also that, $m_{+}(x,t;0) 
= m_{-}(x,t;0) \equiv m(x,t;0)$, since $G(0) = {\rm I}$ 
(Lemma~2.1); hence, $\exp \{- \frac{i \sigma_{3}}{2} \int_{x_{
0}}^{+ \infty} \vert Q(\xi,t) \vert^{2} d \xi\} = {\rm I}$. 
\ \ \rule{6pt}{6pt}

\setcounter{lna}{1}
\begin{lna}
\begin{eqnarray*}
&(\vert \vert Q \vert \vert_{{\cal L}^{2}(\Bbb R)})^{2} 
= \frac{2}{\pi} \left(\int_{0}^{+ \infty} \frac{\ln (1 
+ \vert r(i \mu) \vert^{2})}{\mu} d \mu - \int_{0}^{+ 
\infty} \frac{\ln (1 - \vert r(\mu) \vert^{2})}{\mu} d 
\mu \right) \!.& 
\end{eqnarray*}
\end{lna}

{\em Proof.\/} For $Q(x,0) \in {\cal S}(\Bbb R)$, {}from 
Definition~2.1, Proposition~2.5, and the definitions of 
$m(x,t;\lambda)$ for $\Im(\lambda^{2}) {> \atop <} 0$ given 
in Lemma~2.1, one shows that, $\lim\limits_{x \rightarrow 
- \infty} \! m(x,t;0) = (a^{-}(0))^{\sigma_{3}}$, where 
$a^{-}(0) = \exp \{\int_{\widehat{\Gamma}} \frac{\ln (1 + 
r^{+} (\mu) r^{-} (\mu))}{\mu} \frac{d \mu}{2 \pi i} \}$: 
the proof now follows {}from Proposition~2.5 $(r^{+}(\mu) 
= - \overline{r^{-}(\overline{\mu})})$, Lemma~2.1 $(r(\mu)
\equiv r^{-}(\mu))$, and Proposition~8.1. \ \ \rule{6pt}{6pt}

\setcounter{vly}{0}
\begin{vly}
As $t \! \rightarrow \! + \infty$ and $x \! \rightarrow \! 
- \infty$ such that $\lambda_{0} \! \equiv \! \frac{1}{2} 
\sqrt{- \frac{x}{t}} \! > \! M$,
\begin{eqnarray*}
&((m^{-1}(x,t;0))_{11})^{2} = \exp \{\frac{2 i}{\pi} 
(\int_{0}^{\lambda_{0}} \frac{\ln (1 - \vert r (\mu) 
\vert^{2})}{\mu} d \mu - \int_{0}^{+ \infty} \frac{\ln 
(1 + \vert r (i \mu) \vert^{2})}{\mu} d \mu)\} + {\cal 
O} \! \left( \frac{\underline{c} \ln t}{\sqrt{t}} 
\right) \!.& 
\end{eqnarray*}
\end{vly}

{\em Proof.\/} Writing $\int_{+ \infty}^{x} \vert 
Q(\xi,t) \vert^{2} d \xi = - (\vert \vert Q \vert 
\vert_{{\cal L}^{2}(\Bbb R)})^{2} + \int_{- \infty}^{
x} \vert Q(\xi,t) \vert^{2} d \xi$, the proof follows 
{}from the asymptotic expansion of $Q(x,t)$ given in 
Theorem~3.2, (16), (18), (19), and (21), the fact 
that $r(\lambda) \in {\cal S}(\widehat{\Gamma})$, 
and Proposition~8.2. \ \ \rule{6pt}{6pt}

{\em Remark 8.1.\/} It is possible to prove the estimate 
in Lemma~8.1, ${\cal O} \! \left( \frac{\underline{c} \ln 
t}{\sqrt{t}} \right)$, without reference to the conserved 
quantity in Proposition~8.2 by using only the asymptotic 
results for $Q(x,t)$: this fact was first pointed out by 
Ablowitz and Segur \cite{a9} in connection with their 
studies of the NLSE. To follow this paradigm, one needs the 
asymptotic expansion of $Q(x,t)$ when $\lambda_{0} \! \simeq 
\! 0$; however, we did not prove the aforementioned asymptotics 
for $Q(x,t)$ (Remark~3.3). 

\setcounter{cheterye}{0}
\begin{cheterye}
As $t \! \rightarrow \! + \infty$ and $x \! \rightarrow \! 
- \infty$ such that $\lambda_{0} \! \equiv \! \frac{1}{2} 
\sqrt{- \frac{x}{t}} \! > \! M$, the solution of the DNLSE 
is given in Theorem~3.2, Eqs.~(17)--(19) and (22).
\end{cheterye}

{\em Proof.\/} Consequence of (11) in Lemma~2.1 and Lemma~8.1. 
\ \ \rule{6pt}{6pt}

\setcounter{pyaty}{1}
\begin{pyaty}
As $t \! \rightarrow \! + \infty$ and $x \! \rightarrow \! 
+ \infty$ such that $\widehat{\lambda}_{0} \! \equiv \! 
\sqrt{\frac{1}{2}(\frac{x}{t} \! - \! \frac{1}{s})} \! > \! 
M$ and $\frac{x}{t} \! > \! \frac{1}{s}$, $s \! \in \! \Bbb 
R_{> 0}$, the solution of the MNLSE is given in Theorem~3.3, 
Eqs.~(32) and (33).
\end{pyaty}

{\em Proof.\/} Follows {}from Proposition~2.4 and 
Corollary~8.1. \ \ \rule{6pt}{6pt}
\vspace{1.0cm}
\begin{flushleft}
{\LARGE {\bf Acknowledgements}}
\end{flushleft}
The authors are grateful to the II. Institut f\"{u}r 
Theoretische Physik of Hamburg University, where the 
final stage of this work was completed, for hospitality. 

\end{document}